\documentclass{aa}  

\usepackage{graphicx}
\usepackage{txfonts}
\usepackage[usenames,dvipsnames,svgnames,table]{xcolor}
\usepackage{multirow}
\usepackage[
hyperindex=true,
colorlinks=true,
linkcolor=Sepia,
citecolor=Purple]{hyperref}
\usepackage{caption}
\usepackage{subcaption}
\usepackage{booktabs}
\usepackage{ulem}

\newcommand{\Msun}{\mathrm{M}_\odot}

\newcommand{\E}[1]{\times 10^{#1}}

\makeatletter
\renewcommand*\aa@pageof{, page \thepage{} of \pageref*{LastPage}}
\makeatother

\begin{document} 

   \title{Cloud properties across spatial scales in simulations of the interstellar medium}

   \author{Tine Colman
          \inst{1}\thanks{\email{tine.colman@cea.fr}}
          \and
          Noé Brucy\inst{2}\fnmsep\thanks{\email{noe.brucy@uni-heidelberg.de}}
          \and
          Philipp Girichidis\inst{2}\fnmsep\thanks{\email{philipp@girichidis.com}}
          \and
          Simon~C.~O.~Glover\inst{2}
          \and
          Milena Benedettini\inst{3}
          \and
          Juan~D.~Soler\inst{3}
          \and
          Robin~G.~Tress\inst{4}
          \and
          Alessio Traficante\inst{3}
          \and
          Patrick Hennebelle\inst{1}
          \and
          Ralf S.\ Klessen\inst{2,5}
          \and
          Sergio Molinari\inst{3}
          \and
           Marc-Antoine Miville-Desch\^enes\inst{6}
          }

   \institute{AIM, CEA, CNRS, Université Paris-Saclay, Université Paris Diderot, Sorbonne Paris Cité, 91191 Gif-sur-Yvette, France
         \and
             Universit\"{a}t Heidelberg, Zentrum f\"{u}r Astronomie, Institut f\"{u}r Theoretische Astrophysik, Albert-Ueberle-Str. 2, 69120 Heidelberg, Germany
        \and
            INAF -- Istituto di Astrofisica e Planetologia Spaziali, via Fosso del Cavaliere 100, 00133 Roma, Italy
        \and
            Institute of Physics, Laboratory for Galaxy Evolution and Spectral Modelling, EPFL, Observatoire de Sauverny, Chemin Pegais 51, 1290 Versoix, Switzerland
        \and
            Universit\"{a}t Heidelberg, Interdisziplin\"{a}res Zentrum für Wissenschaftliches Rechnen, Im Neuenheimer Feld 205, 69120 Heidelberg, Germany
        \and
        Laboratoire de Physique de l'École Normale Supérieure, ENS, Université PSL, CNRS, Sorbonne Université, Université de Paris, F-75005 Paris, France
            }

   \date{Received December 17, 2023; accepted xxxx xx, xxxx}

   \authorrunning{Colman, Brucy, Girichidis et al.}

 
  \abstract
   {Molecular clouds (MC) are structures of dense gas in the interstellar medium (ISM), that extend from ten to a few hundred parsecs and form the main gas reservoir available for star formation. Hydrodynamical simulations of varying complexity are a promising way to investigate MC evolution and their properties. However, each simulation typically has a limited range in resolution and different cloud extraction algorithms are used, which complicates the comparison between simulations.}
   {In this work, we aim to extract clouds from different simulations covering a wide range of spatial scales. We compare their properties, such as size, shape, mass, internal velocity dispersion and virial state.}
   {We apply the \textsc{Hop} cloud detection algorithm on (M)HD numerical simulations of stratified ISM boxes and isolated galactic disk simulations that were produced using \textsc{Flash}, \textsc{Ramses} and \textsc{Arepo}.}
   {We find that the extracted clouds are complex in shape ranging from round objects to complex filamentary networks in all setups. Despite the wide range of scales, resolution, and sub-grid physics, we observe surprisingly robust trends in the investigated metrics. The mass spectrum matches in the overlap between simulations without rescaling and with a high-mass power-law index of -1 for logarithmic bins of mass, in accordance with theoretical predictions. The internal velocity dispersion scales with the size of the cloud as $\sigma\propto R^{0.75}$ for large clouds ($R\gtrsim3\,\mathrm{pc}$). For small clouds we find larger $\sigma$ compared to the power-law scaling, as seen in observations, which is due to supernova-driven turbulence. Almost all clouds are gravitationally unbound with the virial parameter scaling as $\alpha_\mathrm{vir}\propto M^{-0.4}$, which is slightly flatter compared to observed scaling, but in agreement given the large scatter. We note that the cloud distribution towards the low-mass end is only complete if aggressive refinement is used that also refines more dilute gas rather than only collapsing regions.}
   {}


   \maketitle
%

\section{Introduction}

Historically, (molecular) clouds have been considered important building blocks in the interstellar medium (ISM). The warm and diffuse gas condenses to form cloud-like structures with masses of $\sim 10^4 - 10^6$ M$_\odot$, typical sizes of $\sim 10-100$ pc, mean densities of $n_{\rm H_2} \sim 100$  $\mathrm{cm}^{-3}$, and temperatures of $T\sim 10$ K \citep[e.g.][or \citealt{KlessenGlover2016} for a review]{Solomon_et_al1987, Scoville87a, Dame87a, Dame_et_al2001}. Most of the mass in these conditions is in the form of molecular hydrogen, hence the term \emph{molecular clouds} (MCs). However, most of the dynamical properties and the connection to hydrodynamical evolution does not require the gas being molecular. One particularly important aspect connected to clouds is that they harbour the coldest ($\sim10\,\mathrm{K}$) regions of the ISM including the collapsing cores that eventually form stars.

For a long time, clouds were considered to be stable and long-lived structures that are gravitationally bound and supported against collapse by e.g. magnetic fields \citep[e.g.][]{Mouschovias76a, Shu87a, Mouschovias99a}. This paradigm has changed. In particular, the idea that clouds are magnetically supported  has given way to a picture in which they are formed by a combination of gravity and turbulence \citep[e.g.][]{BallesterosParedesEtAl1999, BallesterosParedesMacLow2002, Mac-Low04a, Krumholz05a, BallesterosParedes2006, Hennebelle_Chabrier2008, BallesterosParedesEtAl2011, VazquezSemadeniEtAl2017,VazquezSemadeniEtAl2019}, with the magnetic field playing an important but not dominant role \citep[e.g.][]{Crutcher2012,Girichidis2021,whitworthMagneticFieldsNot2023}. Turbulence creates over-densities and large-scale coherent structures, whose shapes are similar to the complex shapes of molecular clouds \citep[e.g.][]{SchneiderEtAl2011, Ebagezio2023}. As a result, the gas in the ISM is constantly reshaped by turbulence without permanent structures on a crossing time $t_\mathrm{cross} = L/v$, where $L$ is the cloud size and $v$ the local velocity dispersion on a scale $L$. In the Solar neighbourhood, representative values for $L$ and $v$ are 30~pc and $3 \: {\rm km \, s^{-1}}$, respectively, yielding $t_{\rm cross} \sim 10$~Myr, a small fraction of the orbital period of the Galaxy.
Clouds can form in regions of converging flows and are dispersed by shear and/or feedback from stars \citep[e.g.][for a recent review]{ChevanceEtAl2023}. Therefore, molecular clouds should not be regarded as well-defined discrete entities with clearly identifiable boundaries.

\citet{Larson1981} was the first to establish power-law scaling relations between molecular cloud properties. He observed that cloud mass and size followed the relation $M \propto R^{1.90}$ while the velocity dispersion scaled as $\sigma_{\rm v} \propto R^{0.38}$. Both of these relations showed significant scatter. In the Milky Way and neighbouring galaxies, MCs seem to adhere to the Larson relations, where the non-thermal line-width increases with cloud size following an approximately 1/2 power law \citep{Larson1981, Solomon_et_al1987, Bolatto08, Fukui08, HeyerEtAl2009, Roman-Duval_et_al2010, Wong11}. This relation extends within clouds \citep{heyer04} and may be attributed to the power law scaling expected for compressible turbulence \citep{McKee_Ostriker07}. \citet{Larson1981} also argued that MCs exhibited similar levels of kinetic and gravitational energy and hence were approximately in virial equilibrium \citep[see also][]{Blitz1993}. This can be quantified through the use of the virial parameter \citep{BertoldiMcKee1992}
\begin{equation}
    \label{eq:virial-parameter}
    \alpha_\mathrm{vir} \equiv \frac{5\sigma_v^2 R}{GM},
\end{equation}
in which $\sigma_v$ is the one-dimensional velocity dispersion, $G$ is the gravitational constant, and $R$ and $M$ are the size and mass of the cloud. Virial equilibrium corresponds to $\alpha_{\rm vir} = 1$.
Although early studies of massive clouds found that $\alpha_{\rm vir} \sim 1$, more recent surveys that are sensitive to a much broader range of cloud masses find a more complex picture \citep{heyer01, oka01, gratier12, DonovanMeyer13, Rice_et_al2016, Miville-Deschenes_et_al2017, colombo19, Rosolowsky_et_al2021}. These surveys show that low mass clouds are unbound with virial ratios $\gg1$ and only massive clouds with masses of order $10^6\,\mathrm{M}_\odot$ or more are marginally bound. However, 
there are large uncertainties involved in determining the mass of MCs from observations \citep{Szucs2016}, and so the characteristic mass at which MCs become gravitationally bound has been difficult to pin down with any precision. Clouds with virial parameters exceeding unity mostly follow a relation with $\alpha_\mathrm{vir}\propto M^{-1/2}$ \citep{ChevanceEtAl2023}.

The formation of MCs and the dynamical evolution of the gas within them has been addressed in hydrodynamical simulations with a range of different complexities. Early studies either focused on simulating isolated clouds \citep{Vazquez-Semadeni95a, Stone1998,Klessen00a,Ostriker2001,Bate03a,Padoan07a} or covered large fractions of the interstellar medium without resolving the interior of clouds \citep{deAvillez2000, deavillez2005, Joung_MacLow2006, KimOstriker2007, HillEtAl2012, GentEtAl2013a, WalchEtAl2012}. However, increasing computational power now allows us to cover both the large-scale environment, including the hot and dilute gas from which clouds condense as well as the coldest collapsing phases. \citet{SeifriedEtAl2017} showed the importance of embedding the clouds into the turbulent large-scale ($\sim100\,\mathrm{pc}$) environment, in order to accurately follow the turbulent cascade and the formation of cold gas. They adopted a zoom-in approach in which they followed the evolution of individual clouds selected from a large-scale simulation of the ISM with successively improving resolution. This approach allowed them to reach resolutions as small as $0.1\,\mathrm{pc}$, but limited them to modelling only a few clouds, thereby preventing them from drawing robust statistical conclusions. 
A companion study at a lower resolution of $0.25\,\mathrm{pc}$ by \citet{Girichidis2021} complements these efforts. Both models only consider environmental conditions similar to the Solar neighbourhood. Similar simulations of stratified boxes but with different turbulent driving recipes and total gas masses have been performed by other groups \citep[e.g.][]{Joung2009,Simpson2016,Martizzi2016,kim2017,brucyLargescaleTurbulentDriving2020,brucyLargescaleTurbulentDriving2023,ColmanEtAl2022}. These simulations cover slightly larger scales and employ different cooling recipes. In addition, it has now become possible to carry out simulations of entire disk galaxies with a resolution sufficient to resolve individual clouds, allowing one to study the roles of global galactic rotation and differential shear and to cover a large range of local surface densities \citep[e.g.][]{TressEtAl2020, Tress2021, Jeffreson2020}.
All of these models have their own strengths and weaknesses in terms of physical processes included, environmental properties of molecular clouds and maximum resolution. Therefore, an important question to ask prior to comparing the results of these simulations to observations is whether the predictions of different simulations for the properties of the MC distribution agree with one other, i.e.\ whether these predictions are robust to changes in the numerical approach. Because the definition of a cloud is not trivial, it is important to use the same cloud extraction method with identical parameters for cloud identification, so that we can be sure that any differences we find are due to differences in the simulations and not in the cloud identification algorithm. A thorough comparison between numerical models in overlapping ranges of the cloud masses is crucial for verifying that the full range of cloud masses and sizes can be compared to observations without numerically induced breaks in the scaling relations. Furthermore, a comparison between simulations can be considerably more detailed than a comparison with observations, as we have the full 3D distribution of all relevant physical variables available for our simulated clouds, which is not the case for real, observed MCs.

In this paper, we investigate the molecular cloud properties in a variety of numerical simulations with different resolutions and different numerical recipes for cooling, star formation and stellar feedback. Our goal is to verify the existence of smooth transitions between the simulations and quantify the clouds in terms of global properties that can be transferred to the observational domain. 
Our study is structured as follows: in Section~\ref{sec:numerics} we discuss the algorithm used to extract the clouds and the numerical simulations on which the algorithm was applied. Then, in Section~\ref{sec:mass_size_shape}, we compare the cloud population's masses, sizes and shapes. 
In Section \ref{sec:internal_properties} we focus on the internal properties of the clouds, looking at the thermal state, the turbulence and the gravitational stability.
We discuss the importance of resolution and of the specifics of the cloud extraction process in Section~\ref{sec:discussion} and conclude in Section~\ref{sec:conclusion}.

\section{Numerical methods, simulation data}
\label{sec:numerics}

\subsection{Structure finding algorithm}
\label{subsec:struct_find_algo}

Our structure finding algorithm is built around the \textsc{Hop} clump finder \citep{EisensteinEtAl1998}.
While originally it was developed for finding groups of particles in N-body simulations, the algorithm and its original implementation are quite general and can be used in many more applications.
One advantage \textsc{Hop} has is that it simply requires a list of particles as input.
While this may seem sub-optimal for analysing regular grid-based hydrodynamics simulations, it allows for a direct comparison between the results of particle-based codes, unstructured meshes as well as regular grid-based (AMR) codes. To achieve this, we transform the AMR grid into a list of cells.

Overall, there are three steps towards obtaining a structure catalogue, which are described in depth in Appendix~\ref{sec:hop_algo}.
First, we use \textsc{Hop} to identify peak patches around local density maxima.
Then, we use the regrouping tool from \textsc{Hop} to merge peaks based on a number of specified criteria.
For clouds, we adopt a density threshold of $30 \: {\rm cm^{-3}}$. Everything below this threshold is not considered to be part of a cloud.
Then we require a cloud to have a peak density of at least $60 \: {\rm cm^{-3}}$, which corresponds to a minimum peak-to-background ratio of 2.
When the saddle density between two structures is larger than $300 \: {\rm cm^{-3}}$, i.e.\ ten times the density threshold, we consider them as being substructures within the same cloud and therefore merge them.
Lastly, we require a structure to be made up of at least 100 cells. 
After this merging step, we know for each cell or particle the structure to which it belongs, or whether it does not belong to any structure.
Based on that, we then finally calculate the properties of all identified structures as detailed in Appendix~\ref{sec:hop_algo}.

\subsection{Numerical simulations of the ISM}

\begin{figure*}

    \begin{subfigure}[b]{0.49 \textwidth}
    \centering
    \includegraphics[width=\textwidth]{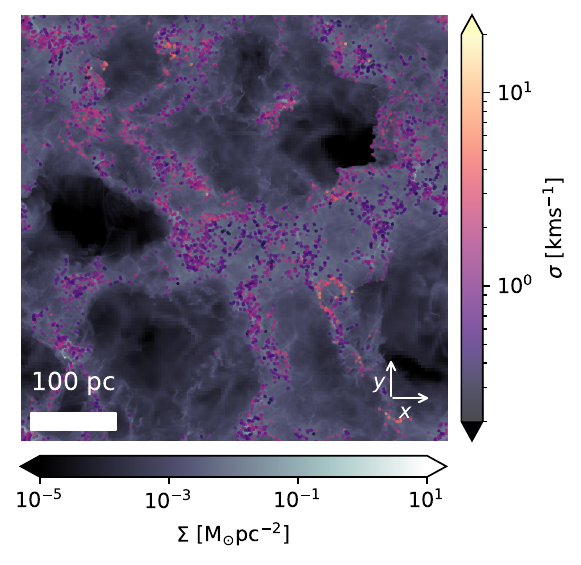}
    \caption{SILCC-0.5pc-3$\mu$G simulation}
    \label{fig:coldens-sigma-SILCC}
    \end{subfigure}
    \begin{subfigure}[b]{0.49 \textwidth}
    \centering
    \includegraphics[width=\textwidth]{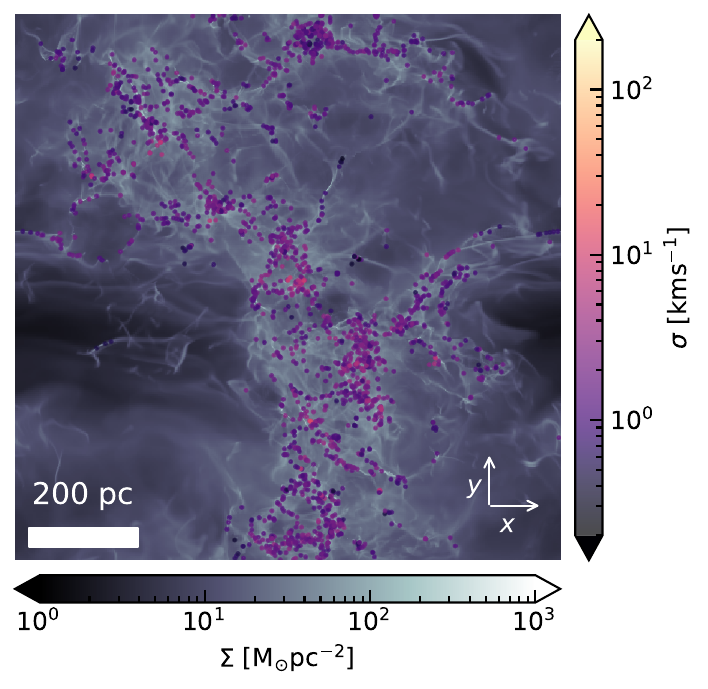}
    \caption{LS simulation (weak driving)}
    \label{fig:coldens-sigma-LS}
    \end{subfigure}

    \begin{subfigure}[b]{0.49 \textwidth}
    \centering
    \includegraphics[width=\textwidth]{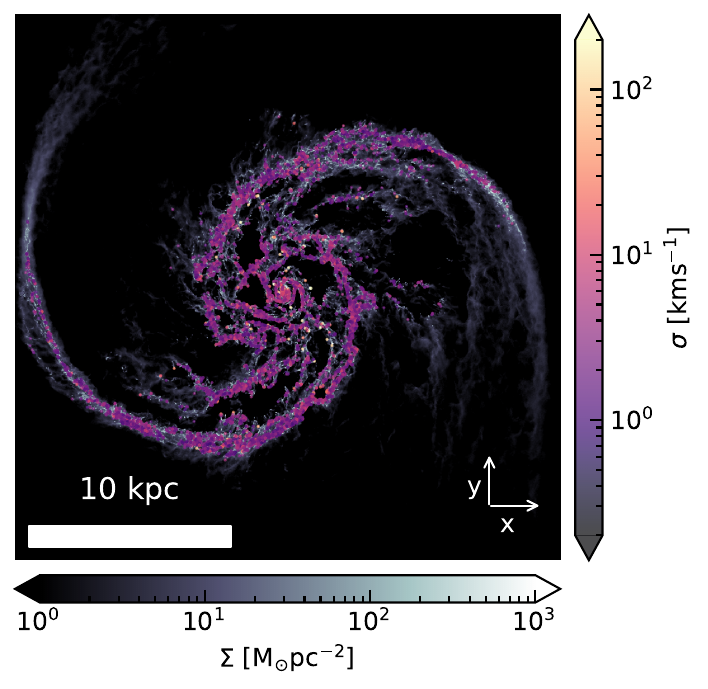}
    \caption{M51 simulation}
    \label{fig:coldens-sigma-M51}
    \end{subfigure}
    \begin{subfigure}[b]{0.49 \textwidth}
    \centering
    \includegraphics[width=\textwidth]{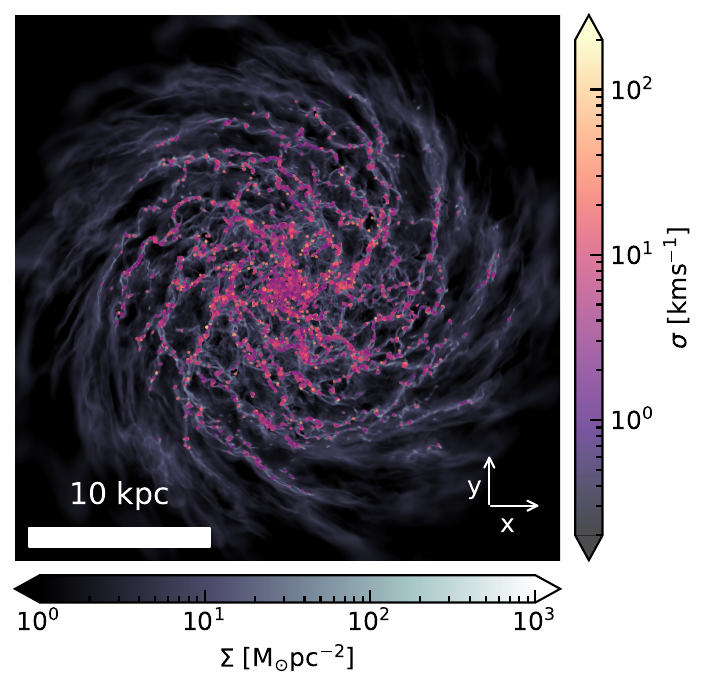}
    \caption{Ramses-F20 simulation}
    \label{fig:coldens-sigma-Ramses-F20}
    \end{subfigure}
    
    \caption{Column density map for a single snapshot from each of the four listed groups of simulations. Details of these simulations can be found in Table~\ref{tab:simulations}. The coloured symbols indicate the locations of the clouds identified by the \textsc{Hop} algorithm, with the colour corresponding to the internal velocity dispersion.} \label{fig:coldens-sigma}
\end{figure*}

\begin{table*}
  \caption{Set of simulations. We list the type of setup, the size of the computational domain, the spatial resolution, the initial mid-plane magnetic field strength, the simulation code used to run the simulation, the reference in which the simulation was first described and the number of snapshots analysed.}
     \label{tab:simulations}
     \centering
     \begin{tabular}{lcrrcclc}
        \toprule
        Name & Setup & size & $\Delta x_\mathrm{min}$ & $B_0$ & Code & Reference & No.\ of\\
              & & (kpc) & (pc) & ($\mu$G) &  & & snapshots \\
        \midrule
        SILCC-1pc-3$\mu$G    & stratified box & 0.5 &  0.98 & 3 & \textsc{Flash} & \citet{GirichidisEtAl2018b} & 4\\
        SILCC-1pc-6$\mu$G    & stratified box & 0.5 &  0.98 & 6 & \textsc{Flash} & \citet{GirichidisEtAl2018b} & 4\\
        SILCC-0.5pc-3$\mu$G  & stratified box & 0.5 & 0.49 & 3 & \textsc{Flash} & \citet{Girichidis2021} & 6\\
        SILCC-0.25pc-3$\mu$G & stratified box & 0.5 & 0.24 & 3 & \textsc{Flash} & \citet{Girichidis2021} & 2\\
        \midrule
        LS-no-driving   & stratified box & 1 & 0.24 & 7.6 & \textsc{Ramses} & \citet{ColmanEtAl2022} & 1\\
        LS-weak-driving & stratified box & 1 & 0.24 & 7.6 & \textsc{Ramses} & \citet{ColmanEtAl2022} & 1\\
        LS-medium-driving & stratified box & 1 & 0.24 & 7.6 & \textsc{Ramses} & \citet{ColmanEtAl2022} & 1\\
        LS-strong-driving & stratified box & 1 & 0.24 & 7.6 & \textsc{Ramses} & \citet{ColmanEtAl2022} & 1\\
        \midrule
        M51 & full galaxy & -- & $\approx$ 1.4 & 0 & \textsc{Arepo} & \citet{TressEtAl2020} & 1\\
        \midrule
        Ramses-F20 & full galaxy & -- & 0.92 & 0 & \textsc{Ramses} & \citet[Chapter 11]{BrucyPhd} & 1 \\
        \bottomrule
     \end{tabular}
\end{table*}

We apply the \textsc{Hop} algorithm to several (M)HD simulations of the ISM. We combine simulations with a wide range of simulation techniques, resolution, box size and physics included.
An overview is given in Table~\ref{tab:simulations}.
A face-on projection of one snapshot of each simulation type is featured in Fig. \ref{fig:coldens-sigma}, where we over-plot the positions of the extracted clouds colour-coded by their derived velocity dispersion for illustration.

\subsubsection{SILCC: Stratified boxes with \textsc{Flash}}

We include high-resolution magnetised simulations from the SILCC collaboration \citep{WalchEtAl2015,GirichidisEtAl2016b}, which are described in detail in \citet{GirichidisEtAl2018b} and 
\citet{Girichidis2021}. The setup consists of a stratified box with a volume of $0.5\times0.5\times0.5\,\mathrm{kpc}^3$. The gas is initially at rest and in hydrostatic equilibrium with a total gas mass surface density of $10\,\mathrm{M}_\odot\,\mathrm{pc}^{-2}$. The two different magnetic field models considered here have central magnetic field strengths at $z=0$ of $B_{x,0} = 3\,\mu\mathrm{G}$ and $B_{x,0}=6\,\mu\mathrm{G}$ and are oriented along the $x$ direction. The field strength scales vertically with the gas density as $B_x(z)=B_{x,0}\,[\rho(z)/\rho(z=0)]^{1/2}$. The initial setup is magnetically supercritical, i.e.\ the field cannot support the gas against collapse.

The MHD equations are solved using the HLLR5 solver \citep{Bouchut2007, Bouchut2010,
Waagan2009, Waagan2011}, which is implemented in \textsc{Flash} in Version 4 (\citealt{FLASH00,DubeyEtAl2008}). Heating of the gas includes spatially clustered supernovae (SNe), as well as cosmic ray \citep{GoldsmithLanger1978} and X-ray heating \citep{WolfireEtAl1995}. The CR ionisation and heating rates are $\zeta_\mathrm{CR}=3\times10^{-17}\,\mathrm{s}^{-1}$ and $\Gamma_\mathrm{CR}=3.2\times10^{-11}\zeta_\mathrm{CR}\,n\,\mathrm{erg\,s}^{-1}\,\mathrm{cm}^{-3}$, respectively. Photoelectric heating follows \citet{BakesTielens1994}, \citet{Bergin2004} and \citet{WolfireEtAl2003}. We assume a spatially constant interstellar radiation field (ISRF) with a strength of $1.7$ in the units of the Habing field $G_0$ \citep{Habing1968, Draine1978} that is then locally attenuated in dense, shielded gas using the TreeCol algorithm \citep{ClarkGloverKlessen2012,WuenschEtAl2018}. We fix the dust-to-gas mass ratio to 0.01, and adopt dust opacities from \citet{MathisMezgerPanagia1983} for wavelengths shorter than $1 \mu$m and \citet{OssenkopfHenning1994} for longer wavelengths.

Heating and radiative cooling processes are computed using a chemical network that tracks the non-equilibrium concentrations of ionised hydrogen (H$^+$), atomic hydrogen (H), molecular hydrogen (H$_2$), as well as singly ionised carbon (C$^+$) and carbon monoxide (CO). The hydrogen chemistry is based on the models by \citet{Hollenbach89, GloverMacLow2007a, GloverMacLow2007b} and \citet{MicicEtAl2012}. The reactions connected to CO follow the model developed by \citet{NelsonLanger1997}. Radiative cooling incorporates contributions from the fine structure lines of C$^+$, O, and Si$^+$. Furthermore, we consider rotational and vibrational lines of H$_2$ and CO, as well as Lyman-$\alpha$ cooling. The energy transfer from the gas to the dust follows \citet{GloverEtAl2010} and \citet{GloverClark2012b}. Above $10^4\,\mathrm{K}$, we assume collisional ionisation equilibrium (CIE) for helium and all heavy elements, and adopt the corresponding CIE cooling rates from \citet{GnatFerland2012}. Hydrogen is not assumed to be in collisional ionisation equilibrium, and its contribution to the cooling process is calculated self-consistently at all temperatures, accounting for any deviations from chemical equilibrium.

Gravitational forces include self-gravity as well as an external potential based on an isothermal sheet \citep{Spitzer1942} with a surface density of $30\,\mathrm{M}_\odot \, \mathrm{pc}^{-2}$, representing the stellar mass distribution. The vertical scale height of this sheet is set to $100\,\mathrm{pc}$. The Poisson equation is solved using the tree-based method as described in \citet{WuenschEtAl2018}.

Star formation and the related supernova feedback are included at a fixed rate. We derive a star formation rate based on the Kennicutt-Schmidt relation \citep{KennicuttSchmidt1998} and convert it to a SN rate using the \citet{Chabrier2003} IMF. This yields 15 SNe per Myr for the volume simulated here. Each explosion injects $10^{51}\,\mathrm{erg}$ of thermal energy. For the positioning of the SNe we distinguish between a type~Ia (20 per cent) and a type~II component (80 per cent). The type~Ia SNe are individual explosions placed at random (uniformly chosen) positions for $x$ and $y$. The $z$ position is randomly sampled from a Gaussian distribution with a vertical scale height of $300\,\mathrm{pc}$ \citep{BahcallSoneira1980, Heiles1987}, where we ensure that SNe that are placed outside the box are re-drawn to be placed inside the box. The type~II SNe are further split into isolated SNe (40 per cent of the type~II) and clustered counterparts (60 per cent) \citep{Heiles1987, KennicuttEtAl1989, McKeeWilliams1997, ClarkeOey2002}. For the positioning of the individual type~II SNe and the clusters we also chose random positions for $x$ and $y$. The vertical distribution is drawn from a Gaussian with a smaller scale height of $90\,\mathrm{pc}$. All SNe within a cluster explode at the same cluster position. The positions of the SNe are computed beforehand and stored in a table to ensure an identical feedback configuration for the different magnetic field strengths and resolutions.

We consider three different resolutions for the simulations. In all cases, the base resolution is $128^3$, corresponding to a maximum cubic cell with a side length of $3.9\,\mathrm{pc}$. On top we add two additional levels of refinement for simulations SILCC-1pc-3$\mu$G and SILCC-1pc-6$\mu$G, which reach a minimum cell size of $0.98\,\mathrm{pc}$. For simulations SILCC-0.5pc-3$\mu$G and SILCC-0.25pc-3$\mu$G we restart simulation SILCC-1pc-3$\mu$G at a time of $20\,\mathrm{Myr}$ and add one and two additional levels of refinement, which yields minimum cell sizes of $0.49$ and $0.24\,\mathrm{pc}$, respectively. The simulations with a resolution of $1\,\mathrm{pc}$ run for a total time of $60\,\mathrm{Myr}$; the higher resolution ones were evolved for a shorter time of a few Myr \citep{Girichidis2021}.

\subsubsection{LS: Stratified boxes with \textsc{Ramses}}
\label{subsubsec:ls_sims}

For our comparison, we also use \textsc{Ramses} \citep{Teyssier2002} simulations of a stratified piece of a galactic disk. The simulations are fully described in \cite{ColmanEtAl2022} and references therein. The computational domain is cubic with a box size of 1 kpc. The base grid with a resolution of 3.9 pc is further refined using a mass-based criterion up to a minimum cell size of 0.24 pc in the densest regions. 
Compared to the SILCC simulations, the computational domain is larger, but the different refinement strategy can results in fewer AMR grids (see Table~\ref{tab:refinement}).

The initial density profile is Gaussian $n(z)=n_{0}\, \exp [-0.5 \, (z/z_0)^2]$ with a mid-plane particle density $n_0 = 1.5$ cm$^{-3}$ and a thickness $z_0 = 150$ pc, corresponding to a column density of $\Sigma=19.1 \, \mathrm{M}_\odot \, \mathrm{pc}^{-2}$.
An initial level of turbulence is introduced by adding a turbulent velocity field with a root mean square dispersion of 5 km s$^{-1}$ and a Kolmogorov power spectrum $E(k) \propto k^{-5/3}$ with random phase.
The initial temperature is 5333 K, which is a typical value for the warm neutral medium phase in the ISM.
We also include a magnetic field orientated along the $x$-axis with  $B_x(z)=B_{0}\, \exp [-0.5 \, (z/z_0)^2]$ initially, where $B_0 = 7.62 \, \mu$G is the mid-plane field strength.
The magnetic field is solved using the ideal MHD approximation, as implemented by \citet{Fromang_et_al2006}.
Aside from self-gravity, we also apply an external gravitational disk potential as prescribed by \cite{Kuijken_Gilmore1989} and  \citet{Joung_MacLow2006}, which accounts for the profile of (old) stars and dark matter.

We use an ISM cooling model based on \cite{Audit_Hennebelle2005} that includes the most important processes responsible for regulating the thermal balance of the atomic ISM. At low temperatures, cooling is provided by the fine structure transitions of C$^{+}$ and O, while at high temperatures this is supplemented by contributions from grain surface recombination of atomic hydrogen and from Lyman-$\alpha$ cooling. Heating is provided primarily by the photoelectric effect, calculated assuming a constant uniform UV background with a strength equal to the solar neighbourhood value. In contrast to the SILCC and \textsc{Arepo} simulations, we do not account for local attenuation of this radiation field within dense clouds. We also account for cosmic ray heating, using the rate given in \citet{goldsmithMolecularDepletionThermal2001}. The ionisation of hydrogen is treated by the the RT module from \citet{Rosdahl_et_al2013}.
We use sink particles to track star formation and stellar feedback self-consistently. Sinks represent newly formed star clusters \citep{Bleuler&Teyssier2014}, which accrete gas which is within a 4 cell accretion radius and above the sink formation threshold of $10^4$ cm$^{-3}$. 
Each time a sink has accreted enough mass, it forms a massive star particle with a mass between 8 and 120 $\Msun$ randomly drawn from the Salpeter IMF \citep{Salpeter1955}.
This massive star emits ionising radiation and explodes as a supernova at the end of its lifetime \citep{Rosdahl_et_al2013, Geen_et_al2016,Colling_et_al2018}.
The radiation is treated using a moment-based method \citep{Rosdahl_et_al2013}.
In addition to the turbulence generated by stellar feedback, we also include external driving on scales between 1 box length and 1/3 of the box length with solenoidal fraction 0.75.
Several forcing strengths were explored: weak, medium and strong driving, as well as no driving, which result in mass weighted velocity dispersions of $\sigma_\mathrm{3D}\approx9,~12,~20$ and $8.5\,\mathrm{km\,s^{-1}}$, respectively. The simulations were evolved for 60 Myr.

\subsubsection{M51: Full galaxy with \textsc{Arepo}}

Stratified box set-ups allow one to simulate the ISM at high resolution but miss important aspects linked to large scales: self-consistent large-scale driving, galactic dynamical effects such as rotation or the influence of spiral density waves, and large-scale instabilities such as the Toomre instability \citep{toomre64} or the wiggle instability \citep{sormani17}.
As this may influence the properties of the clouds that form in the ISM, we included two full galaxy simulations in our study.
The first one was carried out using \textsc{Arepo}, a moving-mesh hydrodynamic code \citep{Springel2010}, and was fully presented in \cite{TressEtAl2020}. The clouds from this simulation were already extracted and analysed in \cite{Tress2021}, but with a different cloud extraction algorithm.
The modelled galaxy is interacting with a smaller companion and its properties are chosen to be similar to the M51 ``Whirlpool'' galaxy.
The characteristics of the different components (dark matter halo, stellar bulge, stellar disc and gaseous disc) are summarised in Table \ref{tab:galaxy_init} in Appendix~\ref{sec:gal_sim_params}. Gravitational interactions between all those components were self-consistently accounted for. 

Resolution in \textsc{Arepo} is defined by the target mass, which is the typical mass of a cell. This is analogous to the particle mass in an smoothed particle hydrodynamics (SPH) computation. For the M51 simulation, the target mass is 300 $\Msun$, with additional refinement in the denser part of the ISM such that the Jeans length is resolved by at least four cells up to a density of $10^{-21} \mathrm{g}\, \mathrm{cm}^{-3}$. As a result of this refinement scheme, most of the cells above the \textsc{Hop} density threshold have a radius of 1 pc or less, with the smallest cells having a radius of only 0.4~pc. For comparison with the minimum $\Delta x$ in the AMR simulations, we quote in Table~\ref{tab:simulations} the typical diameter of a cell at the sink creation density threshold, which is $\sim 1.4$~pc. The effective resolution of this simulation in dense regions is thus slightly worse than the low resolution SILCC runs.

Unlike the previous sets of simulations, the magnetic field is not modelled. The thermal and chemical evolution of the gas are solved simultaneously, using the NL97 chemical network described in \cite{GloverClark2012a} and the atomic and molecular cooling function described in \cite{clarkTracingFormationMolecular2019}.
Star formation is modelled using sink particles that are created within gravitationally bound regions of radius 2.5 pc when the gas density exceeds a threshold of $10^{-21}\,\mathrm{g}\,\mathrm{cm}^{-3}$, if the region satisfies several additional criteria: it must be a local minimum of the gravitational potential, and both the velocity and the acceleration must be converging. 
Once formed, sink particles can accrete gas from within an accretion radius of 2.5~pc, with the accretion limited to gas with a density exceeding the sink creation density. Two feedback processes are modelled: type II and Type Ia SNe. The type II SNe are attached to the sinks, via a process similar to the one described in Section \ref{subsubsec:ls_sims}. Type Ia SNe are exploded at the position of a randomly selected old star with a rate of one explosion every 250 years. These are the stellar particles of the disc and bulge components of the N-body model of the galaxy. Supernovae were modelled by injecting energy in a region containing $40$ resolution elements surrounding the injection region. Momentum or thermal energy is injected based on whether the Sedov-Taylor phase is resolved.

\subsubsection{Ramses-F20: Full galaxy with \textsc{Ramses}}

To complete the study, we also ran the algorithm on a isolated disk simulation performed with \textsc{Ramses} that was initially presented in \citet[Chapter 11]{BrucyPhd}.
The model is an isolated disk with a radius of 15 kpc, within a box with a side length of 120 kpc. 
Initial conditions are generated with the code \textsc{Dice}, which allow one to specify the different components of a galaxy. 
\textsc{Dice} generates particles representing dark matter, stars and gas. 
The two first are directly used in \textsc{Ramses} while the gas distribution was mapped from the particles onto the AMR grid.
The parameters for the generation of the initial conditions are summed up in Table \ref{tab:galaxy_init}.

Using the adaptive mesh refinement capabilities of \textsc{Ramses}, the maximal cell size is 1 kpc outside the disk and 7 pc within the disk. The minimal cell size is 0.92 pc within the disk.
Dark matter and initial stars are modelled via particles that only undergo gravity.
The gas follows the law of hydrodynamics (without magnetic field) and gravity.
The simulation is run over a period of 300 Myr.

The simulation also includes a sub-grid model for star formation and stellar feedback, here limited to SN explosions and photo-ionisation-triggered heating.
These models are those developed and described in \cite{kretschmerFormingEarlytypeGalaxies2020}. 
The star formation rate is computed for each cell from its properties. 
First the Mach number $\mathcal{M}$ within the cell is computed thanks to a model of the evolution of the sub-grid turbulent velocity. 
This is then used to estimate the sub-grid density PDF, assuming that it is a log-normal \citep{vazquez-semadeniHierarchicalStructureNearly1994, federrathStarFormationRate2012}, as well as the virial parameter of the cell. The expected star formation rate within the cell is then computed using the multi-free-fall analytical model for the star formation rate adapted from \cite{krumholzFormationStarsGravitational2005} by \cite{hennebelleAnalyticalStarFormation2011} and popularised by \cite{federrathStarFormationRate2012}. 
At each time-step the number of new star particles created is randomly drawn from a Poisson distribution, calibrated so that the mean star formation rate over time is equal to the one given by the analytical model.
The SN injection scheme depend whether the cooling radius the SN is resolved by a least one grid cell. 
If yes, the energy from the explosion is directly injected in the form of thermal energy. 
Otherwise, the terminal momentum of the SN is also injected.
Photo-ionisation is modelled by maintaining the cell where the star sits at a temperature of $10^4$~K until the last SN explodes. 

The cooling method is the same as the one used in the \textsc{Ramses} stratified box simulation presented in Section \ref{subsubsec:ls_sims}.

\subsection{Examples of extracted clouds}
\label{subsec:examples_clouds}

\begin{figure*}[p]
    \begin{subfigure}[b]{\textwidth}
        
        \centering
        \includegraphics[width=0.235\textwidth]{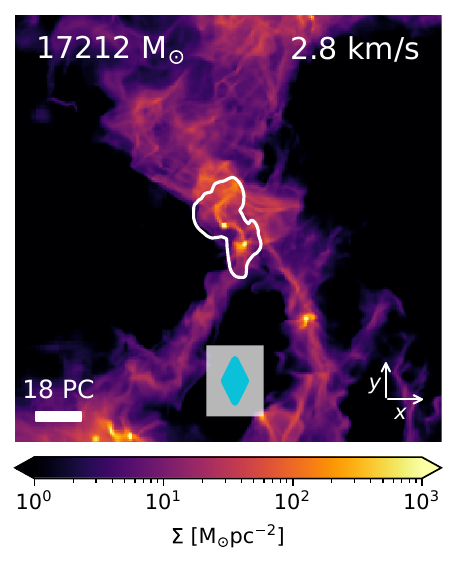}
        \includegraphics[width=0.235\textwidth]{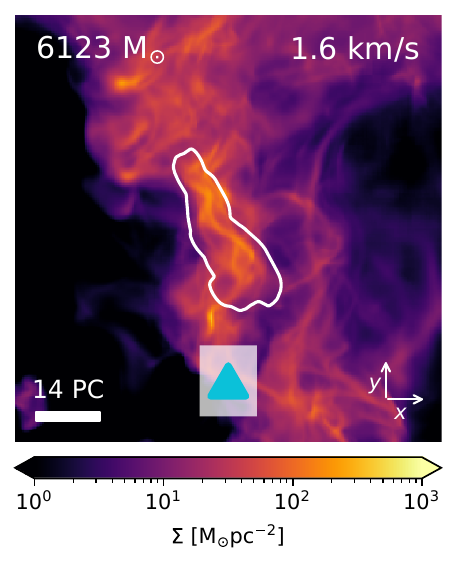}
        \includegraphics[width=0.235\textwidth]{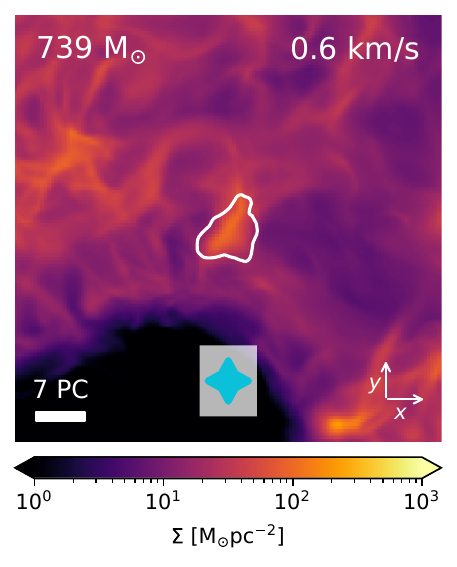}
        \includegraphics[width=0.235\textwidth]{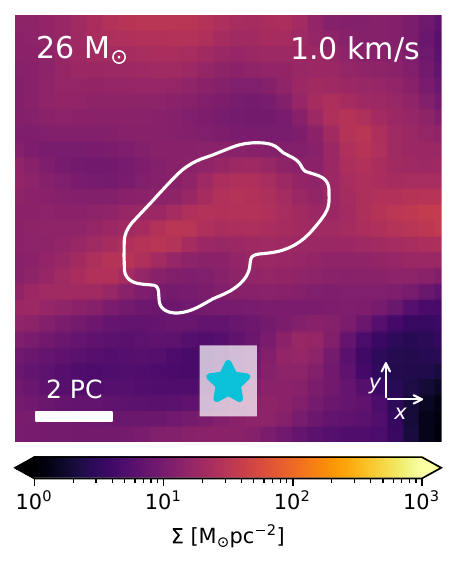}
        \caption{Example clouds from the SILCC-0.5-pc stratified box simulations \citep{Girichidis2021}.}
        \label{fig:example_clouds_silcc}
    \end{subfigure}

    \begin{subfigure}[b]{\textwidth}
        \centering
        \includegraphics[width=0.235\textwidth]{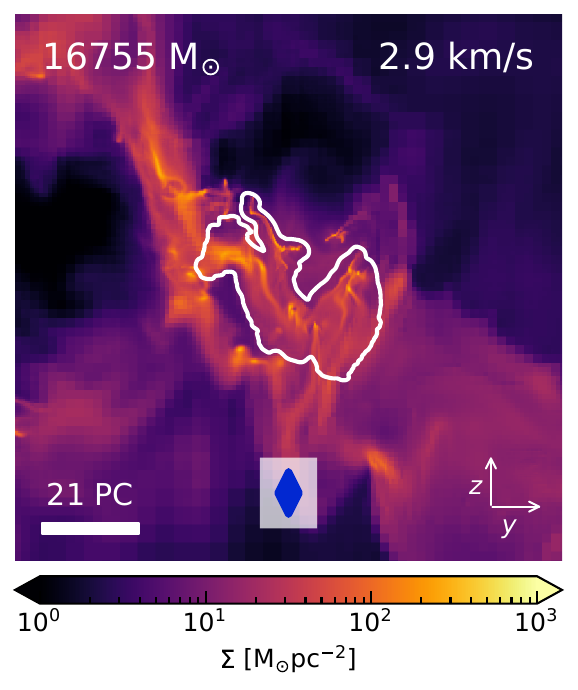}
        \includegraphics[width=0.235\textwidth]{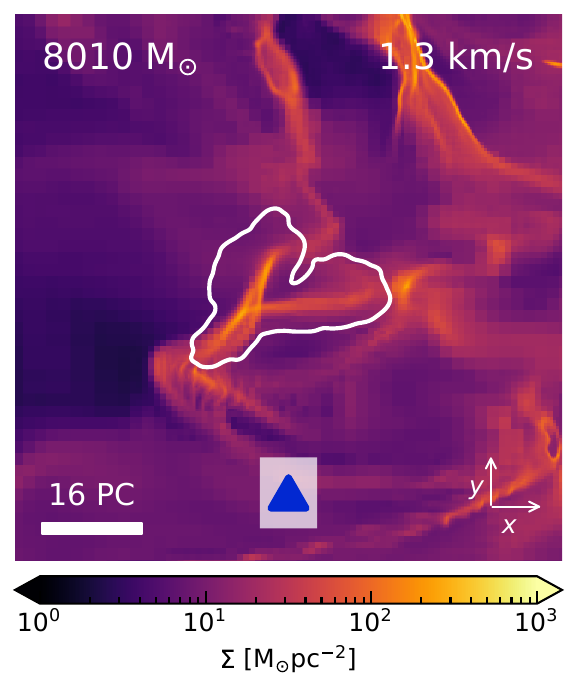}
        \includegraphics[width=0.235\textwidth]{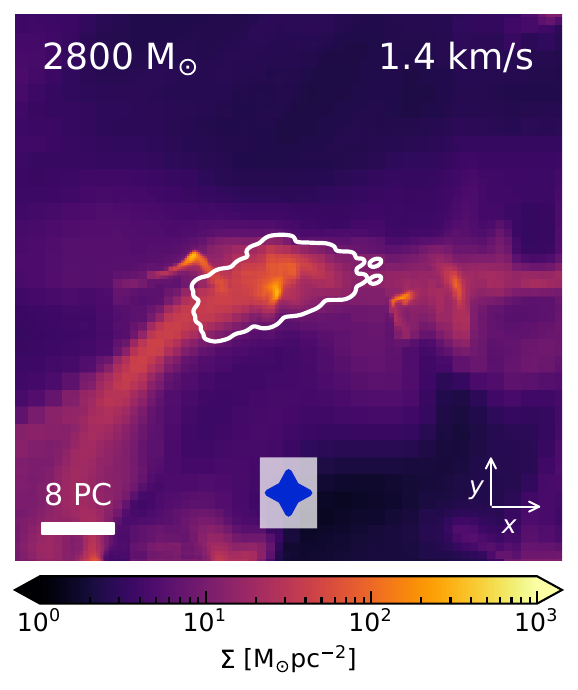}
        \includegraphics[width=0.235\textwidth]{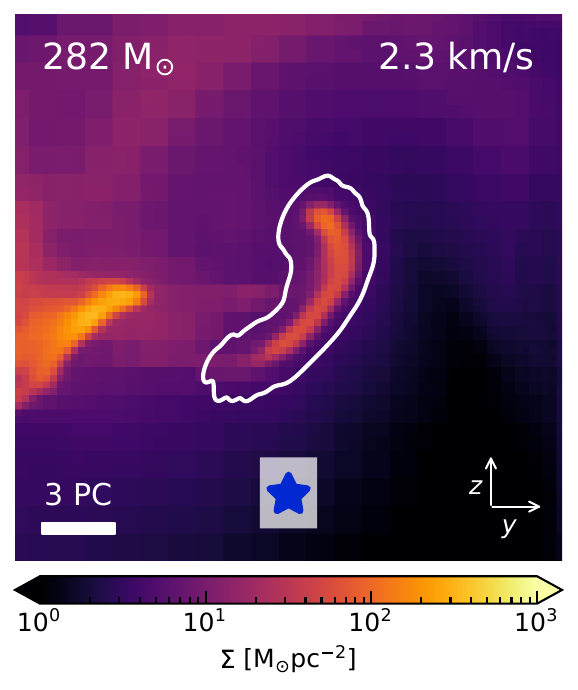}
        \caption{Example clouds from the LS-no-driving stratified box simulations  \citep{ColmanEtAl2022}.}
        \label{fig:example_clouds_LS}
    \end{subfigure}

    \begin{subfigure}[b]{\textwidth}
        \centering
        \includegraphics[width=0.235\textwidth]{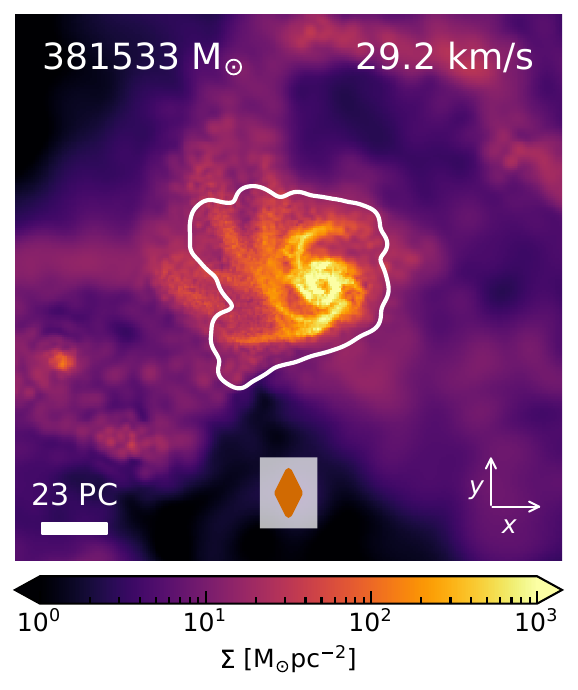}
        \includegraphics[width=0.235\textwidth]{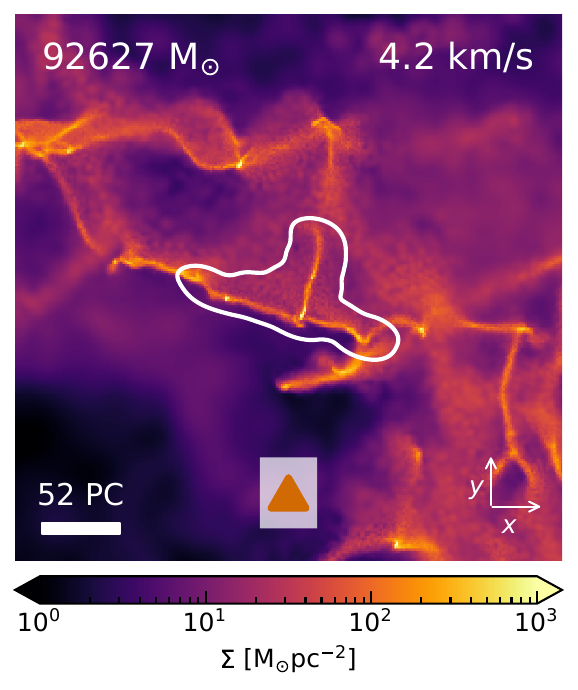}
        \includegraphics[width=0.235\textwidth]{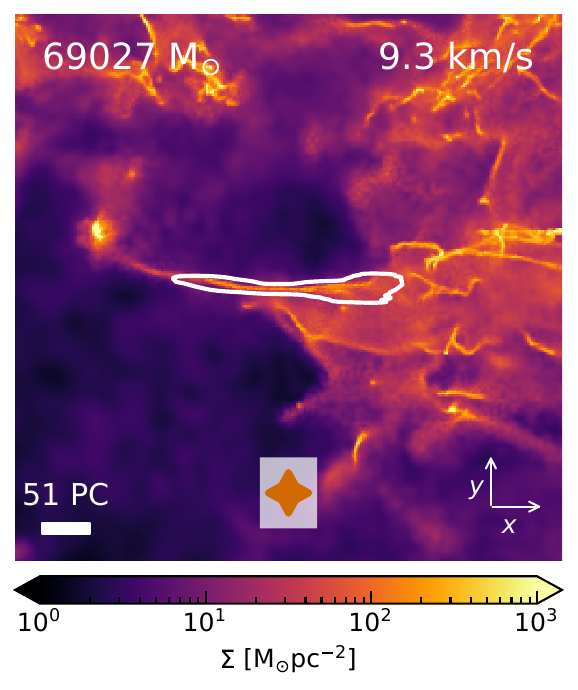}
        \includegraphics[width=0.235\textwidth]{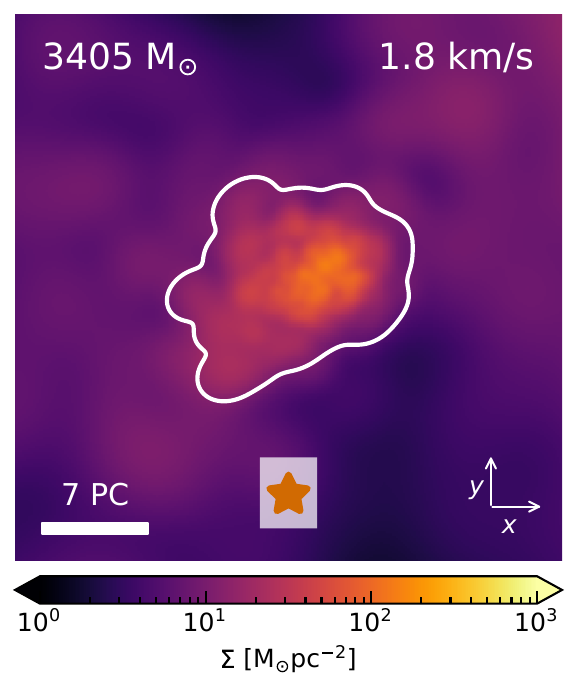}
        \caption{Example clouds from the M51 full galaxy simulation \citep{TressEtAl2020}.}
        \label{fig:example_clouds_M51}
    \end{subfigure}

    \begin{subfigure}[b]{\textwidth}
        \centering
        \includegraphics[width=0.235\textwidth]{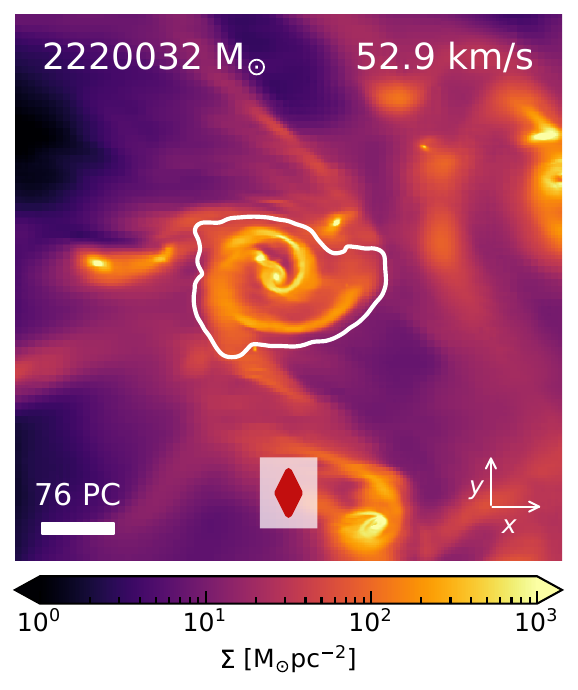}
        \includegraphics[width=0.235\textwidth]{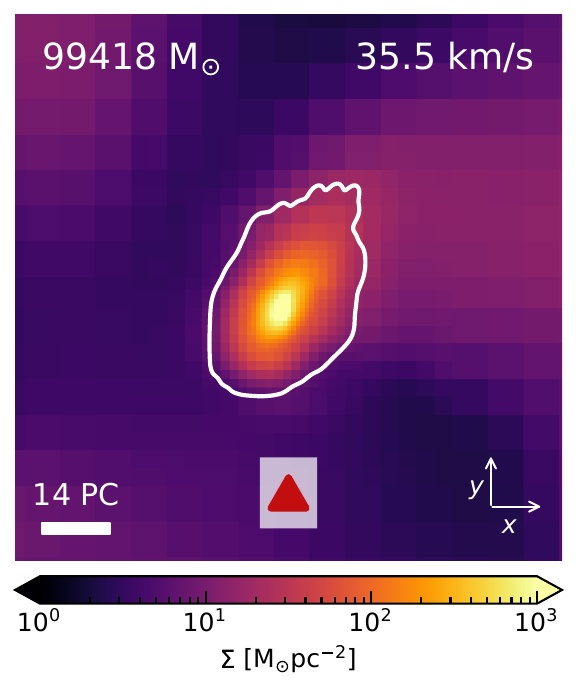}
        \includegraphics[width=0.235\textwidth]{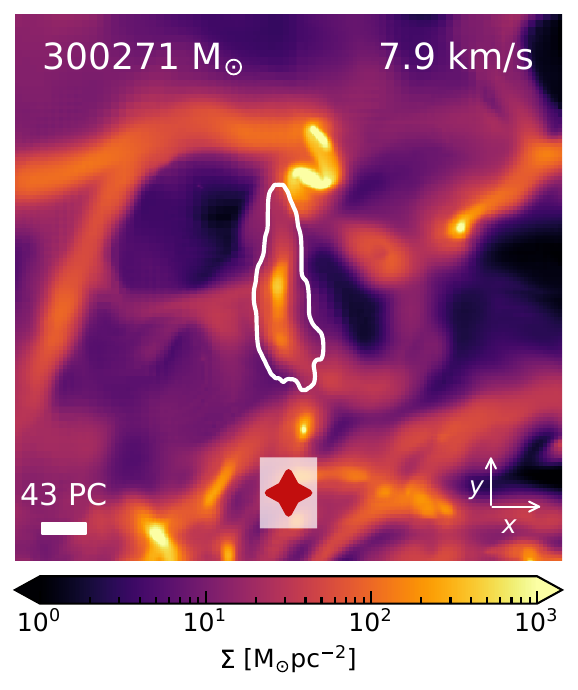}
        \includegraphics[width=0.235\textwidth]{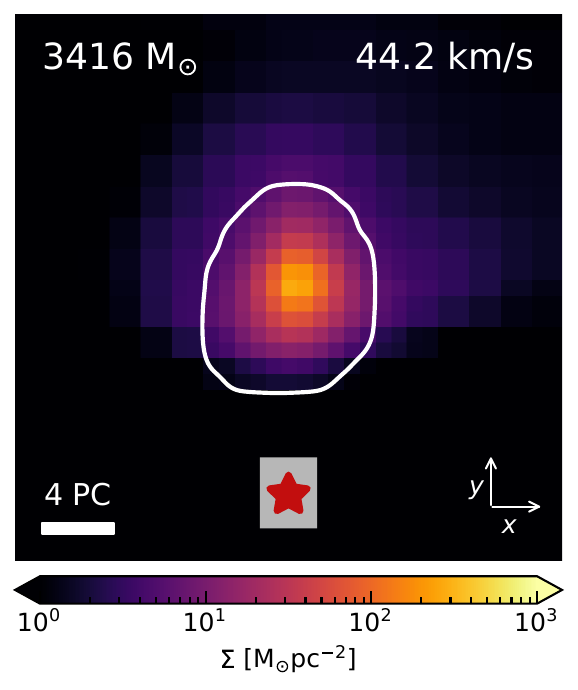}
        \caption{Example clouds from the Ramses-F20 full galaxy simulation \citep[Chapter 11]{BrucyPhd}.}
        \label{fig:example_clouds_ramses}
    \end{subfigure}
    
    \caption{Column density maps of selected clouds extracted from the simulations. Column density is calculated in a box of three times the maximum extent of the object. 
    Labels are added showing the cloud mass and velocity dispersion. The coloured symbol is used to refer to the cloud in the following figures in the paper.}
    \label{fig:example_clouds}
\end{figure*}

Figure~\ref{fig:example_clouds} displays some examples of extracted objects in the various simulations. 
The images show the projected column density in a region around the clouds, so we can also see their environment. The white contours show the projected cloud boundaries. The mass and velocity dispersion are listed at the top of the image. 
A coloured symbol is assigned to each example cloud, which will be used to identify them in further figures.

\begin{figure*}
    \centering
    \includegraphics[width=0.235\textwidth]{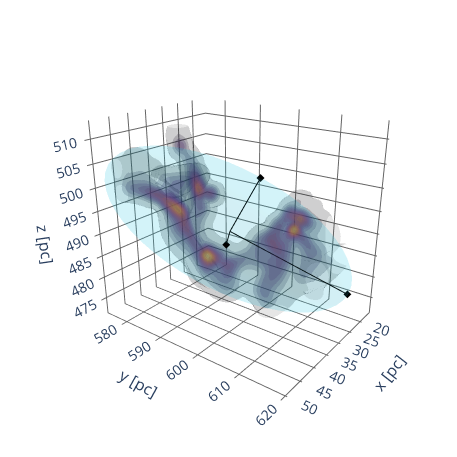}
    \includegraphics[width=0.235\textwidth]{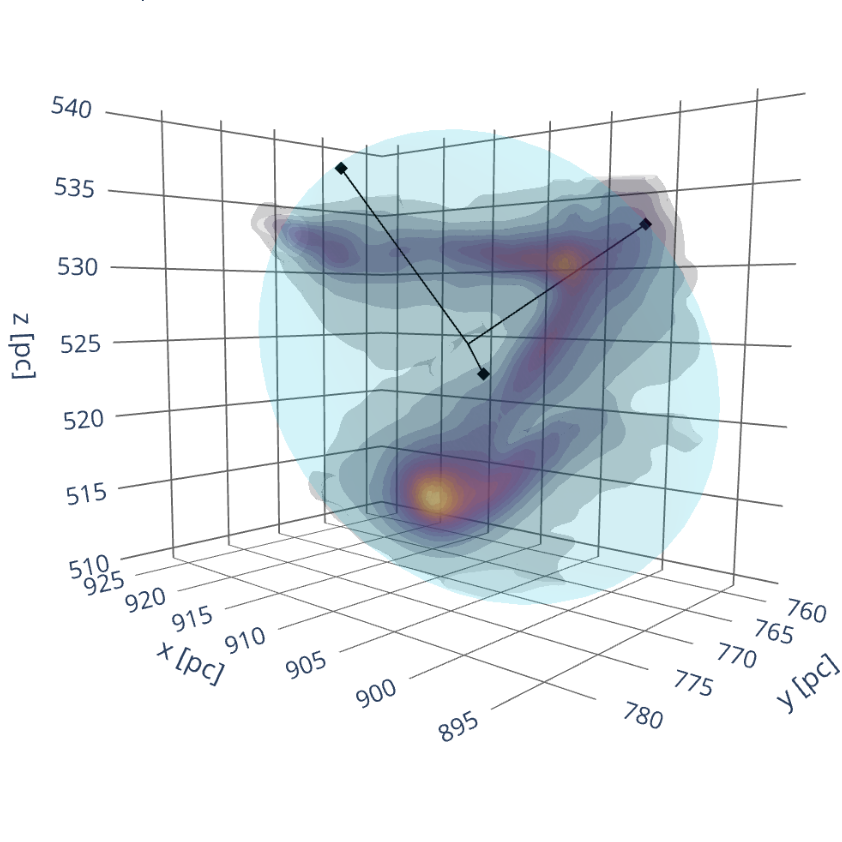}
    \includegraphics[width=0.235\textwidth]{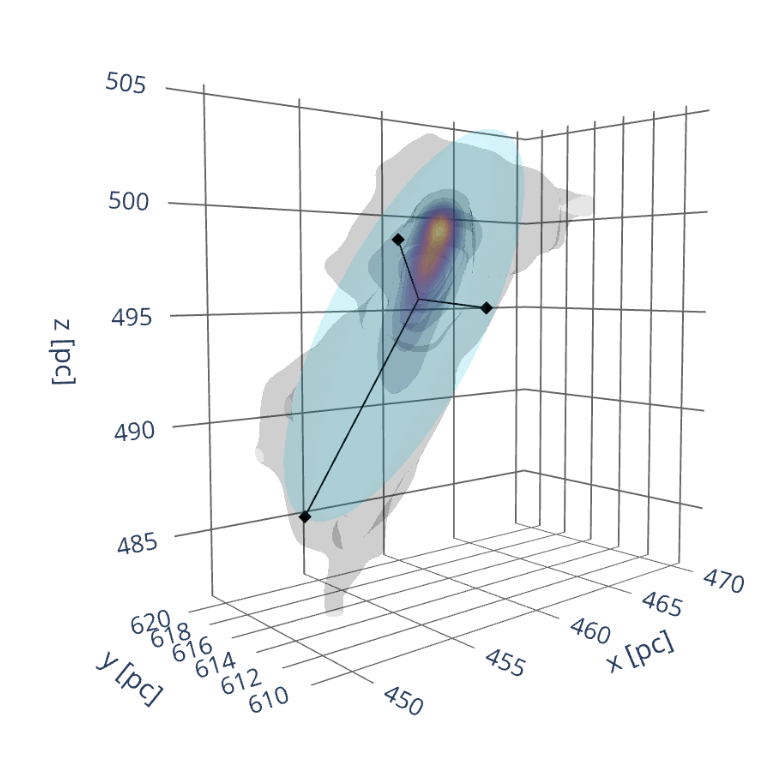}
    \includegraphics[width=0.235\textwidth]{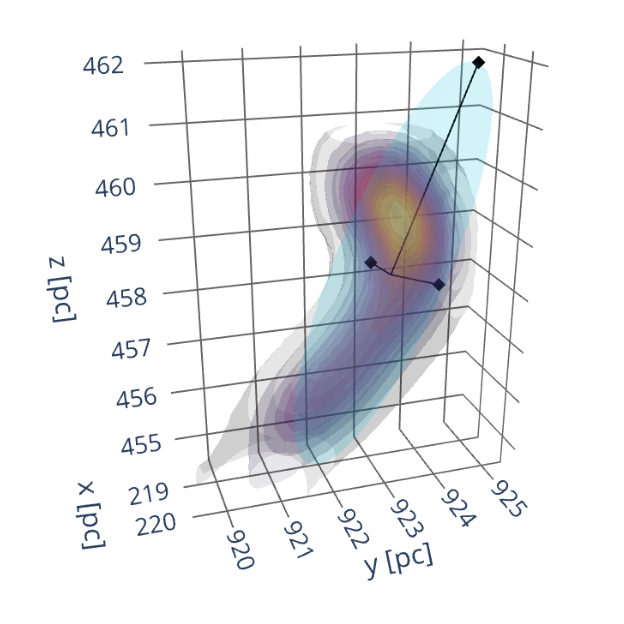}
    \caption{3D visualisations  of the clouds from Fig \ref{fig:example_clouds_LS}, extracted from the LS stratified box simulations. The density scale is linear. The ellipsoid approximation (in cyan) and its half-axes (in black) are also shown. }
    \label{fig:example_clouds_LS_3D}
\end{figure*}

The column density projections illustrate the wide variety in sizes and shapes we find. The clouds range from simple roundish shapes to very complex structures, which cannot be captured with simple measures such as the sphericity or the triaxiality. This is particularly clear when comparing the projected images for the LS clouds in Figure~\ref{fig:example_clouds_LS} with a 3D rendering of the same clouds in Figure~\ref{fig:example_clouds_LS_3D}, where we also show the ellipsoid approximation to the cloud from which its size is estimated (see Section~\ref{sec:mass_size_shape} and Appendix~\ref{appx:struct_prop}). In most cases, the cloud's shape is far from ellipsoid \citep[see also][]{Ebagezio2023}. However the corresponding sizes give a reasonable approximation to the extent of the objects. We also note that projection effects are significant. Especially large clouds can appear quite different when viewing them from different angles. Furthermore, there are often multiple clouds along the same line of sight. Consequently, the projections along the line of sight are limited to the cloud under consideration and its direct environment rather than being the integration throughout the entire simulation box.

In the stratified box simulations, large clouds are filamentary in nature and their internal structure is resolved. Small clouds typically do not have a resolved substructure and appear more regular in shape, which is true for all simulations. In the galaxy simulations, some of the largest clouds have disk-like shapes as illustrated in the left panel of Figure~\ref{fig:example_clouds_M51} and~\ref{fig:example_clouds_ramses}. These are unrealistic and result from a lack of resolution.

\subsection{Notes on comparing with cloud catalogues from observations}
\label{sec:notes_observations}

Many catalogues of ISM structures have been compiled from observations using various techniques.
A complete and robust comparison of simulation results to these catalogues requires the creation of synthetic observations, which could then be processed by the same pipeline as the observations to extract the clouds in the exact same way. This is clearly beyond the scope of this work. Nevertheless, it is useful to compare the trends we find in the simulations, which will be discussed in the following sections, to the ones observed in the Milky Way and nearby galaxies.

The classic molecular cloud catalogues are obtained from CO observations. The main CO isotopologue, $^{12}$CO, is detectable at relatively low densities \citep{2006ARA&A..44..367S}, potentially as low as our cloud extraction density threshold of  30 cm$^{-3}$, provided there is molecular gas at these densities. Emission lines from the $^{13}$CO isotopologue trace somewhat higher densities.
The spectral information can be used to try to disentangle overlapping projected clouds.
The observed line-width of the transition is assumed to be set mainly by turbulence.
While the dimension of the observed clouds is typically derived from the circularised radius of the cloud area in all the catalogues, the mass is derived with different methods that include different assumptions and approximations. A key parameter in the mass and radius estimate is the distance of the cloud. In the Milky Way, this is typically derived using a model for the rotation curve of the Galaxy, a method which is prone to large uncertainties \citep[see e.g.][]{Reid2022}. Thanks to the recent effort to create 3D dust maps \citep[e.g.][]{Chen_et_al2019,Leike_et_al2020}, it has become possible to accurately estimate the distances to clouds detected in dust extinction for structures up to a distance of $\sim$ 400 pc, offering an alternative to CO cloud catalogues.
\citet{Chen_et_al2020} detected clouds directly in PPP space, but cloud properties were still derived from 2D projections.
Very recently, \citet{Dharmawardena_et_al2023} and \citet{Cahlon_et_al2023} calculated the cloud mass and size from the reconstructed 3D dust density distributions.
3D dust maps do not provide kinematic information and so one cannot recover the velocity dispersion with this data alone. 
So far, no one has attempted to combine spectral line and 3D dust information to derive cloud velocity dispersion.
Extra-galactic surveys of MCs in nearby galaxies do not suffer from the same distance uncertainties, but are typically sensitive to only the most massive MCs in all but the closest Local Group galaxies \citep{Rosolowsky_et_al2021}.

All current Milky Way molecular cloud catalogues have specific lower limits in the recoverable masses and radii. In particular, the smallest recoverable radius depends on the spatial resolution of the observations.
The smallest observable mass depends on the sensitivity. Both resolution and sensitivity decrease with increasing cloud distance.
As a consequence, catalogues such as those of \citet{Rice_et_al2016} and \citet{Miville-Deschenes_et_al2017} that were derived from the first generations of CO surveys \citep{Dame_et_al2001} with spatial resolution of 8'5 and spectral resolution of 1.3 km s$^ {-1}$ contain on average larger and more massive clouds than more recent compilations of molecular clouds \citep[e.g.][]{Benedettini_et_al2020,Benedettini_et_al2021, Duarte-Cabral_et_al2021} that are extracted from the new generation of CO surveys with sub-arcminute spatial resolution and spectral resolution well below 1~km~s$^{-1}$. Moreover, in general all the catalogues of the observed molecular clouds are incomplete at the lower masses and radii. The threshold for incompleteness depends on the distance and thus can be variable for catalogues of the entire Milky Way.

It must also be noted that CO, especially $^{12}$CO, becomes optically thick in the denser parts of clouds \cite[see e.g.][]{Tielens10, Draine2011}. This means that we cannot see beyond a certain maximum surface density, leading us to underestimate of the mass if not properly accounted for, which is hard to do. Dust, on the other hand, is an optically thin tracer and thus does not have this problem.
This, and the other remarks made here, must be kept in mind when putting the results described in the next sections into an observational context.

\section{Comparison of cloud mass, size and shape}
\label{sec:mass_size_shape}

We start our comparison of the cloud catalogues by looking at the masses, sizes and shapes of the clouds. The cloud mass is simply the sum of the masses of all of the cells/particles it contains.
The size is determined through the diagonalisation of the inertia matrix. Using the resulting eigenvalues, we construct an ellipsoid of uniform density which has the same moments of inertia as the cloud. The size is then taken to be the geometric average of the axis lengths of this ellipsoid. The corresponding equations can be found in Appendix~\ref{sec:hop_algo}. Figure~\ref{fig:example_clouds_LS_3D} illustrates that the ellipsoid approximation traces well the extent of a cloud.

An alternative definition of the size is the cubic root of the total volume of the cloud, which neglects the shape information. In the following, we use both definitions as they provide different insights.

\subsection{Mass and size distribution}

\begin{figure}
    \centering
    \includegraphics[width=\columnwidth]{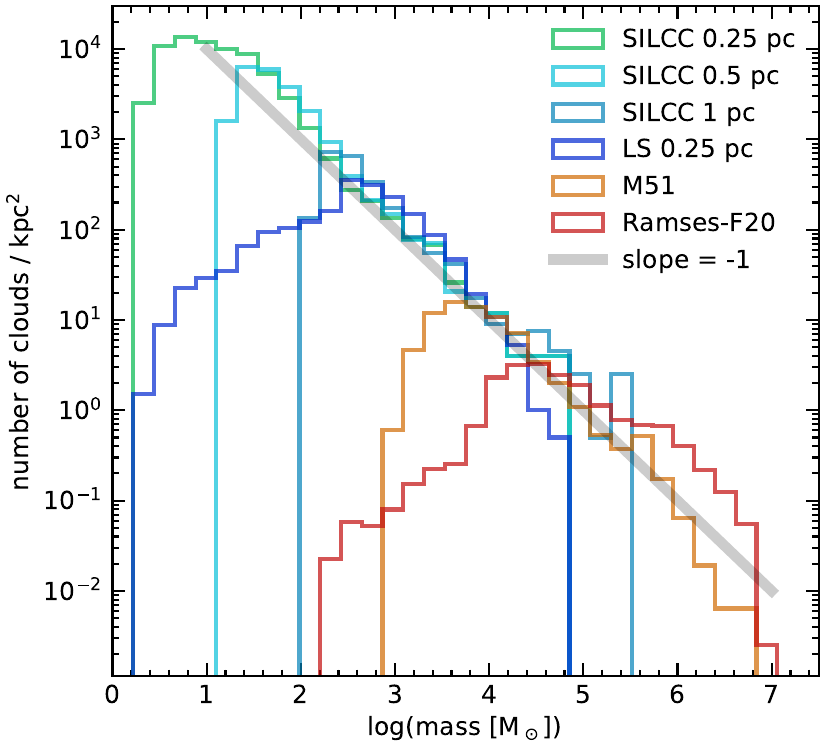}
    \includegraphics[width=\columnwidth]{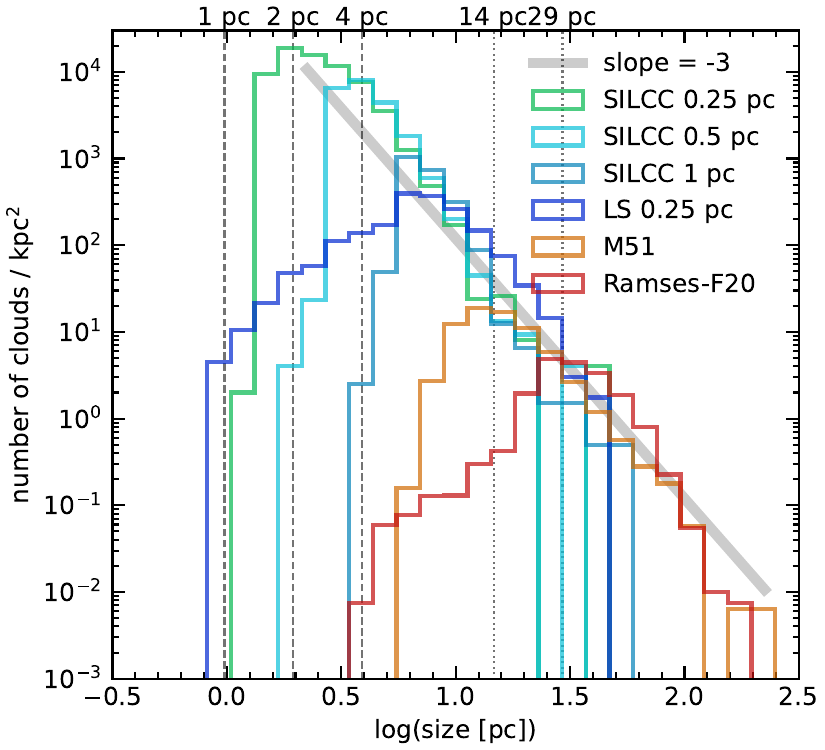}
    \caption{Comparison of the mass (top) and size (bottom) distribution of clouds in the different simulations, normalised according to the relevant surface area of the simulation. To guide the eye, we add a grey line indicating an appropriate power law slopes of -1 for the mass spectrum in the top panel and -3 for the size distribution in the bottom panel. The dashed lines correspond to multiples of the spacial resolution in the SILCC and LS simulations. The dotted lines show multiples of the resolution for the Ramses galaxy.}
    \label{fig:mass_size_distributions}
\end{figure}

The mass and size distributions are shown in Figure~\ref{fig:mass_size_distributions}. To facilitate the comparison, we normalise the histograms according to the surface area covered by the simulation; i.e.\ we show the number of clouds per unit area (here kpc$^2$). For the stratified box simulations (SILCC and LS), this is straightforward as their computational domain defines the area of the region. For the stratified boxes we combine the catalogues of several individual simulations and/or snapshots into one histogram to keep the plot readable. This is justified because the distributions do not differ significantly.
The distribution for SILCC-1pc contains the two SILCC runs with 1 pc resolution and different magnetic field strengths (SILCC-1pc-3$\mu$G and SILCC-1pc-6$\mu$G). The label LS includes all four LS runs with different driving strengths. The number of snapshots used for this compilation is listed in Table~\ref{tab:simulations}. The total surface area is then the box area multiplied by the number of snapshots. For the galaxy simulations, the relevant area is less straightforward to define. To get an estimate, we divided the face-on column density image into 1 kpc$^2$ tiles and counted how many of them have an average column density larger than 3 M$_\odot$ pc$^{-2}$. 

\subsubsection{Power-law tail}
The high mass end of the mass distribution of clouds is typically described by a power-law $\mathrm{d}N/\mathrm{d} \log M \propto M^{\alpha}$, or alternatively $\mathrm{d}N/\mathrm{d}M \propto M^{\gamma}$ with $\gamma = \alpha - 1$, optionally with a truncation. Unless stated differently, listed exponents correspond to $\alpha$.
With the appropriate normalisation, the agreement of the slope of both the mass and size distribution is remarkable. Especially for the mass spectrum which spans six orders of magnitude ranging from giant cloud complexes of $10^6$--$10^7$ M$_\odot$ down to tiny cloudlets of barely 10 M$_\odot$. The value of the power law exponent is close to -1, as indicated by the grey line. The exponent of the size distribution is around -3. For large masses and sizes, we typically see a cut-off in the distributions for the stratified box simulations due to the limited box size and the resulting limited total mass. Since we analyse only a handful of snapshots, values of the histogram below 1 cloud per kpc$^2$ correspond to rare events and are not statistically significant.

The only massive clouds which do not lie on the general mass spectrum are the most massive clouds in the Ramses-F20 simulation, which show an excess compared to the overall trend. We found 286 clouds with a mass above $10^6$ M$_\odot$, indicating that this is not a statistical fluctuation. Furthermore, the cloud mass distribution in this simulation seems to consist of multiple components. Upon closer inspection (cf. Figure~\ref{fig:mass_spectrum_inner_outer} and~\ref{fig:mass_fit}), we see a secondary peak around $5 \times 10^5$ M$_\odot$, indicating an excess of clouds with high masses, in both the Ramses-F20 galaxy and M51. These object are spurious disk-shaped clouds like the examples shown in the left column of Figure~\ref{fig:example_clouds_M51}, and thus not physical because of limited resolution. Once formed, they are very difficult to destroy by regular stellar feedback, which is limited as well by the overall worse resolution.

We fit the mass spectrum slope in the appropriate range for individual simulations (Appendix~\ref{appx:fit_mass_spectrum}). The resulting slopes for the stratified boxed and M51 agree well with one another, ranging from -1.16 to -1.35. Ramses-F20 has a shallower slope of -0.76. This difference is likely related to differences in the numerical feedback recipes used in each simulation.
The SN recipe used in  Ramses-F20 injects energy or momentum only in one of the cell neighbouring the exploding star, which makes it hard to destroy large structures. Meanwhile, because the injection radius is linked to a minimal number of cells, the recipe used in the \textsc{Arepo} simulation tends to overestimate the feedback effect when the resolution is poor. This difference alone could explain why clouds in Ramses-F20 are typically larger than those in M51.

\subsubsection{Turn-over and low-mass end}
The low-mass end of the spectra are shaped by resolution effects. In all cases, the mass and size distribution is incomplete for values below the peak. The SILCC setups show a relatively sharp cut-off slighty above four times the resolution limit, an effect of our requirement that a cloud has to contain at least 100 cells\footnote{the actual minimal size depends of the shape, but we note that $\sqrt[3]{100} \approx 4.64$.}. We note that \textsc{Ramses} simulations (LS and Ramses-F20) have different shapes at the low-mass end, with a much shallower transition between a peak and a low-mass cut-off. We discuss in Section~\ref{sec:resolution} that this stems from a difference in the grid refinement strategy used in \textsc{Ramses} and \textsc{Flash}. For LS, the peak is observed at a value slightly larger than the spatial resolution of the coarse grid (refinement level 8, corresponding to a cell size of 4 pc). For the Ramses-F20 galaxy, it is somewhere between twice and four times the coarse grid resolution. For M51 the cut-off is less straightforward to determine, since both the cell size and the cell mass vary in \textsc{Arepo} simulations. This results in a relatively wide mass and size distribution for the cells. We also typically have overall fewer cells in \textsc{Arepo} runs compared to AMR simulations (\textsc{Flash} and \textsc{Ramses}). The combination of these two aspects result in a low cloud mass cut-off that is largely determined by the minimum number of cells for a structure used by \textsc{Hop}, i.e.\ 100. At the highest densities reached in the M51 simulation, the typical particle mass is around 10-20~M$_\odot$, and so we see a cut-off in the cloud mass distribution at 100 times this value, namely 1000-2000~M$_\odot$. We note that this cut-off is not as sharp as in the SILCC runs because there are some cells in the M51 simulation with masses below 10~M$_\odot$ and hence \textsc{Hop} identifies a few clouds containing these cells with total masses below 1000~M$_\odot$.

It is worth noting that the ranges of masses and radii measured in observational data are in agreement with the ranges derived from simulations with a minimum cell size comparable to the spatial resolution of the observations. When quoting median values for the property distributions, it thus only makes sense to compare with studies which have comparable resolution.

\subsubsection{Predictions from theory}

Turbulent fragmentation theories predict a mass spectrum with a power-law for massive objects on all scales from GMCs to cores.
The self-similarity of gravity, which is argued to be the dominant force, results in $\mathrm{d}N/\mathrm{d}\log M \propto M^{\alpha}$ with $\alpha=-1$. Detailed derivations of turbulent fragmentation theories, which take into account several corrections, predict slopes which are slightly shallower \citep{Hennebelle_Chabrier2008, Hopkins2012}. Also accretion theories predict $\alpha=-1$ when the accretion rate scales as $\dot{M} \propto M^{2}$ \citep{Kuznetsova_et_al2018}. The global slope we find is indeed close to $-1$.

\subsubsection{Slope variations in observations}
\label{sec:slope_variations}

\begin{figure}
    \centering
    \includegraphics[width=\columnwidth]{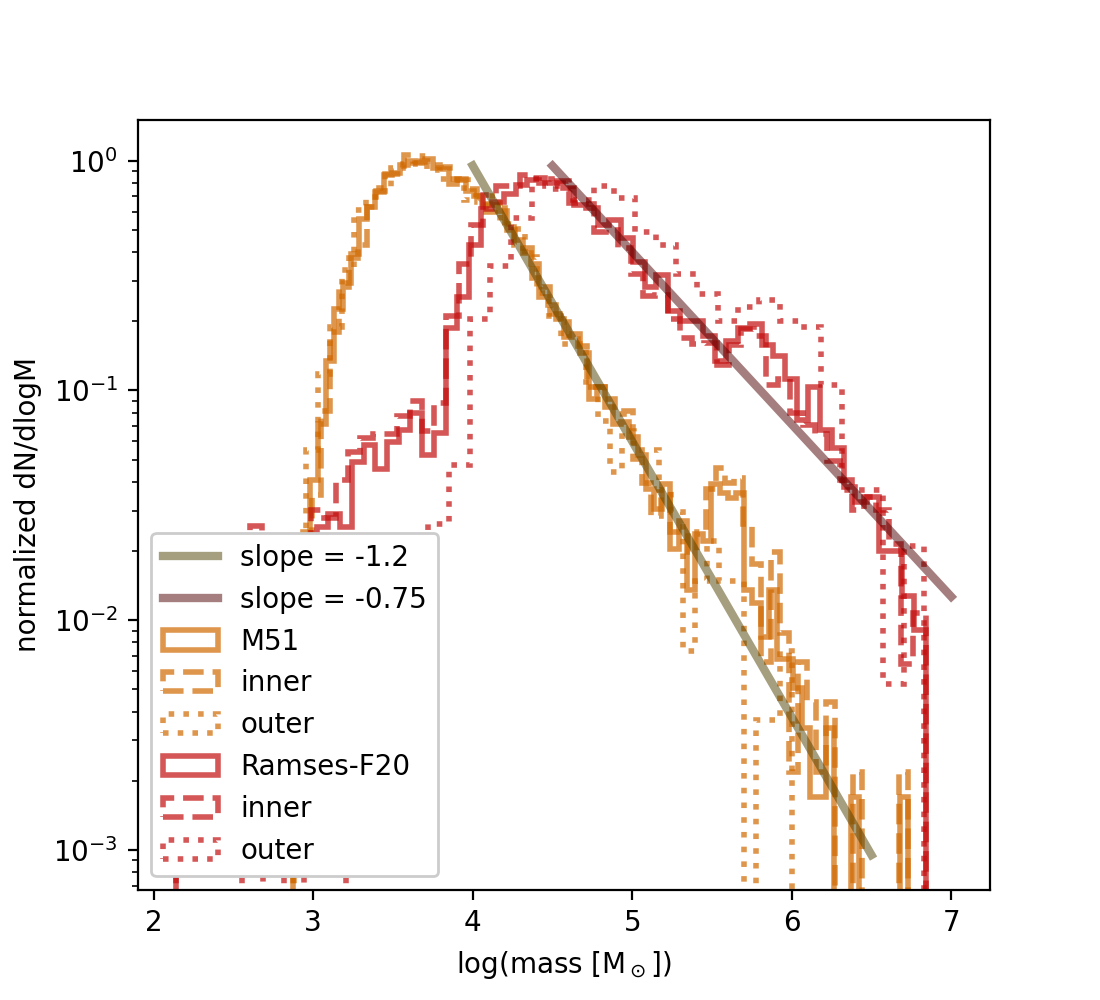}
    \caption{Comparison of the mass distribution of clouds in the inner 7 kpc of a galaxy versus those in the outskirts. While Ramses-F20 overall has a shallower mass spectrum than M51 (and the kpc boxes), there is no difference in the measured slope between the inner and the outer galaxy for either simulation.}
    \label{fig:mass_spectrum_inner_outer}
\end{figure}

Observationally, the obtained slope varies significantly between studies.
\citet{Roman-Duval_et_al2010} find $\alpha = -1.64 \pm 0.25$, consistent with the value of -1.5 obtained in older work \citep{Sanders_et_al1985, Solomon_et_al1987, Williams_McKee1997}.
\citet{Miville-Deschenes_et_al2017} find $\alpha = -2.0 \pm 0.1$ for their full Milky Way catalogue.
\citet{Rice_et_al2016}, who used the same underlying CO data as \citet{Miville-Deschenes_et_al2017}, report a shallower slope and differences between the outer Galaxy for which $\alpha = -1.2 \pm 0.1$, and the inner Galaxy where $\alpha = -0.6 \pm 0.1$.
A similar diversity was obtained by \citet{Rosolowsky2005} who found $\alpha = -0.5 \pm 0.1$ and  $\alpha = -1.1 \pm 0.2$ for the inner and outer Milky Way respectively, $\alpha = -1.9 \pm 0.4$ for the M33 galaxy and $\alpha = -0.7 \pm 0.2$ for the Large Magellanic Cloud.
A study of the actual M51 galaxy also reports variations with environments: $\alpha \approx -0.3$ for the centre and bar, $\alpha \approx$ -0.8 to -0.6 for the molecular ring and spiral arms and  $\alpha \approx -1.5$ for the inter-arm and outer galaxy regions \citep{colomboPdBIArcsecondWhirlpool2014}.
Moreover, the mass spectrum is not well-described by a power-law in all regions, suggesting different dominant mechanisms for the formation and destruction of clouds in the different parts of the galaxy. In summary, the observed slopes range from as shallow as -0.3 to as steep as -2.0.
This variety is in contradiction to the universality of our results which are all fairly close to -1.
Figure~\ref{fig:mass_spectrum_inner_outer} investigates more closely the mass spectrum in our two galaxy simulations, dividing the cloud population into two groups: clouds in the inner and outer galaxy with the boundary being at 7 kpc from the centre-of-mass.
We find no significant difference in the slope.
A small difference in typical cloud mass is obtained for clouds in the very centre of the galaxies, as discussed in Appendix~\ref{appx:radial_dep}. A similar trend was already found by
\citet{Tress2021}, who extracted clouds from their M51 galaxy using a different algorithm than we apply here.
Possibly the measured slope depends on the exact method with which clouds have been extracted from observations. The same holds for simulations as will be discussed in Section~\ref{sec:other_algos}. Note also that it can be difficult to estimate the completeness limit, which may also affect the fit of the slope. The situation is even more complex for catalogues for which the cloud distance is not uniform. Since the effective spatial resolution is lower at greater  distances, small clouds appear blended into larger objects, which results in a shallower mass spectrum slope.
A more rigorous comparison with observations through synthetic observations could provide a more direct conclusion about whether or not the simulation results are in agreement with the observations. 

\subsection{Mass--size relation}

\begin{figure*}
\begin{subfigure}[t]{0.47  \textwidth}{
    \centering
    \includegraphics[width=\textwidth]{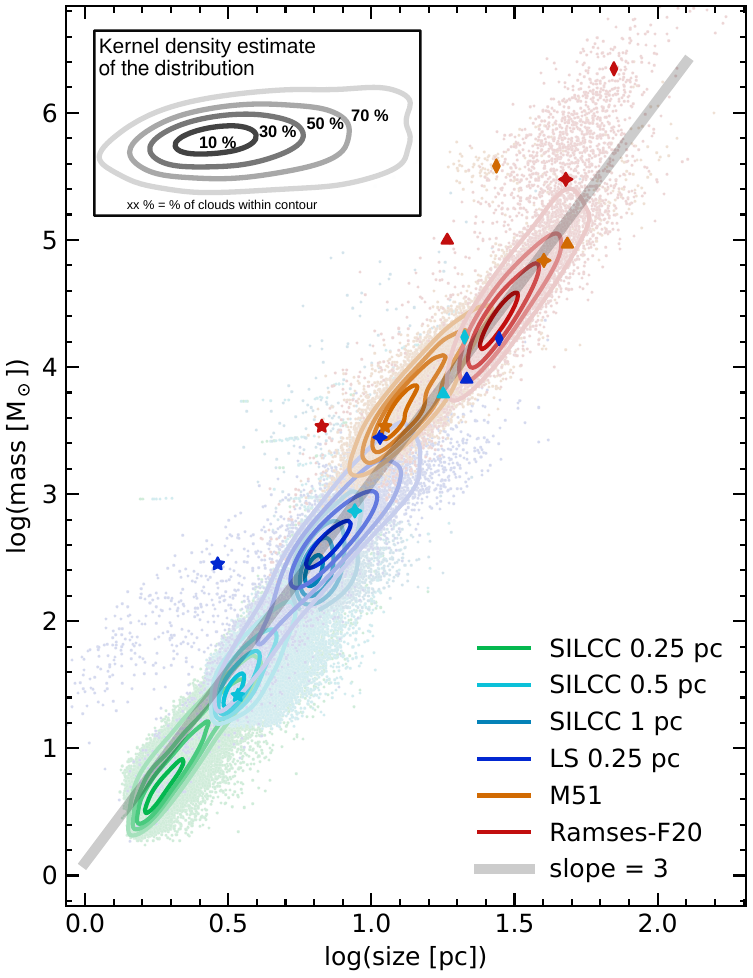}
    \caption{Mass-size relation with the size defined as the average between the three half-axis of the approximating ellipse, as determined by the structure algorithm. The grey line indicates a slope of 3.}
    \label{fig:correlation_mass_size}
}
\end{subfigure}
\hfill
\begin{subfigure}[t]{0.47 \textwidth}{

    \centering
    \includegraphics[width=\textwidth]{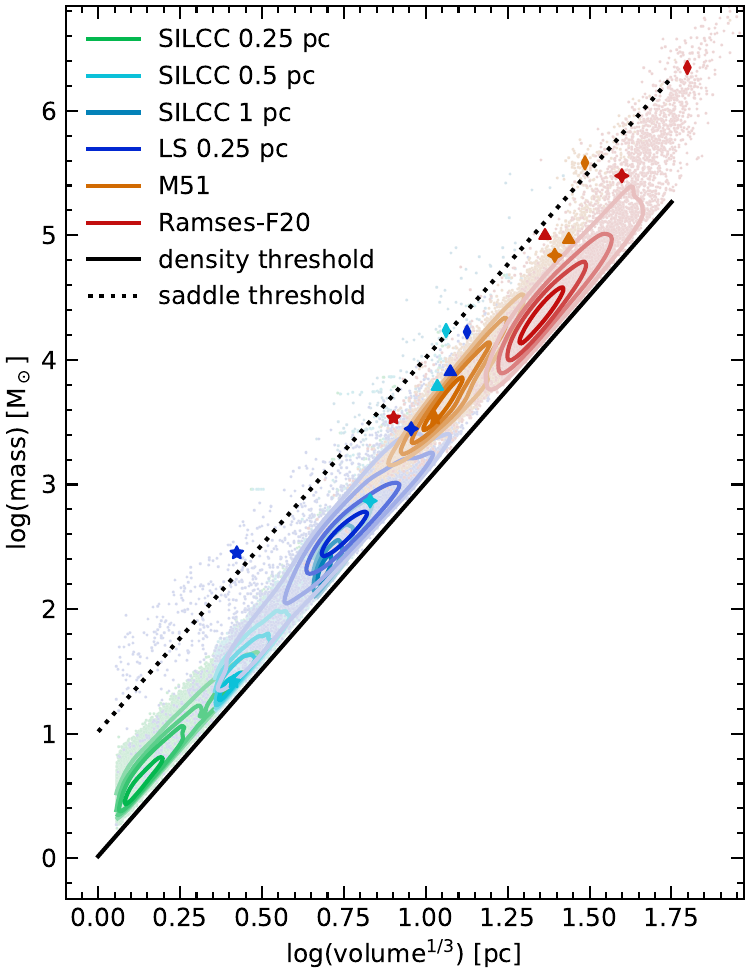}
    \caption{Correlation between the cloud size, this time computed as the cubic root of the volume, and its mass. The solid and dotted line are iso-density lines for the detection and saddle threshold respectively, as defined in Section \ref{subsec:struct_find_algo} and Appendix \ref{sec:hop_algo}. Their slopes are 3. 
    }
    \label{fig:mass_size_off}
}
\end{subfigure}
\caption{Correlation between cloud size and mass, with the size computed in two different ways. Each dot represents one cloud. The contour lines show a kernel density estimate of the distribution. Starting from the innermost line, approximately 10, 30, 50, and 70~\% of the distribution lie within the respective contours. The symbols highlight the example clouds from Fig. \ref{fig:example_clouds}.}
\label{fig:correlation_mass_size_alldef}
\end{figure*}

Observations show that cloud properties are not independent from each other \citep{Hennebelle_Falgarone2012, Heyer&Dame2015}.
A tight relation between cloud mass and size is reported by many studies \citep[e.g.][]{Solomon_et_al1987, Roman-Duval_et_al2010, Kauffmann_et_al2013, Miville-Deschenes_et_al2017}.
In Figure \ref{fig:correlation_mass_size}, we plot the relation between the mass and size of each cloud for the various simulations of our comparison set. We remind the reader that the size of the cloud is defined as the average between the three principal axes, as identified by the structure algorithm.
The tight correlation we see with a power law slope of 3 is expected and is due to the way clouds are extracted. By construction, the mean density of the cloud cannot be below the chosen threshold of 30 cm$^{-3}$ above which cells are considered by the cloud detection algorithm. 
Also, it is rather improbable to have an average density above the saddle density threshold. Unless the density gradient is very sharp, such a high density region will be associated with a less dense envelope which together form a bigger, more massive and on average less dense cloud.
In Figure \ref{fig:mass_size_off}, we replace the size by the cubic root of the volume, and by doing so, remove the information about the shape. 
We can then see clearly that almost all the clouds have indeed an average density between the algorithm density and saddle thresholds.

It is thus interesting to understand the spread and the outliers in both of these figures.
The spread in Fig. \ref{fig:correlation_mass_size} is mainly due to the shape of the clouds. 
Indeed, for a given mass and volume, a spherical cloud will have a smaller effective size than a more elongated one. We study in detail the distribution of shapes in Section \ref{subsec:shape}.
The simulation LS 0.25 pc stands out from the others with a significant number of small clouds with high densities. 
These are dense clumps where the low density envelope has been stripped by stellar feedback, specifically ionising radiation. This is less destructive than SN feedback, and in a cloud with dense substructure, it tends to preferentially blow away the diffuse gas, leaving the isolated dense clumps, which are then picked up by the cloud detection algorithm. This form of feedback is not included in the other simulations, which either include only SN feedback (SILCC, M51) or SN feedback plus localised pre-SN heating (Ramses-F20), which explains why we only see this population of objects in the LS simulations.

\subsubsection{Relation between the number of dimensions and the M-R slope}

\begin{table*}
    \centering
    \caption{Compilation of observational studies, with summary of their underlying data used to produce the molecular cloud catalogue, and their reported mass-size relation.}
    \begin{tabular}{l l c c}
        Reference                       & Data type & Milky Way region          & M-R slope\\
        \midrule
        \citet{Rice_et_al2016}          & $^{12}$CO(1--0)           & full           & 1.98\\
        \citet{Miville-Deschenes_et_al2017}     & $^{12}$CO(1--0)           & full           & 2.2\\
        \citet{Roman-Duval_et_al2010}   & $^{12/13}$CO(1--0)        & Galactic ring  & 2.36\\
        \citet{Benedettini_et_al2020}   & $^{12}$CO(1--0)           & outer Galaxy   & $\approx$ 2\\
        \citet{Benedettini_et_al2021}   & $^{13}$CO(1--0)           & outer Galaxy   & 2.13\\
        \citet{Duarte-Cabral_et_al2021} & $^{13}$CO(2--1)           & inner Galaxy   & $\approx$ 2.3\\
        \citet{Ma_et_al2022}            & $^{12/13}$CO(1--0)        & third quadrant & 2.24\\
        \citet{Chen_et_al2020}          & 3D dust (projected) & solar neighbourhood & 1.96\\
        \citet{Dharmawardena_et_al2023} & 3D dust             & solar neighbourhood & $\approx 3$ \\
        \citet{Cahlon_et_al2023}        & 3D dust             & solar neighbourhood & 2.9\\
        \citet{Cahlon_et_al2023}        & 3D dust (projected) & solar neighbourhood & 2.1
    \end{tabular}
    \label{tab:slope_MR}
\end{table*}

In Table~\ref{tab:slope_MR} we list some of the recent mass-size relation slopes obtained from observed cloud catalogues.
Both CO and dust extinction studies report mass-size relations with an exponent of about 2 when cloud properties are obtained from projections on the sky. This suggests a constant surface density for clouds \citep{Larson1981, Lada_Dame2020}.
Some studies report a slightly steeper slope, but \citet{Ballesteros-Paredes_et_al2019} demonstrated that this is likely an effect of overlapping clouds.
Interestingly, when cloud properties are calculated from 3D data, i.e. the radius from the volume and the mass from the volume density, a scaling relation of $M \propto R^3$ is obtained, in line with the global relation established in our simulations.
\citet{Cahlon_et_al2023} investigated exactly the effect of projection on the mass-size relation in their study of clouds extracted from 3D dust maps. 
They obtained $M \propto R^{2.9}$ for 3D clouds but recovered a scaling of $M \propto R^{2.1}$ for projected clouds, in agreement with the classical observed relation.
This effect was predicted by theory \citep{BallesterosParedesMacLow2002, Ballesteros-Paredes_et_al2012, Ballesteros-Paredes_et_al2019} and previous analysis of hydrodynamics simulations \citep{Shetty_et_al2010}.
These works indicate that the slope is determined by the way the size is measured and the underlying number of dimensions. Using the area results in a slope of 2 while using the volume results in a slope of 3. Note that clouds are thought to be fractal in nature \citep{falgaroneEdgesMolecularClouds1991,elmegreenFractalOriginMass1996,elmegreenFractalStructureGalactic2001} and thus in principle neither of these methods uses a good approximation for the shape.
This behaviour is due to the cloud mass being dominated by low density, volume-filling material. This can be seen clearly in the 3D visualisations of our extracted clouds (Figure~\ref{fig:example_clouds_LS_3D}). This makes the mass sensitive to how the cloud boundaries are defined which relates to the shape.
We thus can expect the mass--size relation to change when studying denser structures, such as dense clumps, for which this condition may no longer hold.
The LS simulations do contain a the population of small objects which have been stripped of their envelope. Interestingly, they indeed follows a relation which is shallower than 3, as can be seen in Figure~\ref{fig:correlation_mass_size}.

\begin{figure*}[p]
\begin{subfigure}{\textwidth}
    \centering
    \includegraphics[width=0.8 \textwidth]{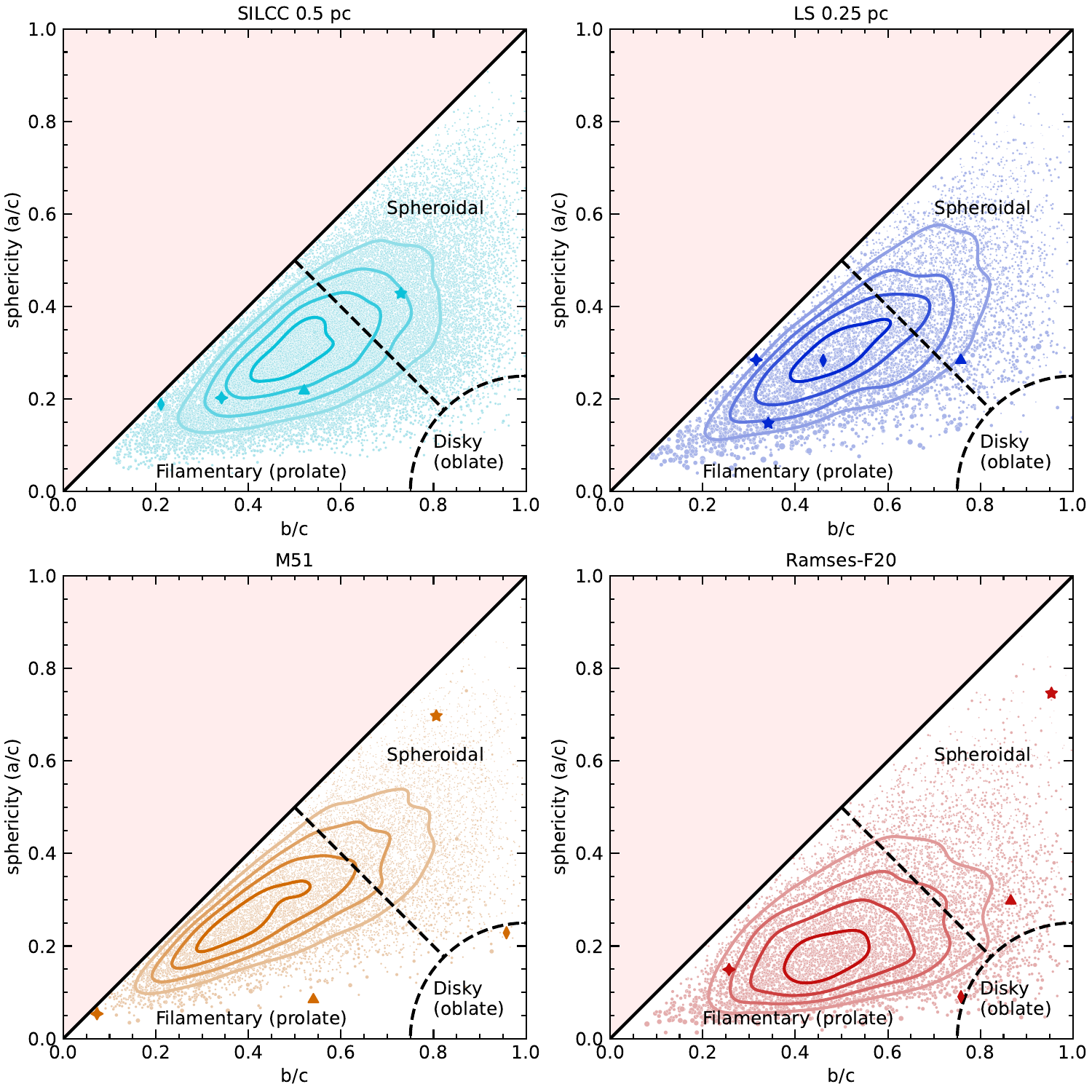}
    \caption{Relation between the ratio of the middle half-axis over the largest $b/c$ and the ratio of the smallest half-axis over the largest $a/c$ assuming clouds have a ellipsoidal shape. The characterisation of the shape the one used in \cite{vanderwelGeometryStarformingGalaxies2014}. The contour lines and symbols are the same as in Fig. \ref{fig:correlation_mass_size}.}
    \label{fig:scatter_shape}
\end{subfigure}

 \vspace{1em}
 
\begin{subfigure}{\textwidth}
    \centering
    \includegraphics[width= 0.82 \textwidth]{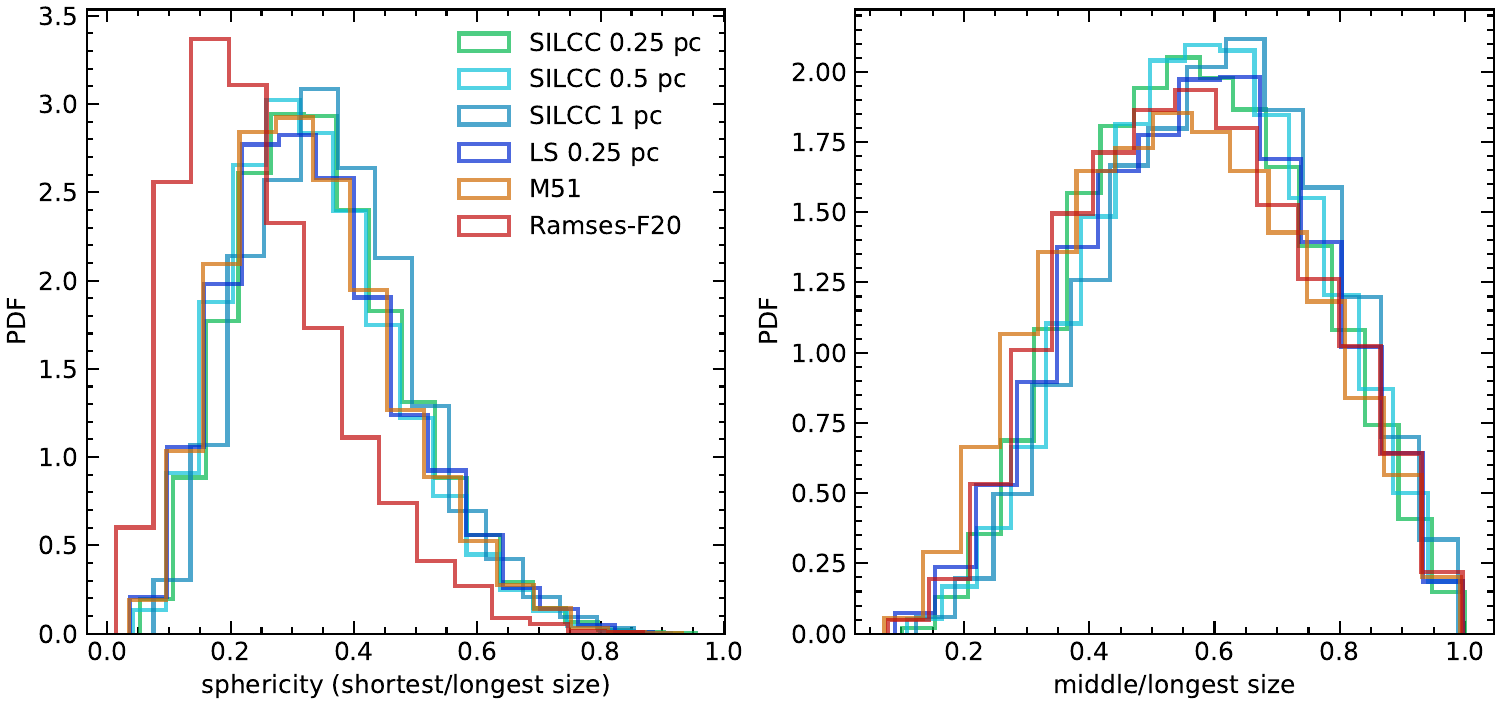}
    \caption{Marginal distributions of the ratio of the ellipsoidal half-axes of the clouds between the simulations.}
    \label{fig:shape_marginal_distribution}
\end{subfigure}

 \caption{General shapes of the clouds within the simulations.} \label{fig:shape_distribution}
\end{figure*}

\subsubsection{Normalisation of M-R relation}

When $M = A R^{2}$ the normalisation $A$ provides a roughly constant surface density for clouds, or alternatively a constant average volume density when $M = A R^{3}$.
\citet{Lada_Dame2020} reported a systematic shift of the M-R relation to higher surface densities when clouds have been extracted from dust extinction versus $^{12}$CO, i.e. the normalisation of the relation is different for different tracers.
In fact, for any area in the sky the mass derived from CO is lower than the mass derived from cold dust.
Small differences can also be observed between results derived from $^{12}$CO and $^{13}$CO \citep{Benedettini_et_al2021}.
This discrepancy could be (partially) due to optical depth effects, which can be severe for $^{12}$CO as mentioned already in Section~\ref{sec:notes_observations}.
If the mass is mainly set by the low density material in the cloud envelop rather than the dense cores and filaments, the effect of optical depth on the mass determination should be limited. In this case, the normalisation is set by the density threshold of the cloud extraction algorithm or the equivalent sensitivity limit of the observations, as we concluded from Figure~\ref{fig:mass_size_off}.
The threshold sets a characteristic average density or column density which is a few times larger than the threshold value. 
\citet{Ballesteros-Paredes_et_al2012} demonstrate that this is due to the power law density PDF of clouds. The amount of mass within a certain density bin decreases rapidly with increasing density. The cumulative mass above a certain density threshold is thus quite sensitive to where you make the cut in the PDF.
This has indeed been demonstrated in observations \citep{Lombardi_et_al2010, beaumont_et_al2012}.

\subsection{Cloud shapes}
\label{subsec:shape}

We characterise the shape of a cloud by considering the ratio of the half-axes $a \leq b \leq c$ of the ellipsoid approximation of the cloud. 
We compute them from the matrix of inertia of the cloud, assuming the cloud is an ellipsoid with an uniform density (see Eq.~\ref{eq:halfaxes}). The ratio of the shortest half-axis $a$ to the longest $c$ defines how spherical the cloud is, and is called the sphericity. The ratio $b/c$ tells us whether the cloud is oblate, that is pancake-shaped or disk-shaped ($a \ll b \sim c$) or prolate, that is filamentary shaped ($a \sim b \ll c$). This is illustrated in Fig. \ref{fig:scatter_shape} where we apply the shape categorisation from \citet{vanderwelGeometryStarformingGalaxies2014}.
Note that, as we already discussed in Section~\ref{subsec:examples_clouds}, the 3D cloud structures are very complex, and the ellipsoidal approximation fails to account for the very different shapes that clouds can actually have. 
However they are a useful simplification to obtain the extents of the clouds and help classifying them. 

\subsubsection{Comparison of the typical shapes between simulations}

In Figure~\ref{fig:shape_distribution} we compare the distributions of the shape quantities, which are very similar for all the simulations, except for Ramses-F20.
The typical sphericity of the clouds is around 0.35 with values ranging from 0.05 to 0.8, and the middle half-axis is typically equal to 0.55 times the large one. This corresponds to rather prolate structures, such as filament hubs, bent filaments, merging disks and compound structures, or unresolved blobs.
We verified that this similarity between all the simulations is not an artefact caused by the cloud finding algorithm: \textsc{Hop} is indeed able to extract objects with largely varying shapes (see Appendix \ref{subsec:hop_test_shape}). 
If we look in more detail, we note that filaments, defined as structures with $a \approx b \ll c$, are common as can be seen in bottom left corner of the diagrams in Figure~\ref{fig:scatter_shape}.
In contrast, spherical clouds are rare as indicated by the emptiness in the top right corner of the diagrams. 
This indicates that most of the structures are connected with their environment, the filamentary ISM. Oblate disk or sheet shaped structures, i.e.\ structures with $a \ll b \approx c$, are also very rare. 
Ramses-F20 shows a clear excess of them compared to the other simulations. The presence of such disk-like structures is most likely linked with a lack of resolution that prevents further collapse in these regions as already discussed in the previous sections.

Overall, while there is some variety in sphericity within cloud populations, the different simulations feature very similar distributions of the shape parameter. 
Since here the shape of the clouds is approximated to match an ellipsoid, we can conclude that the global proportions of the clouds are not affected by the details of the physics and the geometry of the simulations.  
More elaborate measures, such as the fractal index, may provide more insights on the detailed structure of the cloud. This is however beyond the scope of this work.

\subsubsection{Discussion of example clouds}

To aid with the interpretation of the shape diagnostics and to improve our intuition, we discuss here the shapes of the example clouds displayed in Figure~\ref{fig:example_clouds}.
We start with the clouds extracted from the galaxies Ramses-F20 (fourth row) and M51 (third row) which are generally most massive and illustrate some extreme shapes, before moving to those extracted from the stratified boxes LS (second row) and SILCC (first row) which have properties that are more representative for the entire sample.

The most massive example for each galaxy (left panels, and indicated as a flat vertical diamond in the scatter plots) is a spurious disk with spiral arm pattern.
These are amongst the most massive clouds found in their respective simulation.
Their disk-like shape is reflected by the shape measures: the Ramses-F20 cloud has $a/c$ as low as 0.1 with $b/c \approx 0.75$, while the M51 cloud is a little thicker with $a/c \approx 0.2$ and $b/c \approx 0.95$. This indeed classifies their shape as disky in Figure~\ref{fig:scatter_shape}. Both these examples have a velocity dispersion high above the global relation due to ordered rotation.

Next, we take the second and third examples of each galaxy simulations, resulting in a collection of four clouds with comparable mass but a variety of shapes.
The Ramses-F20 cloud indicated by a triangle appears very isolated and has a smooth appearance with a density gradient towards the centre. It is an categorised as an oblate spheroid with $a/c \approx 0.3$ and $b/c \approx 0.85$.
The second example from M51 looks completely different. It consists of a filament hub system where one major filament is perpendicular to the other. The ellipsoidal shape approximation clearly breaks down in this case. While $a/c$ is indeed very low, $b/c \approx 0.55$. This inflates the average size which is almost four times larger than the smooth compact Ramses-F20 cloud of the same mass. This illustrates how the shape alters the mass-size relation in Figure~\ref{fig:correlation_mass_size}. If, on the other hand, we estimate the size as the cubic root of the volume, both clouds appear very similar (see Figure~\ref{fig:mass_size_off}).
The third example from each galaxy is a filament which is connected to a larger filament network. The example from M51 is one of the purest filaments in our entire collection with very low $a/c$ and $b/c$. The Ramses-F20 filament is somewhat thicker with $a/c \approx 0.15$ and $b/c \approx 0.25$~. Both of these seem to be very normal clouds when placed on the mass-size and size-velocity dispersion relation (cf. Figures~\ref{fig:correlation_mass_size} and\ \ref{fig:correlation_size_sigma}).

The final example for each galaxy features a small round cloud. For Ramses-F20 this is one of the smallest clouds recovered in this simulation. With $a/c \approx 0.75$ and $b/c \approx 0.95$ it is also one of the most spherical objects in the entire collection. We note that these are also the least resolved clouds, in which further substructure is subsequently harder to resolve. Extreme non-spherical shapes are therefore less likely to be driven from a natural cascade of turbulent motions.

The example clouds from the stratified box simulations have less extreme shapes. 
We see that clouds are shaped by a network of connecting filaments, which generally do not align, resulting in the range of $a/c$ and $b/c$ values shown in Figure~\ref{fig:shape_marginal_distribution}.
Of note is the fourth example of the LS runs: a comma shapes clump which has been stripped of its low density envelope, making it lie above both versions of the mass-size relation. Generally, a similar argument applies here for the resolution limit and the shape of the cloud. However, this effect is less severe compared to the two disk simulations for two reasons. First, in these simulations we find overall better resolved feedback physics and consequently better resolved turbulent motions. Second, the higher resolution and the overall lower masses for the clouds allows for a stronger impact of sudden external perturbations such as strong shock fronts, which might reshape the cloud to non-spherical objects.

\subsubsection{Dependence of shape on size}

\begin{figure}
\begin{subfigure}{\columnwidth}
    \centering
    \includegraphics[width=\columnwidth]{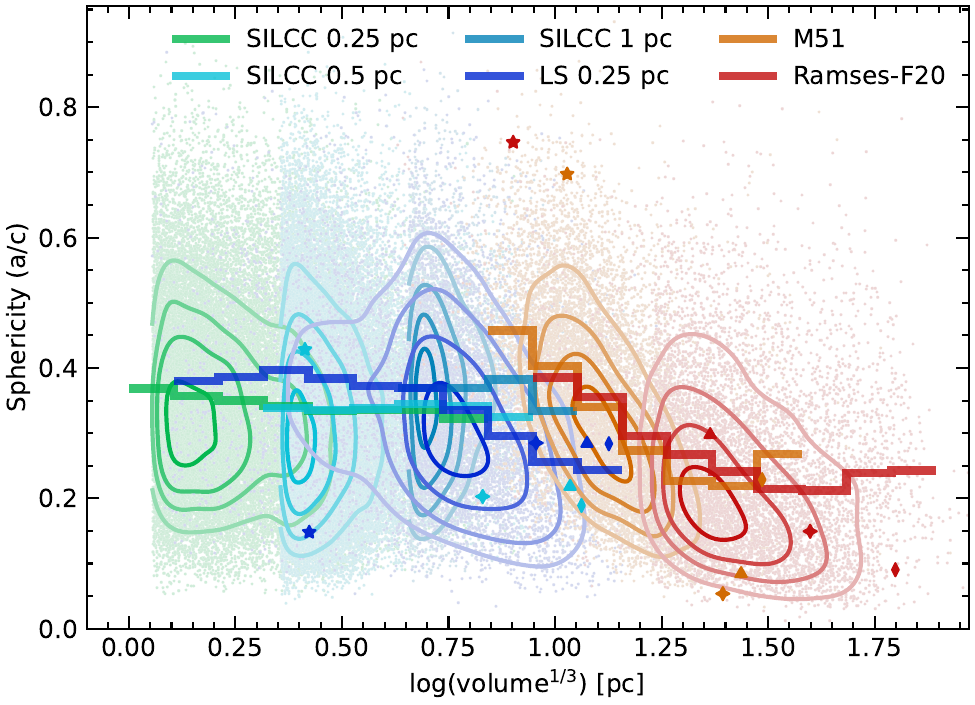}
    \caption{Relation between sphericity and the typical size, computed as the cubic root of the volume.
    Contour lines and symbols are as in Fig. \ref{fig:correlation_mass_size}.}
    \label{fig:sphericity_volume}
\end{subfigure}
\vspace{1em}
\begin{subfigure}{\columnwidth}
    \centering
    \includegraphics[width=\columnwidth]{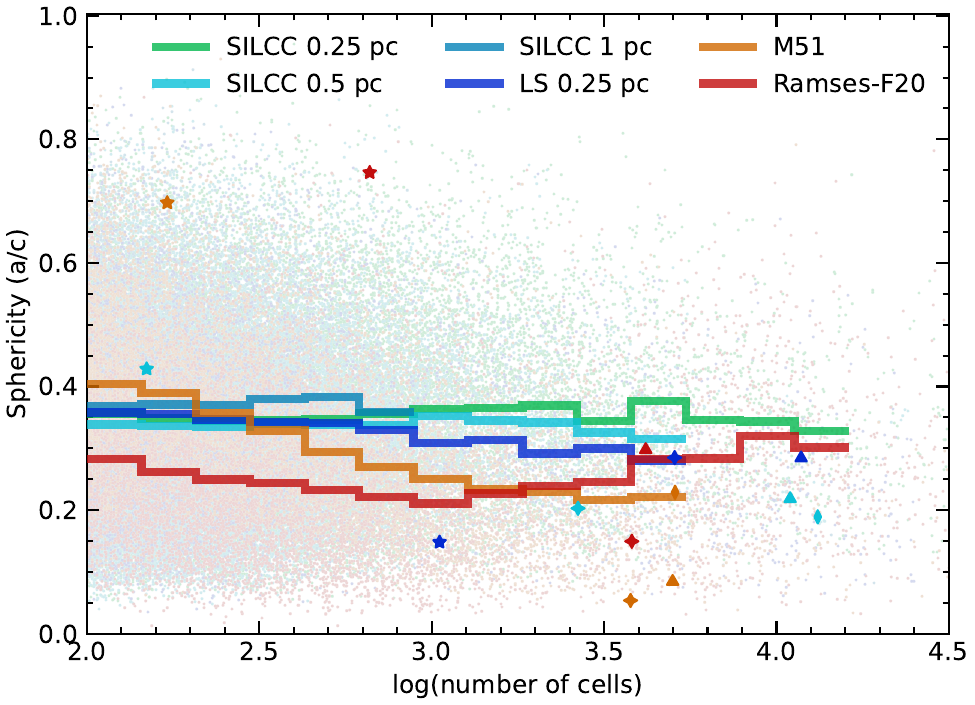}
    \caption{Sphericity distribution as a function of the number of cells. }
    \label{fig:sphericity_nbcell}
\end{subfigure}
    \caption{Sphericity as a function of structure size, computed from the volume (top panel) or the number of cells (bottom panel). The solid line depicts the mean in a logarithmic bin of size, and is computed only if the bin contains at least 100 clouds.}
    \label{fig:sphericity_volume_cells}
\end{figure}

From the examples discussed above, we might have gotten the impression that small clouds tend to be spherical. This is however not generally true.
Figure \ref{fig:sphericity_volume} shows the relation between the sphericity of a cloud and the typical size defined as the cubic root of the volume. We do not use the average size defined in the beginning of this section to avoid introducing any spurious correlation.
For all the stratified box simulations, the distribution stays unchanged for all size bins between 1 and 10 parsec. 
Interestingly, the mean value of the sphericity decreases when the typical size is around 10 pc. 
A possible interpretation is that large clouds are more likely to undergo external influence, like turbulent motion. Also, they may be younger objects which kept their initial filamentary shape. 
It may also be just a resolution effect: clouds with a smaller number of cells tend to be more spherical while a higher number of cells allows for more complex shapes. 
However, Figure \ref{fig:sphericity_nbcell} demonstrates that this is not the case: expect for M51, the distribution of the sphericity does not vary much when looking at objects with a different number of cells.
Although clouds with very high sphericity are found only when the number of cells is small, they represent only a tiny fraction of the sphericity distribution, which is well sampled for all cell numbers. In the stratified box simulations, there is no significant decrease of the sphericity with an increase of the number of cells.
Such a trend is however visible for the the less resolved full-galaxy simulations, especially in M51, and in a lesser extent in Ramses-F20.

\subsubsection{Shapes in observations}
In observations, clouds can be approximated by the 2D equivalent of our ellipsoids, i.e. ellipses, since they appear as 2D projections on the sky. In this case a minor and major axis is reported and their ratio incorporates the shape. Since there is no information about the size along the line of sight\footnote{This information is in principle retrievable from clouds detected in 3D dust. However, so far no study has reported sizes along three orthogonal axes.}, it is unclear whether this ratio is more comparable to our $a/c$ or $b/c$. If we assume that clouds are randomly oriented, then the observed value would represent an average of $a/c$, $b/c$ and $a/b$.
Taking the geometric mean of the peak value of the distribution for each of these ratios results in $(0.35 \times 0.55 \times 0.65)^{1/3} = 0.5$.
The SEDIGISM \citep{Duarte-Cabral_et_al2021}, FQS \citep{Benedettini_et_al2020, Benedettini_et_al2021} and \citet{Miville-Deschenes_et_al2017} catalogues provide the two projected sizes for their objects. Their ratio is typically between 0.2 and 0.8 and the distribution peaks between 0.4 and 0.6. For SEDIGISM and FQS, which have extracted clouds in a similar way, the peak is closer to 0.45, in line with the averaged value we find.
For the \citet{Miville-Deschenes_et_al2017} catalogue, where the sizes were derived differently, the peak is more towards 0.6, close to what we find for $b/c$.
As a test, we can also project our 3D clouds on a 2D plane and calculate the projected size as a geometric average of the major and minor axis of an approximate ellipse, equivalently to our ellipsoid approximations. This results indeed in a broad distribution with a peak around 0.5.

Thus, while an accurate comparison for the shape seems rather impossible, we can confidently say that simulation and observations agree that clouds are typically elongated. 

\section{Analysis of internal properties of clouds}
\label{sec:internal_properties}

\subsection{Size -- velocity dispersion relation}
\label{sec:size-sigma}

\begin{figure}
    \centering
    \includegraphics[width=\columnwidth]{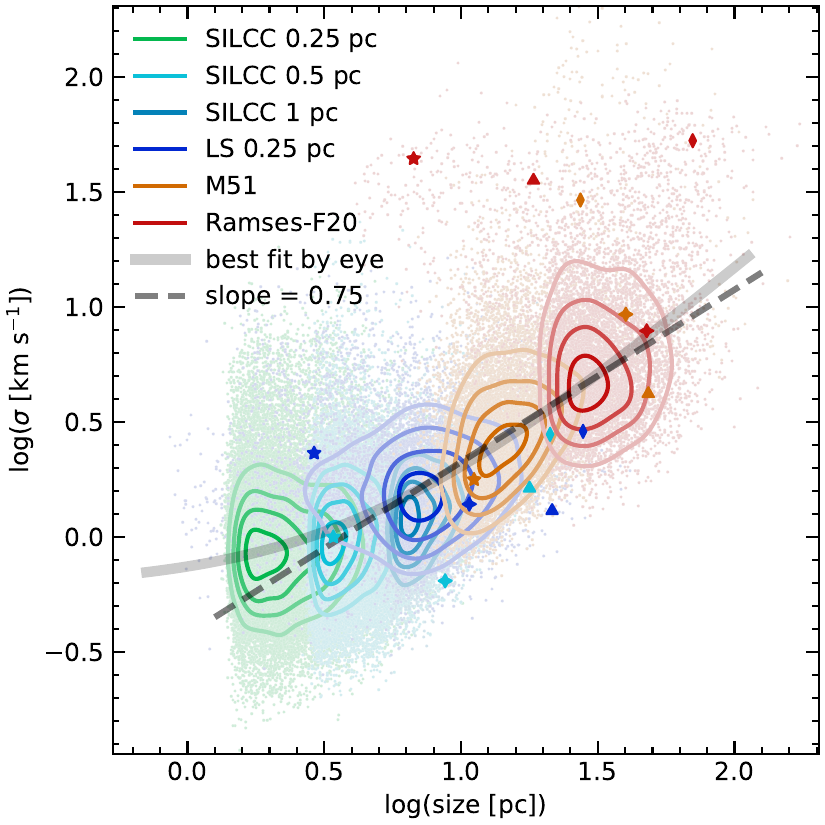}
    \caption{
    Cloud velocity dispersion as a function of the size. The dashed line illustrates a scaling of $\sigma$ with size $R^{0.75}$ while the curved solid line illustrates the trend of the centroids of each distributions. The contour lines and symbols are the same as in Fig. \ref{fig:correlation_mass_size}.}
    \label{fig:correlation_size_sigma}
\end{figure}

\begin{figure}
    \centering
    \includegraphics[width=\columnwidth]{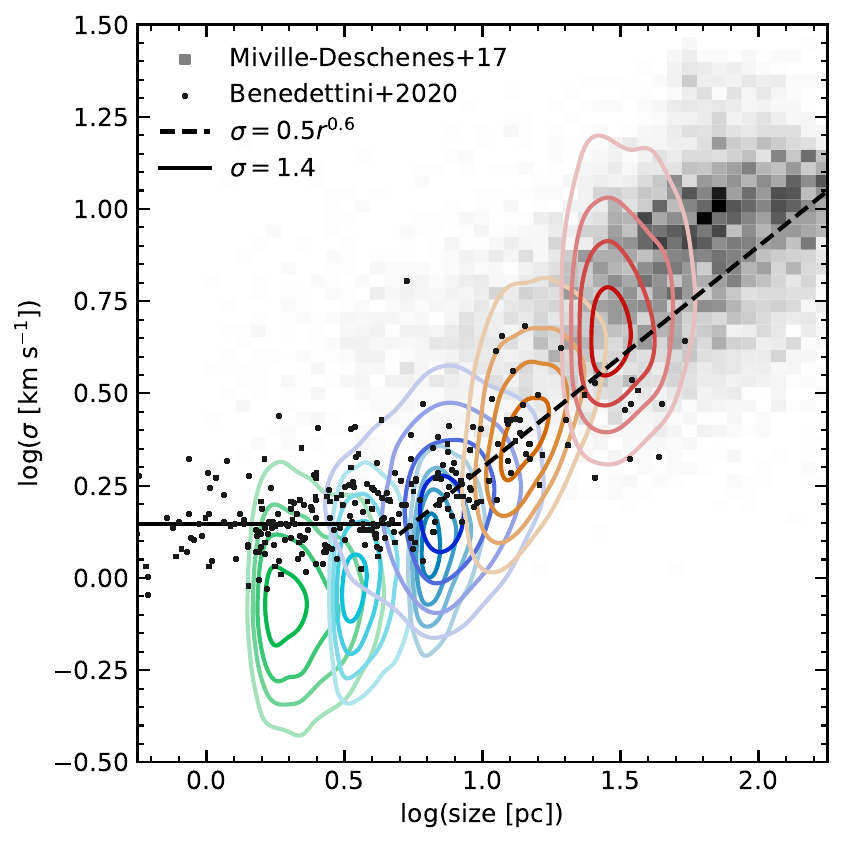}
    \caption{Comparison of the 3D size-velocity dispersion relation found in the simulation with several cloud catalogues from observations. Coloured contours are as in Figure \ref{fig:correlation_mass_size}. Note that observational data feature only the 1D line-of-sight velocity dispersion, which was here multiplied by $\sqrt{3}$ for comparison with the 3D velocity dispersion computed in the simulation. }
    \label{fig:observations_size-sigma}
\end{figure}

\begin{figure*}
    \centering
    \includegraphics[width=\textwidth]{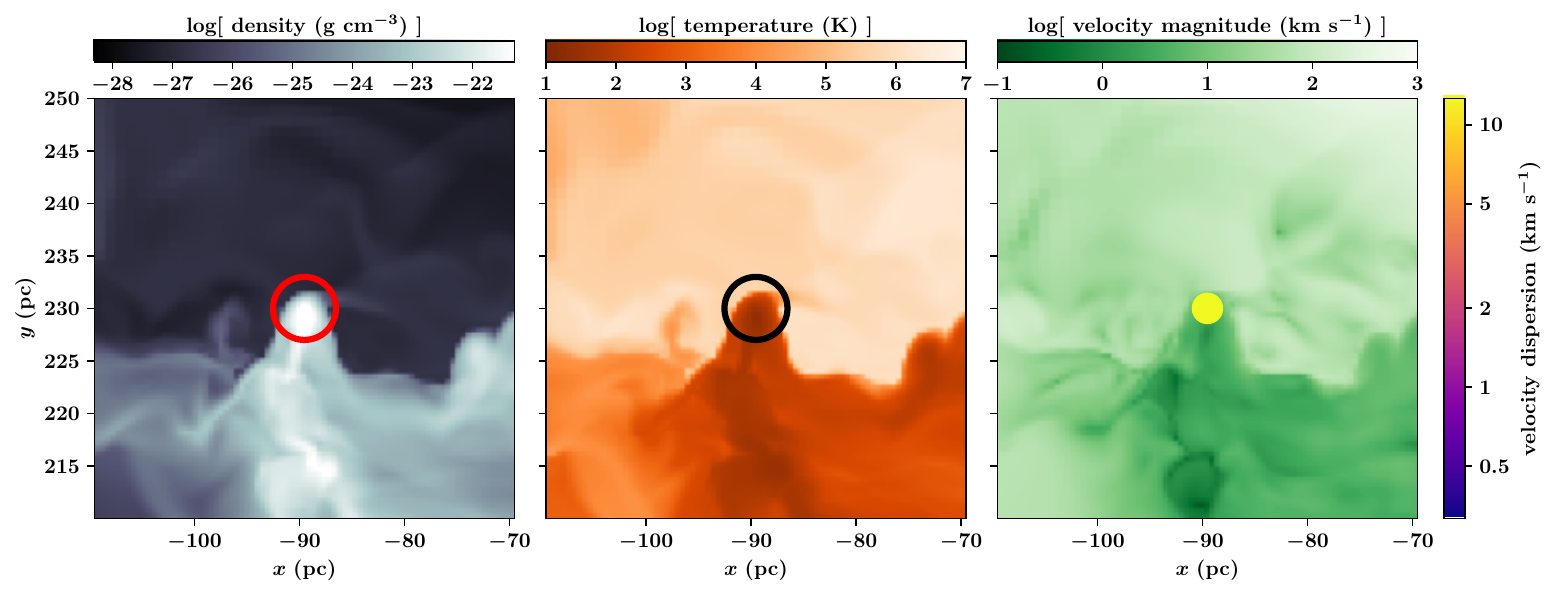}\\
    \includegraphics[width=\textwidth]{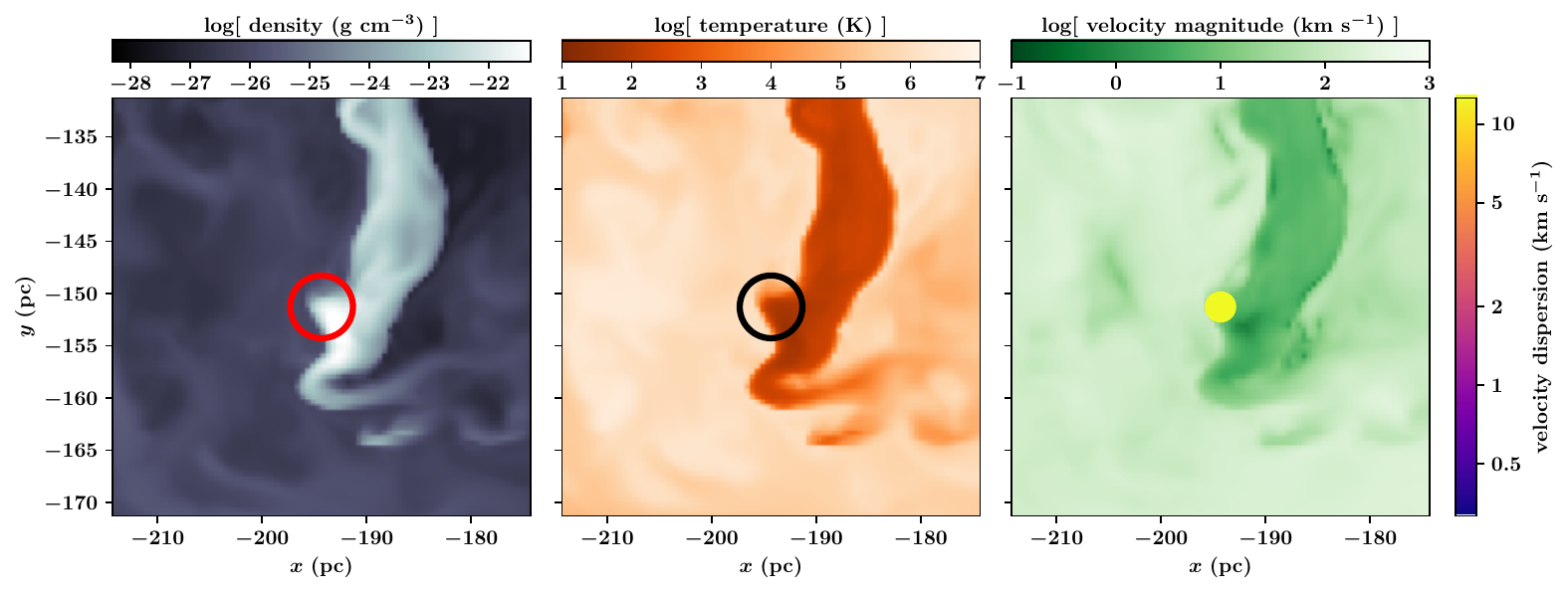}\\
    \caption{Two examples showing the positions of the clouds with their respective density (left), temperature (middle) and velocity environment (right). In both cases the clouds are located at the edge between cold gas and a hot SN-driven bubble. The resulting high velocity dispersion is thus due to SN-driven turbulence that mixes hot gas into the clouds rather than by gravitational collapse.}
    \label{fig:maps-dens-temp-velo}
\end{figure*}

\begin{figure*}
    \centering
    \includegraphics[width=\columnwidth]{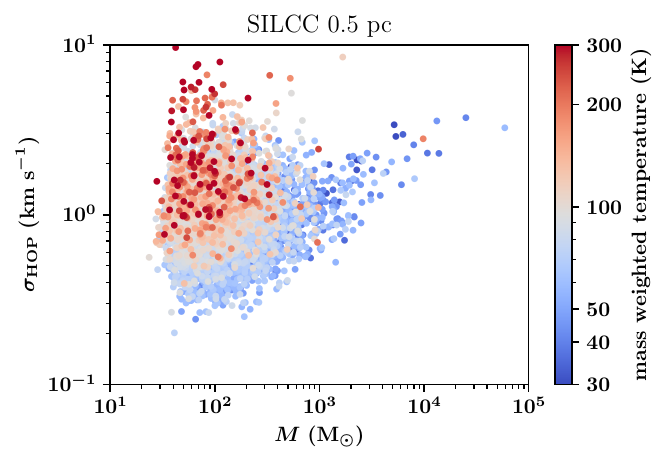}
    \includegraphics[width=\columnwidth]{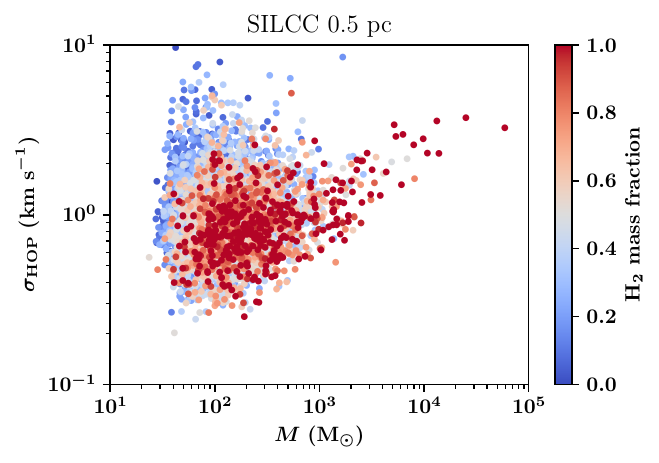}
    \caption{Scatter plot of the velocity distribution as a function of cloud mass for simulation SILCC-0.5pc. Left: colours indicate the mass-weighted temperature in a sphere around the centre of the \textsc{Hop} cloud with a radius corresponding to twice the average size of the \textsc{Hop} cloud. Right: colours indicate the molecular fraction in the surrounding sphere.}
    \label{fig:mass-sigma-temp-chem}
\end{figure*}

The velocity dispersion is also correlated with the cloud size \citep{Hennebelle_Falgarone2012,Heyer&Dame2015}. This is the so-called Larson relation \citep{Larson1981}, which is considered a signature of the turbulent cascade \citep{review_Elmegreen&Scalo2004a}. Indeed, for compressible supersonic turbulence a scaling of $\sigma \propto l^{\beta}$ with $\beta$ between 1/3 and 1/2 is expected \citep{Kritsuk_et_al2007}. Observationally, a large scatter is found in the line-width -- size relation. Some studies agree with this theoretically predicted slope \citep{Solomon_et_al1987, Kauffmann_et_al2013}, but some find steeper slopes \citep{Miville-Deschenes_et_al2017}, which shows that the relation may not be as universal as generally believed.

The velocity dispersion of our extracted clouds has been calculated as a mass averaged quantity (see Eq.~\ref{eq:cloud_sigma}). The size is derived from the ellipsoid approximation as mentioned in the previous section and detailed in Appendix~\ref{sec:hop_algo}.

\subsubsection{Continuous relation across simulations}
Figure~\ref{fig:correlation_size_sigma} shows the cloud internal velocity dispersion as a function of their size for all simulations. The dashed line illustrates the power-law scaling with 0.75. In addition, we show the actual scaling with the solid line that passes through the innermost contour line of all sets of clouds. Despite the overall large scatter, the majority of the clouds (inner contours) form a continuous distribution across the set of simulations. 

We note that the distribution is bounded towards the low-$\sigma$ end approximately half an order of magnitude below the dashed line. There are hardly any clouds with a lower velocity dispersion. This is in line with the turbulent motions in the gas and the gravitational collapse in massive clouds, which increases the internal velocity dispersion. In contrast, there is no clear cut at the high-$\sigma$ end of the distribution. In particular for the small clouds (mainly the SILCC simulations) and the clouds from galaxy Ramses-F20, there is a population of clouds with strong internal motions. Overall, for small clouds the internal velocity dispersion is significantly enhanced compared to the indicated power-law scaling -- in particular the inner contours of the cloud distributions from the SILCC simulations.

\subsubsection{Observed scaling and break in the power-law}

Canonically, the size - velocity dispersion relation has been described by a power-law with slope 0.5. 
However, the simulation results indicate that the relation is somewhat more complex, in a sense that it in fact does not follow a simple power-law.
In Figure~\ref{fig:observations_size-sigma} we compare our results to the catalogues from \citet{Miville-Deschenes_et_al2017} and \citet{Benedettini_et_al2020}.
We keep in mind that the velocity dispersion is determined in a very different way in the simulations compared to the observations.
Nonetheless, the observations of \citet{Benedettini_et_al2020} show a flattening of the relation below sizes of about 5 pc, with the velocity dispersion saturating at a value of about 1.4 km s$^{-1}$ (when converted from 1D to 3D dispersion assuming isotropy), in agreement with the global trend found in the simulations.
The bulk of the SILCC clouds even seem to slightly underestimate the small scale velocity dispersion. This is not surprising since the simulations do not take into account processes that inject turbulence at small scales, such as stellar winds and jets. Another possible interpretation is that the observations are biased towards gravitationally collapsing objects. The outermost contour of LS, simulations which preferentially resolve the collapsing objects within them as discussed in section~\ref{sec:resolution}, indeed seems to follow the observations.

While we do recover a global power-law behaviour for large clouds, the slope from the simulations is somewhat steeper, i.e. about 0.75, compared to the already fairly steep 0.6 found by \citet{Miville-Deschenes_et_al2017}.
For large clouds, the extraction from the simulations measures on average a slightly larger velocity dispersion than the observations from \citet{Miville-Deschenes_et_al2017} for a certain size bin. 
The simulation results may overestimate the dispersion due to several reasons.
We calculate the dispersion in a somewhat naive way which does not take into account organised motions within the cloud such as rotation, radial gravitational inflow or colliding flows. These are not forms of turbulence and should thus be accounted for when one wants to quantify the turbulence velocity dispersion.
An alternative interpretation is a systematically different size estimate, such that the sizes for the observed clouds are larger than the ones from the simulations. This is in fact a plausible explanation given that clouds are overlapping along the line of sight as illustrated in the simulations in Fig~\ref{fig:example_clouds} and cloud blending can be sever especially at large distances. 

What the simulations and observations do agree on is the large scatter. For a specific size, the estimated velocity dispersion can vary by an order of magnitude.

\subsubsection{Origin of high velocity dispersion clouds}
A high velocity dispersion can have several origins, among which externally driven turbulence and self-gravitating collapse are two prominent examples. In the case of gravitational collapse the clouds are likely to be embedded in a dense environment with cold temperatures. In the case of externally driven turbulence, that propagates into the clouds, the gas temperature can be enhanced if the external turbulence is driven by SN feedback and penetrates into the clouds from the warm or hot phase of the ISM.

In order to identify the mechanism at work for the small clouds we investigate the locations of the clouds in Fig.~\ref{fig:coldens-sigma-SILCC} for simulation SILCC-0.5pc. The corresponding SILCC simulations with higher and lower resolution are very similar. The grey-scale colour-map shows the gas column density. The clouds identified by \textsc{Hop} are over-plotted with the colour indicating the internal velocity dispersion. There is no clear correlation between the position of the clouds in the global gas structure and their velocity dispersion. However, the lowest values of $\sigma$ are found for points that tend to be more embedded in large gas structures. On the other hand, clouds closer to the low-density bubbles that have formed as a result of SN feedback have intermediate to high velocity dispersions. We further illustrate this in Fig.~\ref{fig:maps-dens-temp-velo}, where we show slices of density, temperature and velocity at the positions of two example clouds with high velocity dispersion, both located at the interface between cold gas and a hot bubble.

A more systematic analysis is shown in Fig.~\ref{fig:mass-sigma-temp-chem} for the clouds in SILCC-0.5pc (again, different resolutions do not differ) for which we investigate a spherical region around the clouds with a radius twice the average size obtained from the cloud algorithm. Both panels show scatter plots of $\sigma$ as a function of cloud mass. In the left-hand panel the clouds are colour coded by the mass-weighted temperature in the investigated volume. The right-hand counterpart shows the fraction of molecular gas in in the volume. We note that the clouds with higher velocity dispersion clearly show enhanced temperatures and lower fractions of molecular hydrogen. This illustrates that the clouds which are surrounded by warm or hot gas have statistically higher velocity dispersions. In the SILCC setups analysed here, the only source of hot turbulent gas are SN explosions. We therefore conclude that the enhanced motions in the clouds are due to SN-driven turbulence.
We note that these warm clouds with low H$_2$ fractions would also contain little CO and as a result could be missing from classical observational catalogues.

In the full galaxy simulations \textsc{Ramses-F20} and \textsc{M51}, there is also a population of clouds with high-velocity dispersion, but the mechanisms at play are different. Most of these clouds, which are also very large, are rotating disks, which explains the high velocity dispersion. The existence of such disks with radius of around 100\,pc is a spurious effect of the lack of resolution in these regions, as we would expect these structures to collapse and fragment into smaller objects. 

\subsection{Virial parameter}
\begin{figure}
    \centering
    \includegraphics[width=8.4cm]{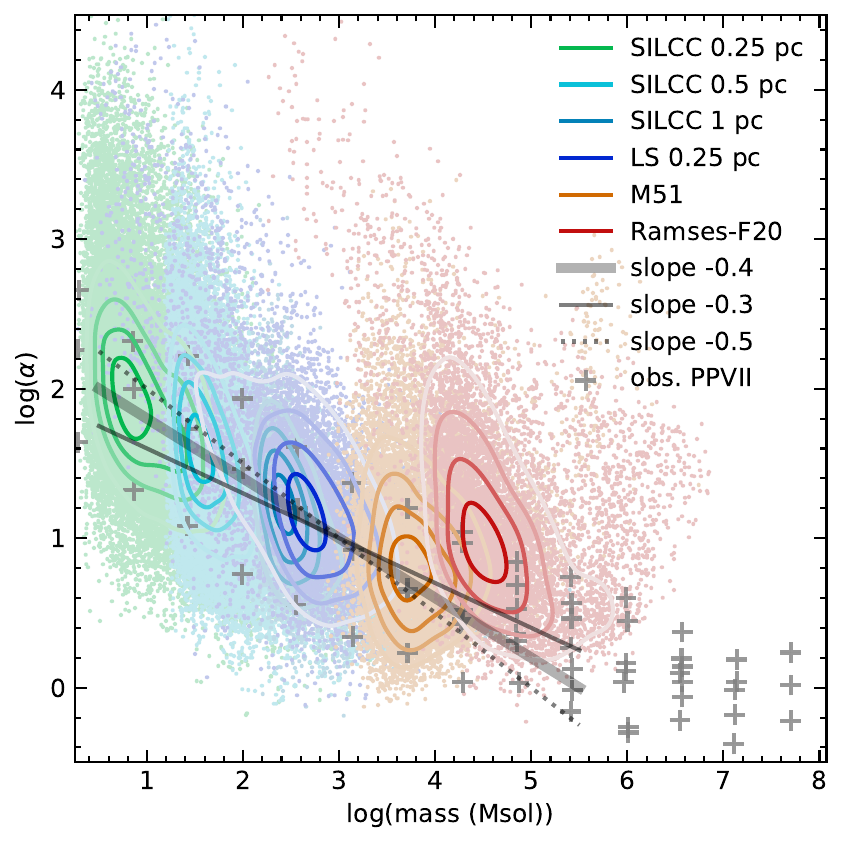}
    \caption{Virial parameter as a function of cloud mass for all simulations. Coloured 
    contour lines are the same as in Fig. \ref{fig:correlation_mass_size}. Over-plotted are the slope of the scaling (thick grey line), as well as the observational data compiled for the PPVII review \citep{ChevanceEtAl2023}. We measure a slightly flatter scaling with mass ($-0.4$ compared to $-0.5$) and lower absolute values compared to \citet{Miville-Deschenes_et_al2017}. The more recent observations agree well with our measured values.}
    \label{fig:virial-parameter}
\end{figure}

In addition to the size-velocity dispersion relationship, we also compute the virial parameter $\alpha_{\rm vir}$ for the clouds in our simulations, using the definition given in Equation~\eqref{eq:virial-parameter}. The results are shown in Figure~\ref{fig:virial-parameter}. We can see immediately that $\alpha_{\rm vir}$ has a clear dependence on mass: $\alpha_{\rm vir} \propto M^{-n}$, with $n = 0.4$. In other words, lower mass clouds on average have much larger values of $\alpha_{\rm vir}$ than higher mass clouds. Therefore, although there is generally good agreement between the simulations for the mass ranges where their cloud populations overlap, the mean value of $\alpha_{\rm vir}$ differs significantly between different simulations. This demonstrates that care must be taken when using simulations to draw conclusions about the gravitational boundedness of interstellar clouds.\footnote{The question of whether the conventional definition of $\alpha_{\rm vir}$ is an adequate diagnostic of whether or not clouds are gravitationally bound is an interesting one \citep[see e.g.][]{RamirezGaleano22}, but is out of the scope of this study.} Different simulations may lead to very different conclusions about whether or not clouds are bound if they differ significantly in terms of resolution or box size, and hence do not probe the same portion of the cloud mass hierarchy. 

We also see that there is substantial scatter in the value of $\alpha_{\rm vir}$ that we recover for any given cloud mass. For the majority of clouds, the value of $\alpha_{\rm vir}$ varies by about one order of magnitude for a specific mass range, but in a small fraction of clouds the value of $\alpha_{\rm vir}$ can be as much as 100 times the typical value for a given mass. 
At fixed mass, $\alpha_{\rm vir}$ depends only on the cloud size $R$ and velocity dispersion $\sigma$. However, we have already seen that at fixed $M$, the cloud size varies by only around a factor of two in most cases (see Figure~\ref{fig:correlation_mass_size_alldef}), and so the majority of this scatter must be driven by variations in the velocity dispersion. This conclusion is consistent with what we have already found for the velocity dispersion-size relation: for most clouds, $\sigma$ varies by around a factor of three at fixed cloud size (see Figure~\ref{fig:correlation_size_sigma}), corresponding to an order of magnitude variation in $\alpha_{\rm vir}$ if we keep $M$ and $R$ fixed, but for a small number of clouds we find much larger values for $\sigma$ and hence for $\alpha_{\rm vir}$. It hence seems likely that the factors responsible for the outliers in the $\alpha_{\rm vir}$--$M$ distribution are the same as those responsible for the outliers in the $\sigma$--$R$ distribution, namely stellar feedback at low cloud masses and lack of resolution leading to artificial rotational support at high cloud masses (see Section~\ref{sec:size-sigma}). 

Finally, it is interesting to compare the $\alpha_{\rm vir}$--$M$ relation that we recover from the simulations with the one measured for real interstellar clouds. As is clear from the compilation of data in \citet{ChevanceEtAl2023}, different galactic and extra-galactic cloud surveys recover a similar scaling $\alpha_{\rm vir} \propto M^{-0.5}$ at low cloud masses, but with a normalisation that varies between surveys by up to an order of magnitude. 
These results are also in agreement with observations of the densest regions within molecular clouds, parsec-scales structures called clumps. Several surveys dedicated to the study of the dynamics at these scales using high density tracers (e.g. NH$_{3}$ or N$_{2}$H$^{+}$) report a scaling $\alpha_{\rm vir} \propto M^{-0.5}$ \citep{Urquhart18,Traficante18_PII}. Although the behaviour of this scaling relationship on the scale of individual clumps is debated \citep{Kauffmann_et_al2013,  Traficante18_PIII, Singh21}, it is remarkable to note that the average slope observed using high-density tracers is in overall agreement with CO cloud surveys and with the results of the simulations in this work. In fact, overall, the range of values found in observations are in good agreement with the location of most of our simulated clouds in the $\alpha_{\rm vir}$--$M$ plane (see gray crosses in Figure~\ref{fig:virial-parameter}), but the small population of clouds we recover that have very high virial parameters do not appear to have observational counterparts in CO-based surveys. This may simply reflect the fact that many of these clouds have been strongly affected by stellar feedback and hence would in reality have little molecular gas associated with them. The slope that we recover for the $\alpha_{\rm vir}$--$M$ relation also agrees well with the observed slope: our best fit value is slightly flatter than that found in the observations, but as we show in Figure~\ref{fig:virial-parameter}, a slope of -0.5 also provides a relatively good fit to the data from all of the simulations apart from Ramses-F20. However, we caution that this agreement may be strongly influenced by selection effects. Observationally, clouds with an integrated intensity close to the detection limit of a survey and a velocity dispersion close to the velocity resolution will all have approximately the same molecular gas surface density, implying that for these clouds, $R \propto M^{0.5}$. For fixed $\sigma$, this then yields $\alpha_{\rm vir} \propto M^{-0.5}$. A similar combination of limited sensitivity and velocity resolution may also explain the similarity between the scaling relationships recovered for dense clumps and for molecular clouds. Our population of simulated clouds does not suffer from this selection effect, but the fact that, as previously discussed, \textsc{Hop} selects clouds with a relatively narrow range of mean densities also strongly biases us against finding clouds with low $\alpha_{\rm vir}$ at low $M$.

\subsection{Thermal state}
\begin{figure}
    \centering
    \includegraphics[width=\columnwidth]{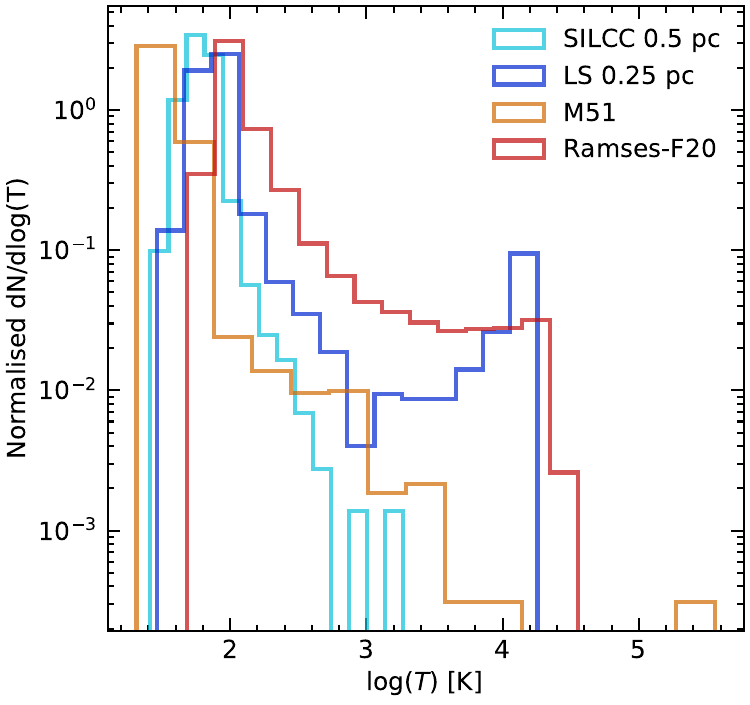}
    \caption{Distribution of cloud-averaged temperatures.
    }
    \label{fig:T_pdf}
\end{figure}

In order to quantify the thermal state of the gas in the clouds identified in the different simulations, we have computed the mean temperature $T$ for each cloud.
We compare the distributions of mass-weighted average cloud temperatures in Figure~\ref{fig:T_pdf}, where we show results from the LS, M51 and Ramses-F20 simulations, plus a representative example from the SILCC simulations.

From the figure, we see that there are some significant differences in the average cloud temperatures in the simulations. All of the simulations show a clear peak at low temperatures, but they do not agree on the location of this peak: it is found at $T \sim 25$~K in the M51 simulation, $T \sim 100$~K in the LS and Ramses-F20 simulations, and at $T \sim 60$~K in the SILCC simulation. This appears to be primarily due to a difference in the effectiveness of photoelectric heating in the various simulations. In the LS and Ramses-F20 simulations, the photoelectric heating rate is computed assuming no local attenuation of the interstellar radiation field. Since much of the cloud mass is located at a density close to the \textsc{Hop} threshold density, we would therefore expect the temperature of this gas to be comparable to that of a patch of unshielded ISM with a density $n \sim 30 \: {\rm cm^{-3}}$. For an interstellar radiation field strength comparable to the Solar neighbourhood value, this is $T \sim 100$~K \citep{WolfireEtAl2003}, in good agreement with the low $T$ peak we find in these simulations. In the SILCC and M51 simulations, we attempt to account approximately for the effects of local attenuation using the TreeCol algorithm \citep{ClarkGloverKlessen2012,WuenschEtAl2018}. The clouds formed in these simulations therefore feel an effective photoelectric heating rate that is lower than in the \text{Ramses} simulations, and hence have lower characteristic temperatures. Finally, the difference between the SILCC and M51 simulations stems from the fact that the two populations of clouds have similar mean densities but different sizes (see Figure~\ref{fig:correlation_mass_size_alldef}). The M51 clouds therefore have mean surface densities that are a factor of a few larger than the clouds in the SILCC~0.5~pc simulation, giving them higher mean extinctions and hence lower photoelectric heating rates. 

The temperature distributions shown in Figure~\ref{fig:T_pdf} also differ significantly at the high $T$ end. In the SILCC and M51 simulations, very few clouds have average temperatures as high as 1000~K, while in the LS and Ramses-F20 simulations, there is a substantial population of clouds with high mean temperatures, extending up to $T \sim 10^{4}$~K. This behaviour is a consequence of differences in the way that stellar feedback is modelled in the simulations. In the LS simulations, the effects of photo-ionisation feedback are included. As \textsc{Hop} identifies clouds based purely on their density structure, without reference to their chemical composition, it therefore extracts some clouds that have been partially or totally photo-ionised, but that have not yet expanded sufficiently to bring their mean density below the cloud identification density threshold. The peak in the temperature distribution at $T \sim 10^4$~K corresponds to fully ionised clouds, while clouds with temperatures between a few 100~K and $10^{4}$~K have only been partially ionised, and hence have a mean temperature that is intermediate between the low temperature of their neutral portion and the $\sim 10^{4}$~K temperature of their ionised portion. In the Ramses-F20 simulations, ionising radiation is not explicitly followed with a radiative transfer code, but is accounted for approximately by maintaining the temperature of the gas in any grid cells containing massive stars at $10^4$~K \citep{kretschmerFormingEarlytypeGalaxies2020}. Therefore, the mean cloud temperature in this simulation depends on the number of cells in the cloud that contain massive stars, with the highest mean temperatures corresponding to clouds in which most or all of the grid cells are affected by the local heating. Finally, we do not see this population of warm clouds in the SILCC or M51 simulations because these simulations do not include any treatment of photo-ionisation feedback. In principle, SN feedback could produce a similar effect if we were to happen to identify a cloud in which a supernova had just exploded, but in practice, clouds in which SNe have exploded are rapidly disrupted and so the chances of this occurring are relatively small. That said, we do find exactly one example of such a cloud in the M51 simulation, with mean temperature of $\sim 3 \times 10^{5}$~K, visible on the far right in Figure~\ref{fig:T_pdf}.

The fact that we see such clear differences between the distributions of cloud-averaged temperatures in the different simulations indicates that this is a quantity that is very sensitive to how we chose to model the thermal physics of the ISM and the impact of stellar feedback in our simulations. However, the fact that many of the other cloud properties that we have examined agree well between different simulations demonstrates that even large differences in cloud temperatures have a relative small effect on the other properties of the clouds, consistent with the idea that these are shaped more strongly by turbulence than by thermal pressure.

\section{Discussion}
\label{sec:discussion}

We already discussed individual results extensively in their corresponding sections.
Here we address further the issue of resolution and explore the dependence on the cloud extraction algorithm.

\subsection{Resolution dependence}
\label{sec:resolution}

\begin{table*}
\centering
    \caption[]{Number of cells and fractions of cells on each refinement level for the stratified box simulations.}
     \label{tab:refinement}
     \begin{tabular}{lrrrrrr}
        \toprule
        Name & $N_\mathrm{tot}$ & $f(4\,\mathrm{pc})$ & $f(2\,\mathrm{pc})$ & $f(1\,\mathrm{pc})$ & $f(0.5\,\mathrm{pc})$ & $f(0.25\,\mathrm{pc})$\\
        \midrule
        SILCC-1pc-3$\mu$G    & $2.9\times10^7$ & 0.04 & 0.16 & 0.80 & -- & --\\
        SILCC-1pc-6$\mu$G    & $3.2\times10^7$ & 0.03 & 0.18 & 0.79 & -- & --\\
        SILCC-0.5pc-3$\mu$G  & $2.0\times10^8$ & 0.004 & 0.02 & 0.09 & 0.89 & -- \\
        SILCC-0.25pc-3$\mu$G & $3.4\times10^8$ & 0.003 & 0.01 & 0.05 & 0.46 & 0.47\\
        LS-no-driving        & $6.9\times10^7$ & 0.22 & 0.13 & 0.19 & 0.25 & 0.21\\
        LS-weak-driving      & $8.0\times10^7$ & 0.19 & 0.11 & 0.20 & 0.27 & 0.22\\
        LS-medium-driving    & $8.1\times10^7$ & 0.19 & 0.12 & 0.20 & 0.27 & 0.22\\
        LS-strong-driving    & $4.6\times10^7$ & 0.34 & 0.20 & 0.17 & 0.16 & 0.13\\
        \bottomrule
     \end{tabular}
\end{table*}

The influence of the resolution of the simulation on the extracted cloud population is most easily seen in the cloud mass spectrum.
We already discussed in Section~\ref{sec:mass_size_shape} that the peak of the mass spectrum depends on the resolution. However, if we normalise according to the number of clouds per unit area, we see that all simulations follow the same distribution. The effect of increasing the resolution is thus purely to be able to resolve an additional population of smaller clouds; the absolute number of large clouds remains the same.
When the resolution is low, it is possible that some clouds become blended. In this case one would expect to find more massive objects. The universality of the obtained mass spectrum slope for simulations with different resolutions shows that this is not an issue, at least when structures are extracted with the algorithm used in this work. Some small clouds may be counted as being part of a larger cloud, but since their mass and size are negligible compared to the properties of the larger cloud, this does not change the statistics.

We also saw that the shape of the low mass part of the mass spectrum depends on the simulation code. The LS and Ramses-F20 simulations show a gradual decline towards smaller and smaller masses, while the SILCC runs show a sharp cut-off at the resolution limit.
The differences between the grid code simulations run with \textsc{Ramses} (LS and Ramses-F20) and \textsc{Flash} (SILCC) are related to the different refinement strategies used in the two codes. While both codes use refinement criteria that trigger grid refinement in dense regions, the SILCC simulations have block refinement, i.e.\ the entire rectangular block of $8^3$ cells around the dense region is also refined in order to keep the grid structure simple. This results in a significant fraction of cells having high refinement levels, as illustrated in Table~\ref{tab:refinement}, which in turn allows for well-resolved turbulence inside the majority of the domain.
By contrast, the \textsc{Ramses} runs were refined in a very localised fashion, and hence although they resolve gravitationally collapsing structures well, they capture less of the turbulent cascade to small spatial scales that occurs in the more diffuse ISM. This results in them missing a population of small structures created by turbulence.

\begin{figure}
    \centering
    \includegraphics[width=\columnwidth]{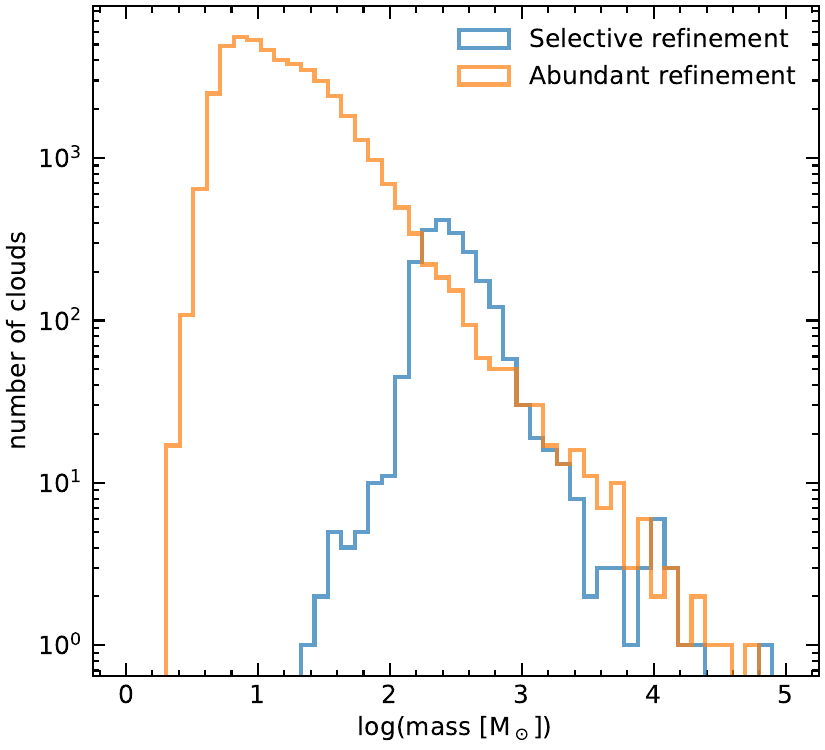}
    \caption{Comparison of the cloud mass spectrum in two \textsc{Ramses} test simulations with a similar setup to simulation LS using different refinement strategies.}
    \label{fig:effect_refinement_strategy}
\end{figure}

To verify this, we compare the results of two \textsc{Ramses} test simulations of the stratified box type which apply different refinement criteria. The first simulation uses a targeted refinement strategy, similar to the one used in the LS simulations listed in Table~\ref{tab:simulations}, which focuses on following the gravitational collapse of dense objects. This is done by requiring the local Jeans length and mass, which vary with scale, to be resolved by a certain number of cells. The second simulation has more abundant refinement, i.e.\ refinement is triggered more frequently in lower density gas. This is achieved by setting a uniform limit on the mass that can be contained in a cell. This results in many more cells and thus significantly longer computation time.
Figure~\ref{fig:effect_refinement_strategy} compares the mass spectra of the clouds extracted in these two runs. We see that the abundant strategy is able to resolve many more low mass structures and has a mass spectrum with a shape comparable to what we found for the SILCC runs.
The run with the selective refinement strategy has a peak at larger values and a low mass tail which becomes more prominent over time as gravitational collapsing structures develop.
This test indeed confirms that the difference in refinement strategy results in a different mass spectrum for SILCC and LS.

\subsection{Cloud extraction parameters and other algorithms}
\label{sec:other_algos}

There are many ways to extract clouds from simulations and the results depend on the implementation and chosen parameters of the extraction algorithm.
In Appendix~\ref{appx:merge_params} we test the dependence of the mass spectrum on the chosen  parameters of the \textsc{Hop} algorithm.
Notably, the saddle merging criterion is found to have a significant effect on the slope.
If we do not use the saddle merging criterion, the mass spectrum is closer to a log-normal rather than having a power-law high-mass tail.
Without saddle merging, the mass spectrum will depend strongly on resolution, since resolved substructures will be counted as individual objects.
When enabling saddle merging, the slope of the power-law decreases with decreasing merging density value.
When merging everything that touches, i.e. the most aggressive merging strategy possible with $\rho_\mathrm{saddle} = \rho_\mathrm{threshold}$, the mass spectrum is significantly shallower than the value of $\rho_\mathrm{saddle} = 10\,\rho_\mathrm{threshold}$ we adopted in this work.
The majority of small objects are absorbed into large cloud complexes.

The mechanics of \textsc{Hop} with the most aggressive saddle merging is similar to how the friends-of-friends (FOF) algorithm operates. For particle simulations, FOF dictates that two particles with a density above the required threshold belong to the same group if they are within a specified distance from one another, typically taken to be some fixed fraction of the mean inter-particle separation. For grid simulations, one typically requires the cells to be neighbours.
\citet{IffrigHennebelle2017} used FOF to identify clouds in kpc-sized boxes similar to the LS simulations used in this work. They applied a slightly higher density threshold of 50 cm$^{-3}$ and
find that the slope of the mass spectrum depends on the strength of the magnetic field. Without magnetic field the fitted slope is around -1 or slightly steeper.
For an initial field of 3 or 6 $\mu$G, comparable to the field strengths used in the stratified box simulations analysed in this work, they obtain shallower slopes, i.e. about -0.7 to -0.8.
These values are shallower than what we recovered with \textsc{Hop}, in line with the difference observed when changing the saddle merge criterion.

\begin{figure}
    \centering
    \includegraphics[width=\columnwidth]{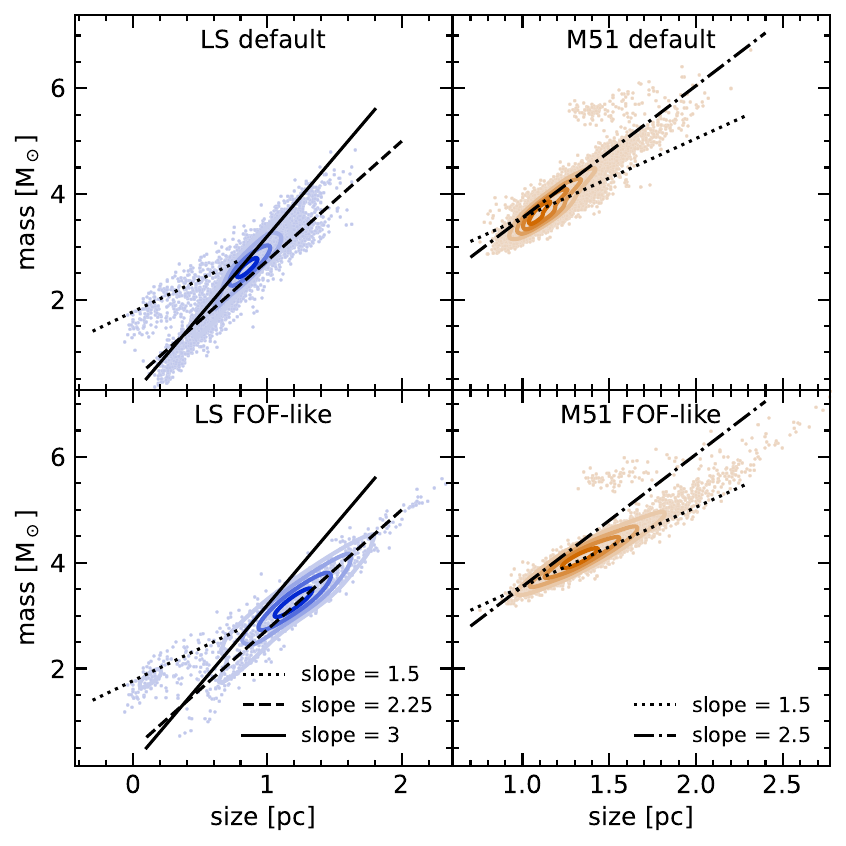}
    \caption{The mass-size relation for the LS stratified boxes and M51 galaxy when using our default \textsc{Hop} parameters versus FOF-like merge criteria. Coloured contour lines are the same as in Fig.~\ref{fig:correlation_mass_size}.}
    \label{fig:FOF-vs-HOP}
\end{figure}

The mass-size relation is also affected.
\citet{IffrigHennebelle2017} report that their results are in line with the observed slope of 2.36, somewhat shallower than the general trend of -3 we found in Section~\ref{sec:mass_size_shape}.
In Figure~\ref{fig:FOF-vs-HOP} we see that we indeed find a shallower mass-size relation when extracting the clouds with a FOF-like algorithm. Small objects are merged into larger ones, tilting the slope.

Another approach is to use dendograms, which describe the hierarchy of structures in the form of a tree.
\citet{Tress2021} extracted clouds from their M51 galaxy, that we also use in the comparison in this work, using \textsc{Scimes} \citep{colomboGraphbasedInterpretationMolecular2015}.
This stands for Spectral Clustering for Interstellar Molecular Emission Segmentation and is a technique mainly used in observations. It is based on the cluster analysis of dendrograms of 3D (in the PPV or PPP space) data cubes that performs particularly well for the identification of large structures such as the molecular clouds in spectroscopical observations of large fields in the Milky Way (e.g. \citealt{colomboGraphbasedInterpretationMolecular2015,schullerSEDIGISMStructureExcitation2017,Benedettini_et_al2020}). \textsc{Scimes} considers the dendrogram tree of the 3D structures in the data cube (using the implementation of \citealt{rosolowskyStructuralAnalysisMolecular2008} for astronomical data-sets) and groups different leaves together into ‘clusters’ of leaves, based on one or more criteria. 
We note that the parameters of the extraction in \citet{Tress2021} are different than ours, which means that we should not over-interpret details in the numbers. They use the extraction based on molecular hydrogen with a threshold number density of $n_\mathrm{H_2,min}=1\,\mathrm{cm}^{-3}$. This does not simply translate into a total number density as shown in detail in e.g. \citet{SeifriedEtAl2017} for the same chemical network. Furthermore, the dendrogram methods works on a grid-based data set, which requires the Arepo cells to be mapped onto a (locally) uniform grid. Whereas, this is not expected to yield systematically different results if the resolution is high enough, it might easily shape the outliers in the distributions. Nonetheless, many of the extracted cloud properties are well in line with our values. The bulk of the cloud masses extracted in \citep{Tress2021} broadly range from $10^3-10^5\,\mathrm{M}_\odot$, slightly higher but well within the scatter given the differences for the parameters. The sizes of the clouds peak at $\sim10\,\mathrm{pc}$, which is close to our sizes. 
Both the mass-size relation and the size-velocity dispersion relation are in line with the trends found in this work.
Furthermore, the virial parameters are very similar ($\alpha_\mathrm{vir}\sim10$) as well as their flat scaling with mass.
This illustrates that different cloud extraction methods do not always have to disagree.

\section{Summary and conclusions}
\label{sec:conclusion}

In this paper we extracted clouds from multi-physics ISM simulations using a set of four different types of (M)HD simulations (SILCC, \citealt{Girichidis2021}; LS, \citealt{ColmanEtAl2022}; M51, \citealt{Tress2021}; Ramses-F20, \citealt{BrucyPhd}). The set covers a large range in spatial scales from 0.25\,pc for the smallest cells in patches of the ISM up to a full galactic disks with a diameter of 30\,kpc. Furthermore, the simulations were performed using different codes (\textsc{Flash}, \textsc{Ramses} and \textsc{Arepo}) with distinctive numerical techniques and physics ingredients of varying complexity. For all data sets, we use a standardised 3D density-based extraction method build around the \textsc{Hop} algorithm.

We find that many properties of the extracted clouds follow robust trends across scales and very smooth transitions between the different simulations despite the vast differences in resolution, physics included, size of the box and setup geometry. In particular the relation between mass, size, shape and velocity dispersion (see Section \ref{sec:mass_size_shape} and \ref{sec:size-sigma} and Figs.~\ref{fig:mass_size_distributions}, \ref{fig:correlation_mass_size_alldef}, \ref{fig:shape_distribution} and \ref{fig:correlation_size_sigma}) show remarkably consistent scaling properties. This suggests some universality of these relations, and that gravity and turbulence, which are equally present in all simulations, are the key drivers of (molecular) cloud properties.

More specifically, our results can be summarised as follows:
\begin{enumerate}
    \item The geometry of the extracted clouds ranges from roundish object to complex filamentary networks. Consequently, the precise shape is only poorly represented by simple quantities such as the sphericity or the triaxiality.
    However these quantities, computed from a ellipsoidal approximation of the shape of the clouds, are useful to characterise their 3D extents.
    The sphericity and triaxiality are very consistent across all simulations \emph{and} across the number of cells per cloud within each simulation with a sphericity of $0.25-0.4$. This is not an artefact of the extraction algorithm, which is able to detect all complex shapes from the simulations.
    \item The mass spectra of the clouds follows a remarkably consistent and universal distribution: the normalised number of clouds per unit area of interstellar gas results in a common shape of the high-mass end of the distributions with a slope of $\mathrm{d} N/\mathrm{d}\ln M\propto -1$ across 6 orders of magnitude. No scaling of individual simulations was required to match this universal high-mass distribution across the simulations.
    \item The slope of the mass distribution in our simulations agrees with theoretical predictions but differs from other numerical work and observations. We show that this discrepancy can easily originate from the extraction method used.
    \item The low mass end of the distributions is sensitive to the applied refinement criterion. An aggressive refinement, such as used in SILCC that does not only act on a small volume around collapsing regions, allows for both higher resolution at intermediate densities and a better representation of the turbulent cascade. Consequently, in this case many more structures form in the presence of and driven by turbulent motions. The resulting mass distributions show the universal scaling down to a small multiple of the resolution limit of the simulation. This naturally comes at a much higher numerical cost. The original LS simulations use a more localised refinement, which results in a peak of the mass spectrum at significantly higher masses. However, tests with similarly aggressive refinement show a consistent evolution with SILCC.
    \item The relation between cloud mass and size follows a smooth power-law of the form $M \propto R^3$ across simulations. The normalisation reflects the density threshold used by the extraction algorithm.
    \item The internal velocity dispersion also scales very smoothly across the setups. For the large clouds, the trend is a power-law $\sigma\propto R^{0.75}$. For clouds smaller than $\sim0.5\,\mathrm{pc}$ we find a deviation from the power-law scaling towards higher velocity dispersions. The median value for $\sigma$ seem to asymptotically approach $\sim0.5\,\mathrm{km\,s}^{-1}$. A detailed analysis shows that this base level of internal velocity dispersion results from supernova-driven turbulence rather than gravitationally driven collapse. The small (low-mass) clouds with higher $\sigma$ show a higher mass-weighted temperature and reduced fraction of molecular gas, which is in line with turbulence driving from the hot and warm phases into the clouds.
    \item A comparison with observational catalogues shows a similar trend of a power-law scaling of $\sigma$ with size and a flat distribution for small clouds with a base level of internal velocity dispersion of approximately $1\,\mathrm{km\,s}^{-1}$. The scaling for the simulated clouds of $\sigma\propto R^{0.75}$ is steeper compared to the observed scaling of $\sigma\propto R^{0.6}$, which does not necessarily mean an inconsistency in the dynamics of clouds, but could also be related to the differently inferred sizes of the clouds between simulations and observations.
    \item The scaling of the virial parameter $\alpha_{\rm vir}$ with mass agrees well with the $\alpha_{\rm vir} \propto M^{-0.5}$ scaling found in observations of both molecular clouds and dense clumps. However, it is difficult to say whether this is a robust physical result or merely a consequence of the selection effects affecting both observational and numerical measurements of $\alpha_{\rm vir}$.
    \item The mean cloud temperatures do not agree well between simulations, primarily due to differences in the treatment of photoelectric heating and stellar feedback in the different models. However, the good agreement we find in other cloud properties despite this difference in temperatures suggests that the thermal pressure of the gas has only a small influence on these properties.
    \item Care must be taken when comparing mean cloud properties between catalogues, as different simulations (or observations) may trace different parts of a universal cloud hierarchy. Catalogue averages reflect the resolution and volume under study.
\end{enumerate}

The catalogues we produced along with the analysis tools are publicly available in the \href{http://www.galactica-simulations.eu}{Galactica} database, under the reference \textsc{cloud\_comp}: \url{http://www.galactica-simulations.eu/db/ISM/CLOUD_COMP}.

\begin{acknowledgements}
We thank the anonymous referee for their useful comments. This project was funded by the European Research Council under ERC Synergy Grant ECOGAL (grant 855130), lead by Patrick Hennebelle, Ralf Klessen, Sergio Molinari and Leonardo Testi. 
The authors acknowledge Interstellar Institute’s program “II6” and the Paris-Saclay University’s Institut Pascal for hosting discussions that nourished the development of the ideas behind this work.
The authors gratefully acknowledge computing resources provided by the Ministry of Science, Research and the Arts (MWK) of the State of Baden-W\"{u}rttemberg through bwHPC and the German Science Foundation (DFG) through grants INST 35/1134-1 FUGG and 35/1597-1 FUGG, and they thanks for data storage at SDS@hd funded through grants INST 35/1314-1 FUGG and INST 35/1503-1 FUGG.
RSK and SCOG furthermore acknowledge financial support from the Heidelberg Cluster of Excellence STRUCTURES in the framework of Germany’s Excellence Strategy (EXC-2181/1 - 390900948), and from the German Ministry for Economic Affairs and Climate Action in project MAINN (funding ID 50OO2206).
\\
\underline{CRediT author statement:}\footnote{More information about the roles defined by the CRediT system can be found here: \url{https://www.elsevier.com/researcher/author/policies-and-guidelines/credit-author-statement}}
\textbf{Tine Colman:} Conceptualisation, Software, Formal Analysis, Investigation, Writing - Original Draft.
\textbf{Noé Brucy:} Conceptualisation, Formal Analysis, Investigation, Writing - Original Draft, Visualisation, Data Curation.
\textbf{Philipp Girichidis:} Conceptualisation, Formal Analysis, Investigation, Writing - Original Draft.
\textbf{Simon~C.~O.~Glover:} Conceptualisation, Writing - Original Draft.
\textbf{Milena Benedettini:} Resources, Writing - Review \& Editing.
\textbf{Juan~D.~Soler:} Writing - Review \& Editing.
\textbf{Robin~G.~Tress:} Resources, Writing - Review \& Editing.
\textbf{Alessio Traficante:} Writing - Review \& Editing.
\textbf{Patrick Hennebelle:} Conceptualisation, Resources, Writing - Review \& Editing, Funding Acquisition.
\textbf{Ralf S. Klessen:} Conceptualisation, Writing - Review \& Editing, Funding Acquisition.
\textbf{Sergio Molinari:} Conceptualisation, Funding Acquisition.
\textbf{Marc-Antoine Miville-Desch\^enes:} Resources, Writing - Review \& Editing
\end{acknowledgements}

%
%

\bibliographystyle{aa} 
\bibliography{astro,astro2,girichidis,global,tine}

\begin{appendix} 
\section{Structure finding algorithm}
\label{sec:hop_algo}
\begin{figure*}
    \centering
    \includegraphics[width=\textwidth]{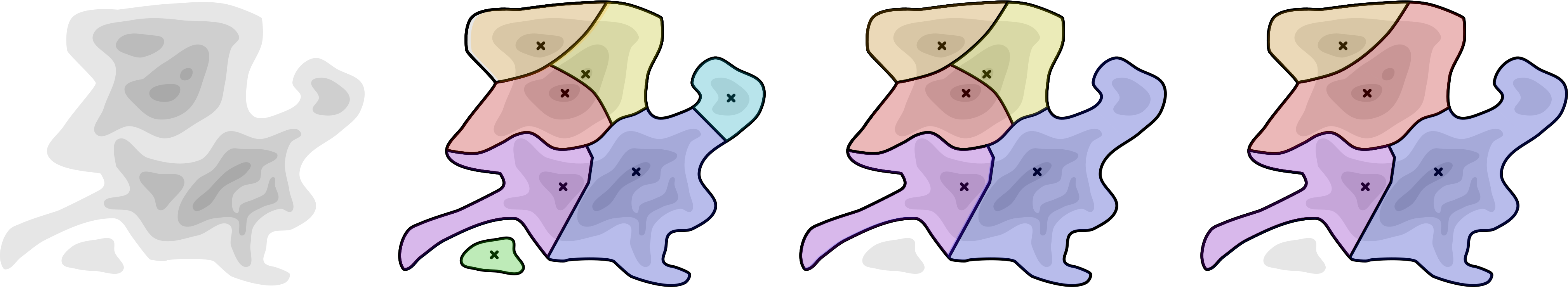}\\
    \caption{Illustration of structure identification and merging by \textsc{Hop} on a hypothetical density field. First a collection of peak patches around local maxima is identified. Then, the green structure is rejected because its peak density is not high enough. The cyan structure is merged into the blue one for the same reason. Next, the yellow structure is merged into the red one because the saddle density between the structures is larger than the saddle threshold, indicating that they are connected structures.}
    \label{fig:hop_how_it_works}
\end{figure*}

The full algorithm, including test setups, can be found in the \href{https://gitlab.com/tinecolman/ecogal_tools}{ecogal\_tools} repository on gitlab.
It is built around the \textsc{Hop} clump finder, as described and implemented by \citet{EisensteinEtAl1998}.
Here we give a summary of their implementation, discuss our adaptation and parameter choices for application to hydrodynamics simulations and present some test results.

\subsection{Finding groups around local peaks}

First, the list of particles is read from the input file.
For this step, we need to provide \textsc{Hop} with the coordinates x, y, and z of each particle or cell\footnote{For regular grids we take the cell centre. For moving mesh codes, one can either pick the centre-of-mass in a cell or the location of the mesh generating point. In this work we opted for the latter. Note that in \textsc{Arepo} a mesh generating point is steered toward the centre-of-mass of the cell, so both are equivalent to first order.}, as well as their mass and density.
If we do not have density information, we can ask \textsc{Hop} to calculate it for us by applying a smoothing kernel. However due to grid codes already providing densities and SPH codes possibly using a variety of kernels, we think it more straightforward to leave the density calculation to the user in the case of hydrodynamics simulations.  
Note that while we typically perform the structure finding on the density field, it can in theory be applied on any field.

A search tree is constructed based on the particle positions to allow for efficient neighbour searching.
For each particle, the highest density neighbour among the \texttt{Nhop} closest neighbours (including the particle itself) is identified.
We keep the default value of \texttt{Nhop}=16. \citet{EisensteinEtAl1998} verified that the result is insensitive to this parameter for particle simulations. For grid simulations, a minimum of 7 is required to probe the direct neighbours of a fully refined region. Adopting large value (e.g. 27 which would probe also the diagonal neighbours in a fully refined region) could lead to artificial merging of structures that are separated by cells that are not at the deepest refinement level. The larger \texttt{Nhop}, the larger the computation time.

Next, for each particle we follow the chain of densest neighbours, i.e. for the highest density neighbour, we find its own highest density neighbour, then find the highest density neighbour of that particle, etc. We proceed in this way along the chain until we eventually reach a particle that is its own densest neighbour. The particle we started with is then assigned to the group of the final particle. 
For example, we start at particle A which has a densest neighbour B. We then go to particle B and see that the densest neighbour of B is C. Arriving at C, we see that there is no particle around C that is denser than C, i.e.\ C is its own densest neighbour.
Particle A is than assigned to the group of C.
The name `hop' actually refers to this process of hopping between neighbours.
The result is a collection of peak patches around local density maxima.
The resulting segmentation is outputted to file as an intermediate result, which can be reused later.

To prepare the field for the next step, \textsc{Hop} also investigates the boundaries of the identified structures.
For each particle, we check whether any of the closest \texttt{Nmerge} neighbours belongs to a different structure.
If this is indeed the case, we found a boundary and its density, i.e.\ the saddle density, is defined as the average between the particle and the neighbour that is in the other group.
For each pair of structures, the highest saddle density is recorded and outputted as an intermediate result.
We set \texttt{Nmerge}=8, larger than the default value of 4, to account for the regularity of the neighbour locations in grid codes.

Note that when provided with information on the size of the computational domain, \textsc{Hop} accounts for periodic boundaries when determining neighbours.

\subsection{Altering and merging groups}

Once \textsc{Hop} has identified all the peak patches, we move to the regrouping phase executed by the \textsc{regroup} program \citep{EisensteinEtAl1998}.
This serves to exclude low density particles, to reduce noise and to merge structures according to user specified criteria.
For example, a cloud could consist of multiple peaks which are blended together.
Or we could have a cell with an unusually high density fluctuation with nothing else around it which should be considered as noise.
Four parameters control this phase.
\begin{itemize}
    \item A density threshold \texttt{rho\_thresh} is set to allow only cells with a density above this value to be included in structures.
    \item A required peak density \texttt{rho\_peak} is set to identify an individual structure, as opposed to noise or low amplitude fluctuations inside another structure.
    \item A maximum allowed saddle density \texttt{rho\_saddle} is set to identify an individual structure as opposed to the patch being part of a larger entity.
    \item The structure has to have a minimum number of cells \texttt{min\_cells}. 
\end{itemize}

First, all particles which have a density below \texttt{rho\_thresh} are removed from peak patches.
Usually the value of this parameter is determined by the type of structure we want to obtain.
We typically adopt the thresholds 1 cm$^{-3}$ for cloud complexes, 30 cm$^{-3}$ for clouds and 1000 cm$^{-3}$ for clumps. It is important to note that generally these thresholds are arbitrary and that varying them can give interesting additional information. From an observational point of view, we can set the threshold to approximate the point where we can detect a certain tracer, e.g. CO. This, in turn is determined by the chemical species under study and the sensitivity of the detector.

Next, the algorithm merges structures based on the remaining criteria.
If the peak density is not sufficient, observations would not pick it up as an individual structure.
There are now several possibilities depending on whether it has neighbours or not. If the structure has a neighbour that does qualify as an individual entity and the saddle is above the saddle threshold, the structure is merged with this neighbour. If there are multiple neighbours, the one with the largest saddle density is chosen. When there are no suitable neighbours, the structure does not stand out enough from the background and is discarded.

The peak density criterion can be thought of as a required signal-to-noise ratio. For example, we could require the peak of the structure to be at least twice as strong as the background density \texttt{rho\_thresh} to be distinguishable.
The saddle threshold criterion is responsible for merging substructures back together.
If the saddle density is \textit{higher} than \texttt{rho\_saddle}, we decide the structure is part of the neighbour with which it shares this saddle, rather than being an individual structure. When looking at clouds for example, we indeed expect a cloud to have substructure with multiple local density peaks inside one cloud. This criterion can also help to reduce differences when comparing simulations with different resolutions. When the resolution is higher, naturally more local density peaks are found and so we need to group them to recover the full cloud which envelopes multiple clumps. This also implies that the simulations must be able to resolve densities of at least this chosen saddle density.

Finally, we discard structures which consist of fewer than the minimum required number of cells. By using this criterion we can select structures that are sufficiently resolved.

The full process is illustrated in Figure~\ref{fig:hop_how_it_works}.

\subsection{Calculating structure properties}
\label{appx:struct_prop}

Once we have identified which cells/particles belong to which structure, we calculate the structure properties. Users of our wrapper can easily select what type of properties they want to calculate. One can also easily extend the code to add additional properties of interest. Here we list some elemental ones that are studied in this work.

\subsubsection{Position, mass, density and magnetic field}

The total mass $M$ is given by the sum of the masses of each particle or grid cell:
\begin{equation}
    M = \sum_i m_i  = \sum_i \rho_i V_i
\end{equation}
where $\rho_i$ is the density and $V_i$ is the volume of the cell $i$.
The average density is defined as 
\begin{equation}
    \rho = \frac{\sum_i m_i}{\sum_i V_i}.
\end{equation}
We also determine the maximum density inside the structure and the corresponding peak coordinates.
The centre-of-mass coordinates are given by
\begin{equation}
    \vec{x} = \sum_i m_i \vec{x}_{o,i} / M
\end{equation}
where we take into account periodic boundaries by first re-centring around the peak location (and assuming the structure is significantly smaller than the box size of the simulation).

\subsubsection{Size and shape}

Once we know the centre-of-mass, we construct the inertia matrix using the centre-of-mass coordinates
\begin{equation}
I = 
    \begin{bmatrix}
    \sum_i m_i (y_i^2 + z_i^2) & 
    - \sum_i m_i x_i y_i       &
    - \sum_i m_i x_i z_i\\
    - \sum_i m_i x_i y_i       &
    \sum_i m_i (x_i^2 + z_i^2) &
    - \sum_i m_i y_i z_i\\
    - \sum_i m_i x_i z_i       &
    - \sum_i m_i y_i z_i       &
    \sum_i m_i (x_i^2 + y_i^2)
    \end{bmatrix}
\end{equation}
and find its eigenvalues $\lambda_1 \leq \lambda_2 \leq \lambda_3$ through diagonalisation.
We approximate the shape of the clouds by considering the ellipsoid with an uniform density that has the same inertia matrix eigenvalues.
The half-axes $a \leq b \leq c$ of this ellipsoid are given by 
\begin{align}
\label{eq:halfaxes}
a &= \sqrt{\dfrac{5}{2 M} \left(\lambda_1 + \lambda_2 - \lambda_3\right)}\\
b &= \sqrt{\dfrac{5}{2 M} \left(\lambda_1 + \lambda_3 - \lambda_2\right)}\nonumber\\
c &= \sqrt{\dfrac{5}{2 M} \left(\lambda_2 + \lambda_3 - \lambda_1\right)}\nonumber
\end{align}
We tested visually that the corresponding ellipsoid matches well with the extent of the clouds.
While this gives good results, it is worth recalling that clouds have a much more complex shape compared to a simple ellipsoid.
We define the average size as the geometric mean of the three half-axes multiplied by 2,
\begin{equation}
\label{eq:cloud_size}
    L = 2 \left(a b c \right)^{1/3}.
\end{equation}
Note that this definition corresponds to a diameter, rather than a radius.

\subsubsection{Kinematic quantities}
The centre-of-mass velocity (a.k.a.\ the bulk velocity) is simply the mass-weighted average velocity of the cells in the structure:
\begin{equation}
    \vec{\Bar{v}} = \dfrac{\sum_i m_i \vec{v}_{i}}{M}.
\end{equation}
The velocity dispersion is obtained using
\begin{equation}
\label{eq:cloud_sigma}
    \sigma = \sqrt{ \dfrac{ \sum_i m_i \left|\vec{v}_{i}-\vec{\Bar{v}}\right|^2}{M} }.
\end{equation}

\subsection{Choosing merge parameters}
\label{appx:merge_params}

\begin{figure}
    \centering
    \includegraphics[width=0.9 \columnwidth]{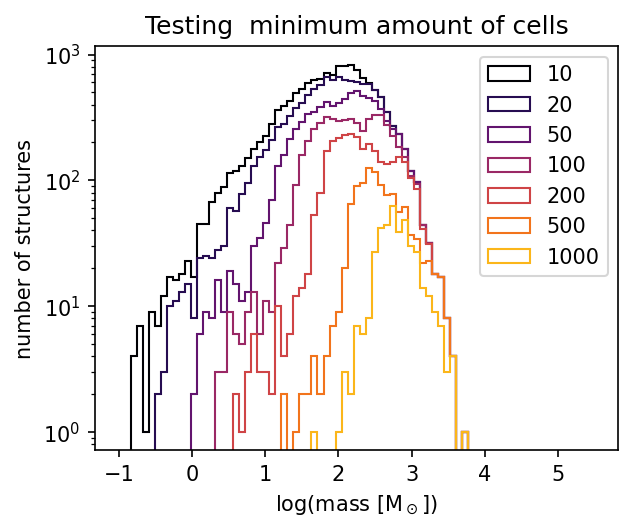}\\
    \includegraphics[width=0.9 \columnwidth]{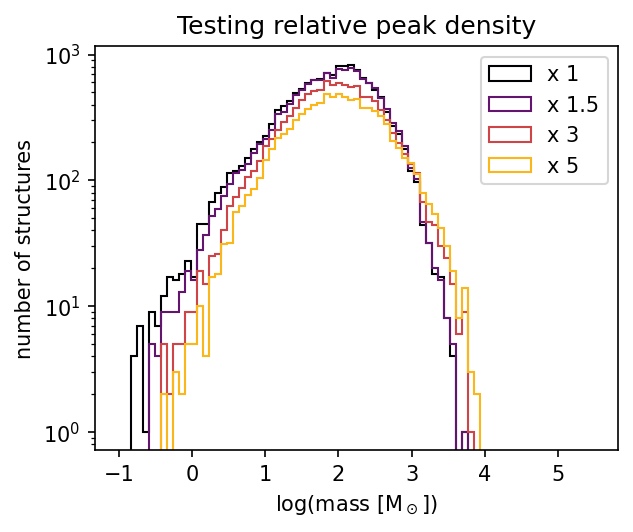}\\
    \includegraphics[width=0.9 \columnwidth]{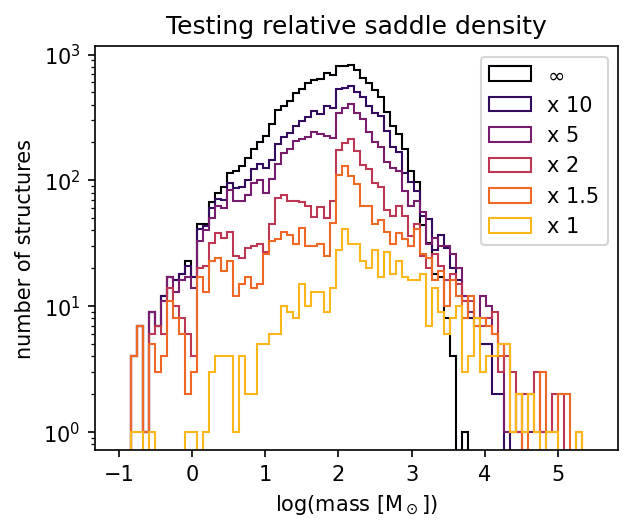}
    \caption{Effect of the merge parameters on the mass spectrum of the extracted clouds in \texttt{LS-weak-driving}.
    Top: Changing \texttt{min\_cells} while disabling peak and saddle merging.
    Middle: Varying the ratio \texttt{rho\_peak}/\texttt{rho\_thresh}, while keeping \texttt{min\_cells}=10 and disabling saddle merging. 
    Bottom: Varying the ratio \texttt{rho\_saddle}/\texttt{rho\_thresh} while keeping \texttt{min\_cells}=10 and disabling peak merging.
    }
    \label{fig:hop_merge_params}
\end{figure}

\begin{figure}
    \centering
    \includegraphics[width=\columnwidth]{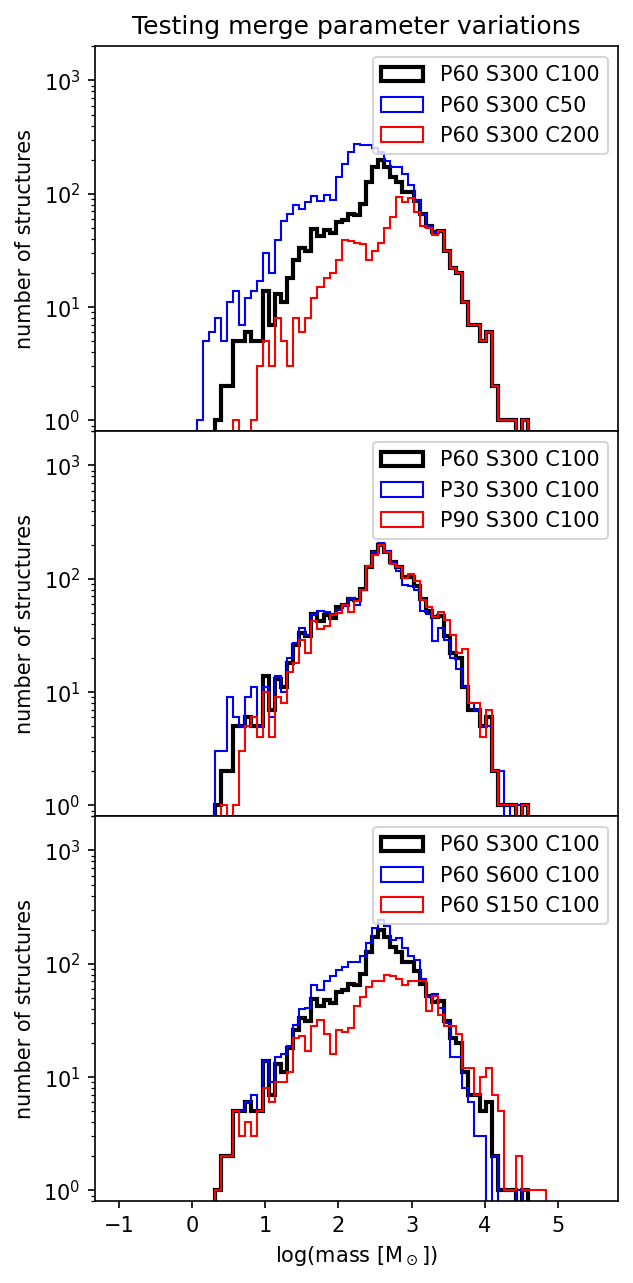}
    \caption{Effect on the mass spectrum of clouds in \texttt{LS-weak-driving} when varying the merge parameters around the chosen values: \texttt{min\_cells} (C) = 100, \texttt{rho\_peak} (P) = 2 x \texttt{rho\_thresh},   \texttt{rho\_saddle} (S) = 10 x \texttt{rho\_thresh}.}
    \label{fig:hop_merge_params_final}
\end{figure}

To determine suitable merge parameters, we first investigate the effect of each individual criterion on the extraction results. More specifically, we look at their influence on the cloud mass spectrum.
Figure~\ref{fig:hop_merge_params} shows the results of our tests on \texttt{LS-weak-driving}, a stratified box simulation run with \textsc{Ramses}.
Requiring a minimum number of cells for a structure means we only select the larger objects, as seen in the top panel. As long as we are not too demanding, we recover the same high mass tail. The middle panel shows that peak merging on its own has a minor effect. It merges small structures with low peak-to-background ratios into larger structures, producing a spectrum which is somewhat more top heavy.
Finally, the saddle merging has the largest effect on the high mass slope, as seen in the bottom panel. Setting the ratio to 1 means merging everything that touches. Almost all of the small and intermediate mass structures have been merged into a handful of massive cloud complexes. Disabling saddle merging altogether (i.e.\ choosing a very large value) results in a sharp cut-off which is basically determined by the resolution. Adding an appropriate saddle merging criterion with respect to the density threshold allows one to recover large structures while keeping a significant population of lower mass objects.

Now that we understand the impact of the merge parameters, we can make an educated choice. We want to obtain a catalogue of structures that are resolved, to avoid having to calculate the properties of unresolved structures close to the resolution limit. We therefore choose \texttt{min\_cells} = 100, which accurately recovers the high mass slope while neglecting a large amount of unresolved objects.
Since the peak density criterion plays only a minor role, we can apply any desired signal-to-noise ratio. As a default value, we set \texttt{rho\_peak} to be twice \texttt{rho\_thresh}.
The saddle threshold is the most difficult parameter to choose, and one might want to vary it depending on the application. Our tests show that a value of ten times \texttt{rho\_thresh} gives good results. This selects structures in a specified density range and allows for the comparison of simulations with different resolutions, provided the saddle density is resolved in the lowest resolution run.
Figure~\ref{fig:hop_merge_params_final} verifies the mass distribution does not change dramatically when slightly varying the chosen merge parameters.

\subsection{Testing resolution}

\begin{figure}
    \centering
    \includegraphics[width=\columnwidth]{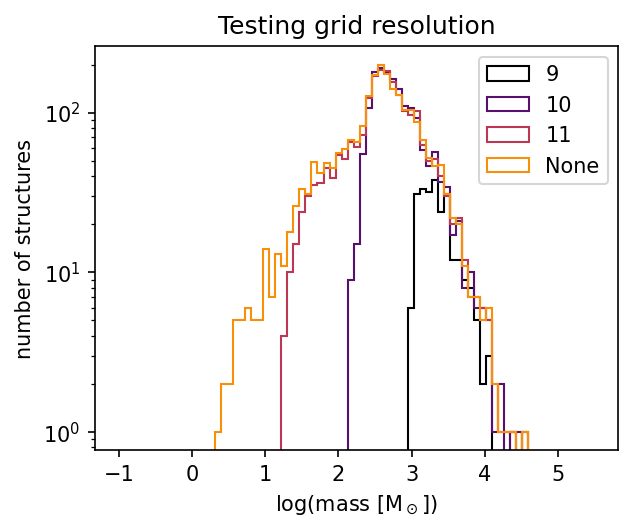}\\
    \caption{Study of the influence of resolution on the structure mass spectrum for \texttt{LS-weak-driving}. We limit the loading of the AMR tree to a certain refinement level, as indicated in the legend. Increasing the refinement by one level means the smallest spatial scale resolved becomes a factor two smaller and the volume of a cell is a factor 8 smaller.}
    \label{fig:hop_resolution_effect}
\end{figure}

In Figure~\ref{fig:hop_resolution_effect} we demonstrate that the high mass part of the cloud mass spectrum does not depend on the resolution.
The cut-off at low masses does depend on resolution, especially since the minimum requested number of cells is kept the same.

\subsection{Testing low density cutoff}

\begin{figure}
    \centering
    \includegraphics[width=\columnwidth]{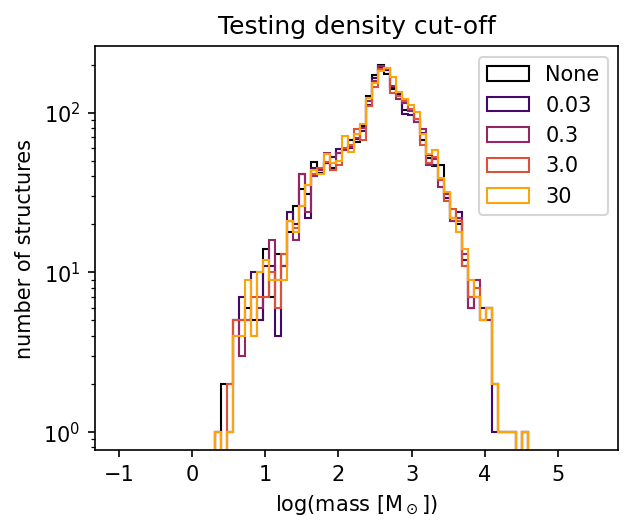}\\
    \caption{Study where we limit the input for \textsc{Hop} to cells above a certain density threshold relative to the extraction threshold. Resulting mass spectrum for \texttt{LS-weak-driving}.}
    \label{fig:hop_cutoff_effect}
\end{figure}

In Figure~\ref{fig:hop_cutoff_effect} we demonstrate that it is possible to recover the same cloud mass spectrum when neglecting low density parts of the simulation in the group finding phase of \textsc{Hop}. This is particularly interesting for simulations which contain many particles or cells and are thus memory heavy.

\subsection{Tests on idealised structures}
\label{subsec:hop_test_shape}

\begin{figure}
    \centering
    \includegraphics[width=0.7\columnwidth]{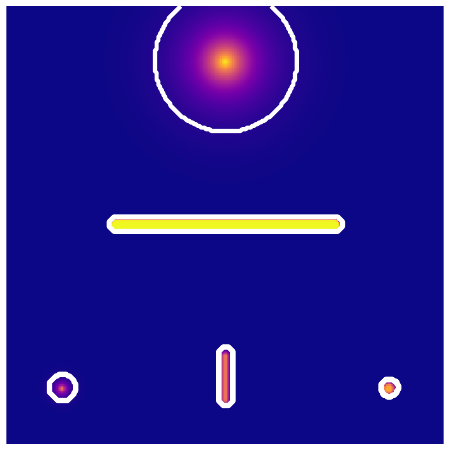}\\
    \caption{\textsc{Hop} detection of simple structures in a synthetic volume. Various spherical or filamentary shapes were correctly detected by \textsc{Hop}.}
    \label{fig:test_synthetic}
\end{figure}

To test how good \textsc{Hop} is at identifying structures and recovering shape information, we run the extraction algorithm on a synthetic volume populated by simple structures (spheres and filaments). \textsc{Hop} was able to correctly identify all structures and recover the shape parameters (see Section \ref{subsec:shape}). This is illustrated by Fig. \ref{fig:test_synthetic}. 

\section{Parameters of the galactic simulations}
\label{sec:gal_sim_params}

\begin{table*}
  \begin{center}
  \caption{Initial conditions for the full galaxy simulations, with $N_{\mathrm{part}}$  the resolution in number of particles, $H_i$ the scale height, $\sigma_i$ the initial velocity dispersion at a scale of $0.1 H_i$ (when relevant), $T_i$ the initial temperature, and $Q_{\mathrm{min}}$ the minimal value for the Toomre parameter. The gas particles are then used to generate the initial gas density, and the other are used directly.}\label{tab:galaxy_init}
	\begin{tabular}{|c|l|c|c|}
	\hline
	\multicolumn{2}{|c|}{Name}  & Ramses-F20 & M51   \\
	\hline
	\multicolumn{2}{|c|}{Total mass}  & $112 \E{10} \Msun$ & $71 \E{10} \Msun$ \\
	\hline
	\multirow{3}{2.6cm}{Dark matter halo} 
		& Profile & NFW sphere \citep{navarroStructureColdDark1996} & Spheroidal \citep{hernquistAnalyticalModelSpherical1990} \\
		& Mass & $107 \E{10}\ \Msun$ & $60.4 \E{10}\ \Msun$\\
		& $N_{\mathrm{part,DM}}$ &  $10^6$ &  $10^6$\\
\hline
	\multirow{4}{2.6cm}{Thin stellar disk}
	 & Mass & $3.4 \E{10}\ \Msun$  & $4.77 \E{10}\ \Msun $\\
	 & $N_{\mathrm{part,\star}}$ & $10^6$& $10^6$\\
	 & $H_{i,\star}$ & 3.4 kpc & 2.26 kpc\\
	 & $Q_{\mathrm{min}}$ & $1.5$ & $0.7$ \\
\hline
\multirow{6}{2.6cm}{Gaseous disk}
	 & Mass & $8.5 \E{9}  \Msun$  & $5.3 \E{9}\ \Msun $\ \\
	 & $N_{\mathrm{part,g}}$ & $10^6$ & $1.8 \E{8}$\\
	 & $T_i$ & $10^4$ K &  $10^4$ K \\
	 & $H_{i,g}$ & 3.4 kpc & 2.26 kpc \\
	 & $\sigma_i$ & $20$ km~s$^{-1}$ & $-$\\
\hline
\multirow{3}{2.6cm}{Stellar bulge}
	 & Mass & $4.2  \E{9}\  \Msun$ &  $5.3 \E{9}\  \Msun$  \\
	 & $N_{\mathrm{part,b\star}}$ & $125000$ & $10^5$\\
	 & $H_{i,b\star}$ & 0.34 kpc & 0.09 kpc\\
\hline

	\end{tabular} 

  \end{center}
\end{table*}

Table \ref{tab:galaxy_init} summarises the initial conditions used for the Ramses-F20 and the M51 simulations.

\section{Fit of the mass spectrum}
\label{appx:fit_mass_spectrum}

\begin{figure*}
    \centering
    \includegraphics[width=\textwidth]{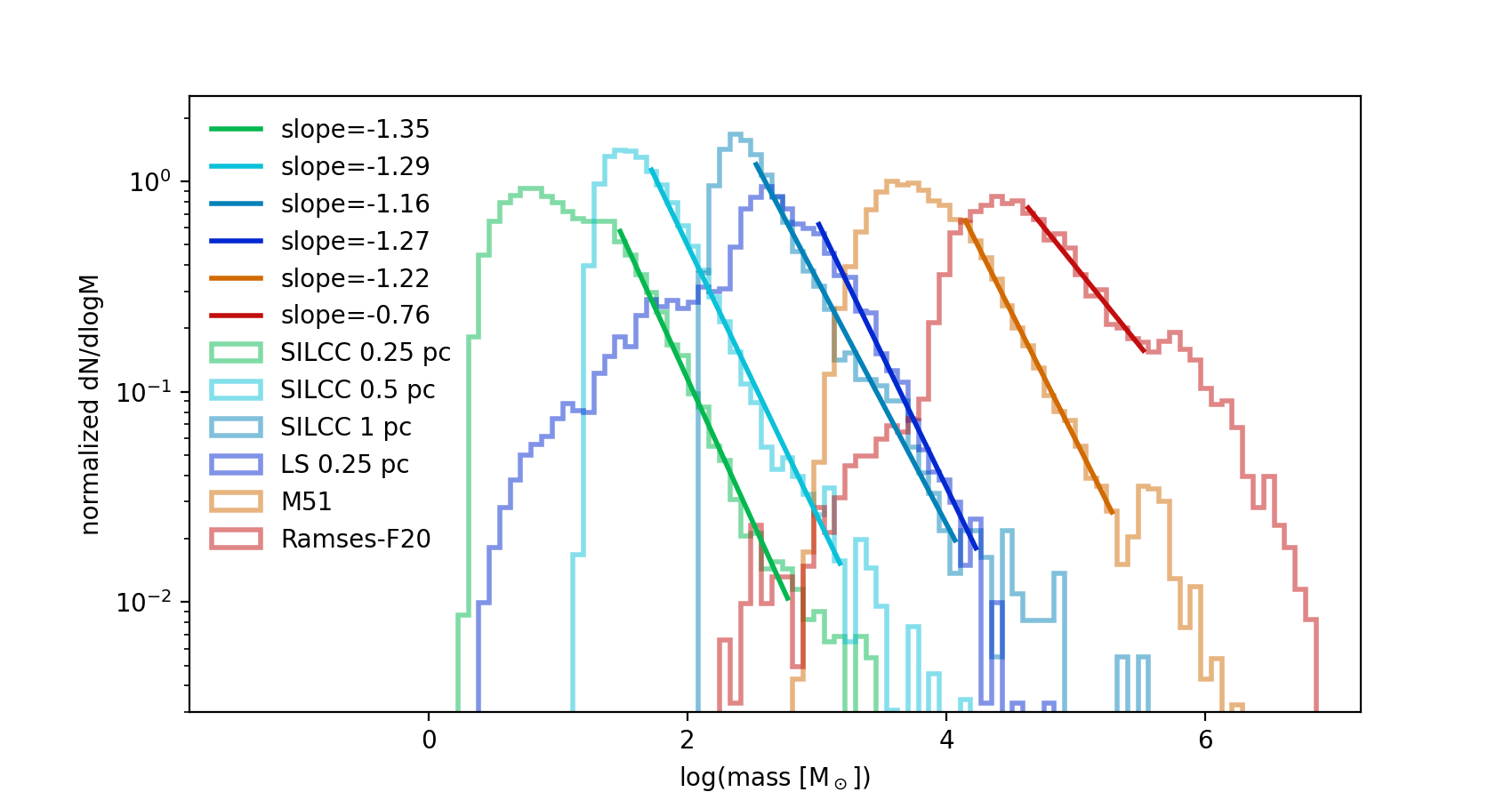}
    \caption{Fits of the individual cloud mass spectra}
    \label{fig:mass_fit}
\end{figure*}

We fitted the mass spectrum in the appropriate range for each individual simulation.

\section{Variations with galactic radius}
\label{appx:radial_dep}
\begin{figure}
    \centering
    \includegraphics[width=\columnwidth]{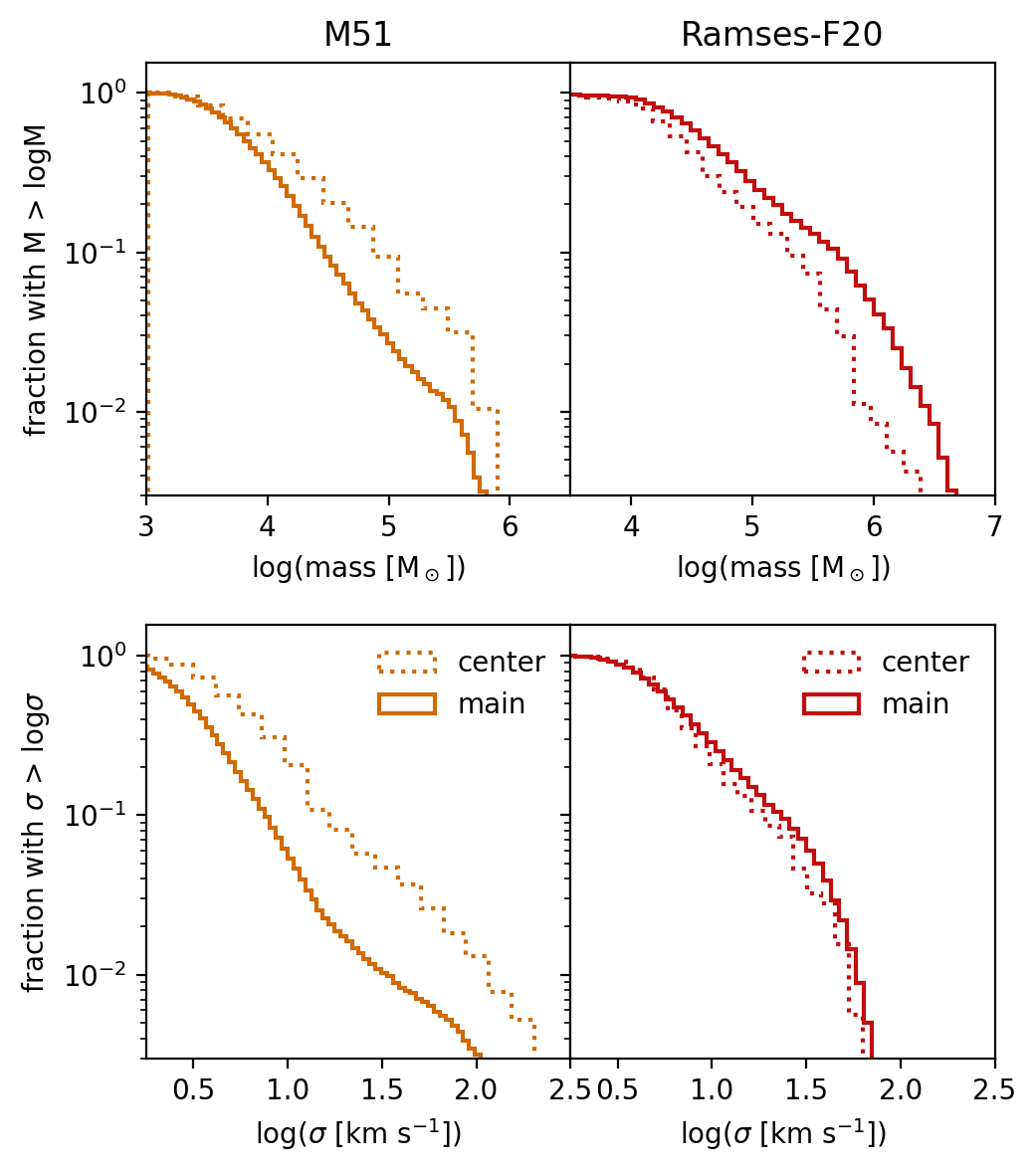}
    \caption{Comparison between key cloud properties in the centre of the galaxies versus those in the main body. The threshold between the two regions is taken at a galactic radius of 1 kpc. 
    We show the cumulative distribution of cloud mass (top) and cloud velocity dispersion (bottom).}
    \label{fig:radial_dependence}
\end{figure}

As mentioned in Section~\ref{sec:slope_variations} and shown in Figure~\ref{fig:mass_spectrum_inner_outer}, we did not find a difference in the mass spectrum for the inner versus outer galaxy in our two galaxy simulations, with a boundary between inner and outer zone set at 7 kpc.
However, \citet{Tress2021} did find that the masses of clouds in the very centre of the simulated M51 galaxy are typically larger.
In Figure~\ref{fig:radial_dependence} we compare the cumulative distribution for clouds in the inner 1 kpc of our galaxies versus that of clouds located in the remainder of the galaxy.
For the cloud mass (top panel) in M51, we recover the qualitative result of \citet{Tress2021}. Clouds also have higher velocity dispersions, as can be seen in the bottom panel.
The Ramses-F20 galaxy on the other hand, has slightly less massive clouds in the centre and the distribution of velocity dispersions is very similar in both sub-regions.
We caution that the statistics in the central region of both galaxies are rather poor.

The weight of the stellar and dark matter component is higher in the centre than in the outskirts. We would expect this to influence the mean statistical properties of the clouds. 
A more efficient channelling of the gas along the deeper galactic potential in the centre could lead to more massive structures. However this effect is modulated by other aspects, such as higher turbulence and shear which can easily disrupt large clouds.
The gas fraction, i.e. $M_\mathrm{gas}/(M_\mathrm{gas}+M_\mathrm{stars})$, of the M51 simulation is overall lower than that of Ramses-F20 (see Table~\ref{tab:galaxy_init}). As a consequence, the impact of the weight of the non-gaseous components is higher in M51. This could provide a possible explanation for the different behaviour of the mass spectrum in both galaxies.

Another key difference between the two galaxies is their morphology.
In particular, M51 has a central bar while Ramses-F20 does not (see Figure~\ref{fig:coldens-sigma}).
Observations have shown that gas in the centre of barred galaxies has an increased velocity disperion \citep{Sun_et_al2020}. 
We indeed recover this behaviour in the bottom left panel of Figure~\ref{fig:radial_dependence}.

\end{appendix}

\end{document}